\documentclass[10pt,twoside]{article}
\usepackage[american]{babel}
\usepackage[utf8]{inputenc}
\usepackage{authblk}

\usepackage{amsmath,amsfonts,amssymb,amsthm}
\usepackage{paralist}
\usepackage{url,xspace}
\usepackage[ruled,vlined]{algorithm2e}
\usepackage{subcaption}
\usepackage{graphicx,color,psfrag}
\graphicspath{{fig/}}
\usepackage{pstool}
\usepackage{multirow}
\usepackage{verbatim,ifthen}
\usepackage{tikz}
\usepackage{todonotes}
\usepackage{cite}

\newtheorem{theorem}{Theorem}

\newcommand{\elim}{\mathit{elim}}
\newcommand{\colelim}{\mathit{\textit{col-elim}}}

\newcommand{\QRTT}{\textsc{QRFlatTT}\xspace}
\newcommand{\QRTS}{\textsc{QRFlatTS}\xspace}
\newcommand{\QRGreedy}{\textsc{QRGreedy}\xspace}
\newcommand{\QRBinomial}{\textsc{QRBinomial}\xspace}
\newcommand{\bidiagBinomial}{\textsc{BiDiagBinomial}\xspace}
\newcommand{\binomial}{\textsc{Binomial}\xspace}
\newcommand{\ratio}{\ensuremath{\beta}\xspace}
\newcommand{\rratio}{\ensuremath{\gamma}\xspace}
\newcommand{\nbcores}{\ensuremath{\mathit{nb}_{\mathit{cores}}}\xspace}
\newcommand{\nbnodes}{\ensuremath{\mathit{nb}_{\mathit{nodes}}}\xspace}

\newcommand{\GEQRT}{\ensuremath{\mathit{GEQRT}}\xspace}
\newcommand{\TSQRT}{\ensuremath{\mathit{TSQRT}}\xspace}

\newcommand{\UNMQR}{\ensuremath{\mathit{UNMQR}}\xspace}
\newcommand{\TSMQR}{\ensuremath{\mathit{TSMQR}}\xspace}
\newcommand{\TTQRT}{\ensuremath{\mathit{TTQRT}}\xspace}
\newcommand{\TTMQR}{\ensuremath{\mathit{TTMQR}}\xspace}

\newcommand{\Greedy}{\textsc{Greedy}\xspace}
\newcommand{\FlatTS}{\textsc{FlatTS}\xspace}
\newcommand{\FlatTT}{\textsc{FlatTT}\xspace}
\newcommand{\Auto}{\textsc{Auto}\xspace}
\newcommand{\bidiagTS}{\textsc{BiDiagFlatTS}\xspace}
\newcommand{\bidiagTT}{\textsc{BiDiagFlatTT}\xspace}
\newcommand{\bidiagGreedy}{\textsc{BiDiagGreedy}\xspace}
\newcommand{\rbidiagTS}{\textsc{R-BiDiagFlatTS}\xspace}
\newcommand{\rbidiagTT}{\textsc{R-BiDiagFlatTT}\xspace}
\newcommand{\rbidiagGreedy}{\textsc{R-BiDiagGreedy}\xspace}

\newcommand{\scalapack}{\textsc{ScaLAPACK}\xspace}
\newcommand{\lapack}{\textsc{LAPACK}\xspace}
\newcommand{\bidiag}{\textsc{BiDiag}\xspace}
\newcommand{\rbidiag}{\textsc{R-BiDiag}\xspace}
\newcommand{\parsec}{\textsc{PaRSEC}\xspace}
\newcommand{\dplasma}{\textsc{DPLASMA}\xspace}
\newcommand{\Grasap}{\textsc{Grasap}\xspace}
\DeclareUnicodeCharacter{00A0}{ }

\newboolean{withproof}
\setboolean{withproof}{false}
\newboolean{withoutproof}
\setboolean{withoutproof}{true}
\newcommand{\forRR}[1]{\ifthenelse{\boolean{withproof}}{{#1}}{}}
\newcommand{\forSCpaper}[1]{\ifthenelse{\boolean{withoutproof}}{{#1}}{}}

\author[1,2]{Mathieu Faverge}
\author[3]{Julien Langou}
\author[2,4]{Yves Robert}
\author[2,5]{Jack Dongarra}
\affil[1]{\footnotesize Bordeaux INP, CNRS, INRIA et Universit\'e de Bordeaux, France}
\affil[2]{\footnotesize University of Tennessee, Knoxville TN, USA}
\affil[3]{\footnotesize University of Colorado Denver, USA}
\affil[4]{\footnotesize Laboratoire LIP, \'Ecole Normale Sup\'erieure de Lyon et INRIA, France}
\affil[5]{\footnotesize University of Manchester, UK}

\title{Bidiagonalization with Parallel Tiled Algorithms}

\begin{document}
\maketitle

\begin{abstract}
We consider algorithms for going from a ``full'' matrix to a condensed
``band bidiagonal'' form using orthogonal transformations.
We use the framework of ``algorithms by tiles''.
Within this framework, we study: (i) the tiled
bidiagonalization algorithm \bidiag, which is a tiled version of the
standard scalar bidiagonalization algorithm; and
(ii) the R-bidiagonalization algorithm \rbidiag, which
is a tiled version of the algorithm which
consists in first performing the QR factorization of
the initial matrix, then performing the band-bidiagonalization of the R-factor.
For both bidiagonalization algorithms \bidiag and \rbidiag,
we use four main
types of reduction trees, namely \FlatTS, \FlatTT, \Greedy,
and a newly introduced auto-adaptive tree, \Auto.
We provide a study of critical path lengths for these tiled algorithms, which shows
that (i) \rbidiag has a shorter critical path length than
\bidiag for tall and skinny matrices, and (ii) \Greedy based schemes are 
much better than earlier proposed variants with unbounded resources.
We provide experiments on a single multicore node, and on a few multicore nodes of a parallel distributed
shared-memory system, to show the superiority of the new algorithms on a
variety of matrix sizes, matrix shapes and core counts.
\end{abstract}

\section{Introduction}

This work is devoted to the design and comparison of tiled algorithms for the
bidiagonalization of large matrices. Bidiagonalization is a widely used kernel
that transforms a full matrix into bidiagonal form using orthogonal
transformations. The importance of bidiagonalization stems from the fact that,
in many algorithms, the bidiagonal form is a critical step to compute the
singular value decomposition (SVD) of a matrix. The necessity of computing the
SVD is present in many computational science and engineering areas. Based on
the Eckart–Young theorem~\cite{EckartYoung36}, we know that the singular
vectors associated with the largest singular values represent the best way (in
the 2-norm sense) to approximate the matrix.  This approximation result leads
to many applications, since it means that SVD can be used to extract the ``most
important'' information of a matrix. We can use the SVD for compressing data or
making sense of data. In this era of Big Data, we are interested in very large
matrices. To reference one out of many application, SVD is needed for
principal component analysis (PCA) in Statistics, a widely used method in
applied multivariate data analysis.

We consider algorithms for going from a ``full'' matrix to a condensed ``band
bidiagonal'' form using orthogonal transformations.  We use the framework of
``algorithms by tiles''.  Within this framework, we study: (i) the tiled
bidiagonalization algorithm \bidiag, which is a tiled version of the standard
scalar bidiagonalization algorithm; and (ii) the R-bidiagonalization algorithm
\rbidiag, which is a tiled version of the algorithm which consists in first
performing the QR factorization of the initial matrix, then performing the
band-bidiagonalization of the R-factor.  For both bidiagonalization algorithms
\bidiag and \rbidiag, we use HQR-based reduction trees, where HQR stands for the Hierarchical
QR factorization of a tiled matrix~\cite{j125}. Considering various
reduction trees gives us the flexibility to adapt to matrix shape and machine
architecture.
In this work, we
consider many types of reduction trees. In shared memory, they are named \FlatTS, \FlatTT, \Greedy,
and a newly introduced auto-adaptive tree, \Auto. In distributed memory, they are somewhat more complex and take into account the topology of the machine.
The main contributions are the following:
\begin{itemize}

\item The design and comparison of the \bidiag and \rbidiag tiled algorithms with many types of reduction trees.
There is considerable novelty in this. 
Previous work~\cite{4967575,Ltaief:2013:HBR:2450153.2450154,6267821,Haidar:2013:IPS:2503210.2503292}
on tiled bidiagonalization has only considered one type of tree (\FlatTS tree)
with no \rbidiag.  Previous work~\cite{Ltaief2012} has considered \Greedy trees
only for the QR steps in  \bidiag (so $m$ of the $m+n$ steps), it considers \FlatTS
tree for the LQ steps ($n$ of the $m+n$ steps), and it does not consider
\rbidiag. This paper is the first to study \rbidiag for tiled bidiagonalization
algorithm and to study \Greedy trees for both steps (LQ and QR) of the tiled
bidiagonalization algorithm.
\item A detailed study of critical path lengths for 
\FlatTS, \FlatTT, \Greedy with \bidiag and \rbidiag (so six different algorithms in total), which shows that:
\begin{itemize}
\item The newly-introduced \Greedy based schemes (\bidiag and \rbidiag) are 
much better than earlier proposed variants with unbounded resources
and no communication: for matrices of $p \times q$ tiles, $p\geq q$, their critical paths have a length $\Theta(q \log_{2}(p))$ instead of $\Theta(p q)$ for \FlatTS and \FlatTT
\item On the one hand, \bidiagGreedy  has a shorter critical path length than \rbidiagGreedy for square matrices; on the other hand, 
\rbidiagGreedy has a shorter critical path length than
\bidiagGreedy for ``tall and skinny'' matrices.
For example, for a $p \times q$ tile matrix, when $q$ go to infinity, and $p=\ratio q^{1+\alpha}$, with $0 \leq \alpha < 1$, then
the asymptotic ratio between \bidiagGreedy  and \rbidiagGreedy
is $\frac{1}{1+\frac{\alpha}{2}}$.
\end{itemize}

\item Implementation of our algorithms within the 
\dplasma framework~\cite{dplasma-6008998}, 
which runs on top of
the \parsec runtime system~\cite{dague-engine-Bosilca201237}, and
which enables parallel
distributed experiments on multicore nodes.  All previous tiled
bidiagonalization study~\cite{4967575,Ltaief:2013:HBR:2450153.2450154,6267821,Haidar:2013:IPS:2503210.2503292,Ltaief2012}
were limited to shared memory implementation.

\item A practical auto-adaptative tree (\Auto) that self-tunes for increased performance.
This tree is especially useful in many practical situations. For example, it is appropriate
(1) 
when the critical path length is not a major consideration due to limited number of resources or (2) when the intra-node ``communication'' is expensive
in which TS kernels are much faster than TT kernels or (3) both.

\item Experiments on a single multicore node, and on a few multicore nodes of a
parallel distributed shared-memory system, show the superiority of the new
algorithms on a variety of matrix sizes, matrix shapes and core counts. The new
\Auto algorithm outperforms its competitors in almost every test case, hence
standing as the best algorithmic choice for most users.

\end{itemize}

The rest of the paper is organized as follows. Section~\ref{sec.related}
provides a detailed overview of related work. Section~\ref{sec.algos} describes
the \bidiag and \rbidiag algorithms with the \FlatTS, \FlatTT and \Greedy
trees.  Section~\ref{sec.critical} is devoted to the analysis of the critical
paths of all variants.  Section~\ref{sec.dague} outlines our implementation,
and introduces the new \Auto reduction tree.  Experimental results are reported
in Section~\ref{sec.expes}. Finally, conclusion and hints for future work are
given in Section~\ref{sec.conclusion}.

\section{Related Work}
\label{sec.related}

This section provides an overview of the various approaches to compute the
singular values of a matrix, and positions our new algorithm with respect to existing 
numerical software kernels.\\
  
{\bf Computing the SVD.}
The SVD of a matrix is a fundamental matrix
factorization and is a basic tool used in many applications. Computing
the SVD of large matrices in an efficient and scalable way,  is an important
problem that has gathered much attention. The matrices considered here are
rectangular $m$-by-$n$. Without loss of generality, we consider $m\geq n$.
We call GE2VAL the problem of computing (only) the singular values of
a matrix, and GESVD the problem of computing the singular values and the
associated singular vectors.\\

{\bf From full to bidiagonal form.}
There are few algorithms to compute the singular value decomposition. A large
class of these algorithms consists in first reducing the matrix to bidiagonal
form with orthogonal transformations (GE2BD step), then processing the bidiagonal
matrix to obtain the sought singular values (BD2VAL step).
These two steps (GE2BD and BD2VAL) are very different in nature.  GE2BD can be
done in a known number of operations and has no numerical difficulties. On the
other hand, BD2VAL requires the convergence of an iterative process and is
prone to numerical difficulties.  This paper mostly focuses on GE2BD: reduction
from full to bidiagonal form.
Clearly, GE2BD+BD2VAL solves GE2VAL: computing (only) the
singular value of a matrix.  If the singular vectors are desired (GESVD), one
can also compute them by accumulating the ``backward'' transformations; in this
example, this would consist in a VAL2BD step followed by a BD2GE step.\\

In 1965, Golub and Kahan~\cite{doi:10.1137/0702016} provides a singular value
solver based on an initial reduction to bidiagonal form. In~\cite[Th.
  1]{doi:10.1137/0702016}, the GE2BD
step is done using a QR step on the first
column, then an LQ step on the first row, then a QR step on the second column,
etc. The steps are done one column at a time using Householder transformation.
This algorithm is implemented as a Level-2 BLAS algorithm in \lapack as~\texttt{xGEBD2}.
For an $m$-by-$n$ matrix, the cost of this algorithm is (approximately) $4m n^{2} - \frac{4}{3}n^3$.\\

{\bf Level 3 BLAS for GE2BD.} 
In 1989, Dongarra, Sorensen and Hammarling~\cite{DONGARRA1989215} explains how
to incorporate Level-3 BLAS in \lapack~\texttt{xGEBD2}. The idea is to compute few
Householder transformations in advance, and then to accumulate and apply
them in block using the WY transform~\cite{doi:10.1137/0908009}.  This
algorithm is available in \lapack (using the compact WY
transform~\cite{doi:10.1137/0910005}) as~\texttt{xGEBRD}.  Gro{\ss}er and
Lang~\cite[Table 1]{GroBer:1999:EPR:330506.330509} explain that this algorithm
performs (approximately) 50\% of flops in Level 2 BLAS (computing and accumulating Householder
vectors) and 50\% in Level 3 BLAS (applying Householder vectors).
In 1995, Choi, Dongarra and Walker~\cite{Choi1995} presents the \scalapack
version, \texttt{PxGEBRD}, of the \lapack \texttt{xGEBRD} algorithm of~\cite{DONGARRA1989215}.\\

{\bf Multi-step approach.} Further improvements for GE2BD (detailed thereafter) are
possible.  These improvements rely on combining multiple steps. These
multi-step methods will perform in general much better for GE2VAL
(when only singular
values are sought) than for GESVD (when singular values and singular vectors are sought).
When singular values and singular vectors are sought, all the ``multi'' steps have to be
performed in ``reverse'' on the singular vectors adding a non-negligible
overhead to the singular vector computation.\\

{\bf Preprocessing the bidiagonalization with a QR factorization (preQR step).}
In 1982, Chan~\cite{Chan:1982:IAC:355984.355990} explains that, for
tall-and-skinny matrices, in order to perform less flops, one can pre-process
the bidiagonalization step (GE2BD) with a QR factorization.  In other words,
Chan propose to do preQR($m$,$n$)+GE2BD($n$,$n$) instead of GE2BD($m$,$n$).  A
curiosity of this algorithm is that it introduces nonzeros where zeros were
previously introduced; yet, there is a gain in term of flops. Chan proves that
the crossover points when preQR($m$,$n$)+GE2BD($n$,$n$) performs less flops
than GE2BD($m$,$n$) is when $m$ is greater than $\frac{5}{3}n$. Chan also
proved that, asymptotically, preQR($m$,$n$)+GE2BD($n$,$n$) will perform half
the flops than GE2BD($m$,$n$) for a fixed $n$ and $m$ going to infinity.  If
the singular vectors are sought, preQR has more overhead: (1) the crossover
point is moved to more tall-and-skinny matrices, and there is less gain; also
(2) there is some complication as far as storage goes.

The preQR step was somewhat known by the community and an earlier reference is
for example the book of Lawson and Hanson
1974~\cite{doi:10.1137/1.9781611971217}.  For this reason, the fact of
pre-processing the bidiagonalization by a QR factorization (preQR) is referred
to as the LHC (Lawson, Hanson and Chan) algorithm in~\cite{Trefethen:1997:NLA}.\\

{\bf Three-step approach: partialGE2BD+preQR+GE2BD.} In 1997, Trefethen and
Bau~\cite{Trefethen:1997:NLA} present a ``three-step'' approach. The idea is
to first start a bidiagonalization; then whenever the ratio $\frac{m}{n}$ passes the
$\frac{5}{3}$ mark, switch to a QR factorization; and then finish up with GE2BD.\\

{\bf Two-step approach: GE2BND+BND2BD.} 
In 1999, Gro{\ss}er and Lang~\cite{GroBer:1999:EPR:330506.330509} studied a
two-step approach for GE2BD: (1) go from full to band (GE2BND), (2) then go
from band to bidiagonal (BND2BD).  In this scenario, GE2BND has most of the
flops and performs using Level-3 BLAS kernels; BND2BD is not using Level-3 BLAS
but it executes much less flops and operates on a smaller data footprint that
might fit better in cache.  There is a trade-off for the bandwidth to be
chosen. If the bandwidth is too small, then the first step (GE2BND) will have
the same issues as GE2BD. If the bandwidth is too large, then the second step
BND2BD will have many flops and dominates the run time. 

As mentioned earlier, when the singular vectors are sought (GESVD), the
overhead of the method in term of flops is quite large. In 2013, Haidar,
Kurzak, and Luszczek~\cite{Haidar:2013:IPS:2503210.2503292} reported that
despite the extra flops a two-step approach leads to speedup even when
singular vectors are sought (GESVD).\\

{\bf Tiled Algorithms for the SVD.} In the context of massive parallelism, 
and of reducing data movement, many dense linear algebra algorithms have been moved
to so-called ``tiled algorithms''. In the tiled algorithm framework, algorithms
operates on \emph{tiles} of the matrix, and tasks are scheduled thanks to a
runtime. In the context of the SVD, tiled algorithms naturally leads to band
bidiagonal form.  In 2010, Ltaief, Kurzak and Dongarra~\cite{4967575} present
a tiled algorithm for GE2BND (to go from full to band bidiagonal form).
In 2013 (technical report in 2011), Ltaief, Luszczek,
Dongarra~\cite{Ltaief:2013:HBR:2450153.2450154} add the second step (BND2BD)
and present a tiled algorithm for GE2VAL using GE2BND+BND2BD+BD2VAL.  
In 2012, Ltaief, Luszczek, and Dongarra~\cite{Ltaief2012} improve the algorithm
for tall and skinny matrices by using 
``any'' tree instead of flat trees in the QR steps.
In 2012, Haidar,
Ltaief, Luszczek and Dongarra~\cite{6267821} improve the BND2BD step
of~\cite{Ltaief:2013:HBR:2450153.2450154}.  Finally, in 2013, Haidar,
Kurzak, and Luszczek~\cite{Haidar:2013:IPS:2503210.2503292}  consider the
problem of computing singular vectors (GESVD) by performing\\
\centerline{{\footnotesize GE2BND+BND2BD+BD2VAL+VAL2BD+BD2BND+BND2GE}.}\\
They show that the two-step approach
(from full to band, then band to bidiagonal) can be successfully used not only for
computing singular values, but also for computing singular
vectors.\\

{\bf BND2BD step.} The algorithm in \lapack for BND2BD is xGBBRD.
In 1996, Lang~\cite{LANG19961} improved the sequential version of the algorithm and developed a parallel distributed algorithm.
Recently, PLASMA released an efficient multi-threaded implementation~\cite{6267821,Ltaief:2013:HBR:2450153.2450154}.
We also note that Rajamanickam~\cite{piroband} recently worked on this step.\\

{\bf BD2VAL step.} Much research has been done and is done on this kernel.
Much software exists.
In \lapack, to
compute the singular values and optionally the singular vectors of 
a bidiagonal matrix,  
the routine \texttt{xBDSQR} uses the Golub-Kahan QR algorithm~\cite{doi:10.1137/0702016};
the routine \texttt{xBDSDC} uses the divide-and-conquer algorithm~\cite{doi:10.1137/S0895479892242232}; and 
the routine \texttt{xBDSVX} uses bisection and inverse iteration algorithm.
Recent research was trying to apply the MRRR (Multiple Relatively Robust Representations) method~\cite{Willems2006} to the problem.\\

{\bf BND2BD+BD2VAL steps in this paper.} 
This paper does not focus neither on BND2BD nor BD2VAL.  As far as we
are concerned, we can use any of the methods mentioned above. The faster these
two steps are, the better for us.  For this study, during the experimental
section, for BND2BD, we use the PLASMA multi-threaded
implementation~\cite{6267821,Ltaief:2013:HBR:2450153.2450154} and, for BD2VAL, we use LAPACK
\texttt{xBDSQR}.\\

{\bf Intel MKL.}
We note that, since 2014 and Intel MKL 11.2, the Intel MKL library
is much faster to compute GESVD~\cite{intel}. Much of the literature on tiled
bidiagonalization algorithm was before 2014, and so it was comparing to much
slower version of MKL that is now available.  The reported speedup in this
manuscript for the tiled algorithms over MKL, are therefore much less impressive
than previously reported.

We also note that, while there was a great performance improvement for
computing GESVD~\cite{intel} from Intel MKL 11.1 to Intel MKL 11.2, there was
no performance improvement for GEBRD. The reason is that the interface for
GEBRD prevents a two-step approach, while the interface of GESVD allows for a
two-step approach.  Similarly, our two-step algorithm does not have the same
interface as LAPACK GEBRD. Consequently, it would not be fair to compare our
algorithm with LAPACK GEBRD or Intel MKL GEBRD because our algorithm does not
have a native LAPACK GEBRD interface. (I.e., it will not produce the
Householder vector column by column as requested by the LAPACK interface.) For
this reason, we will always compare with GESVD, and not GEBRD.\\

{\bf Approach to compute the SVD without going by bidiagonal form.}
We note that going to bidiagonal form is not a necessary step to compute the
singular value decomposition of a matrix.  There are approaches to compute the
SVD of a matrix which avoid the reduction to bidiagonal form.  In the ``direct
method'' category, we can quote Jacobi
SVD~\cite{doi:10.1137/050639193,doi:10.1137/05063920X},
QDWHSVD~\cite{doi:10.1137/120876605}; one can also think to think the singular
value decomposition directly from the block bidiagonal form although not much
work has been done in this direction so far.  Also,  we can also refer to all
``iterative methods'' for computing the singular values. The goal of these
iterative methods is in general to compute a few singular triplets.
As far as software packages, we can
quote
SVDPACK~\cite{SVDPACK},
PROPACK~\cite{PROPACK}, and
PRIMME SVD~\cite{PRIMME_SVDS}. As far methods, we can quote~\cite{doi:10.1137/140979381,doi:10.1137/S1064827500372973,doi:10.1137/S0895479802404192}.\\

{\bf Connection with Symmetric Eigenvalue Problem.} Many ideas (but not all)
can be applied (and have been applied) to symmetric tridiagonalization. This
paper only focuses on bidiagonalization algorithm but we believe that some of
the tricks we show (but not all) could be used for creating an efficient
symmetric tridiagonalization.

\section{Tiled Bidiagonalization Algorithms}
\label{sec.algos}

In this section, we provide background on tiled algorithms for QR factorization,
and then explain how to use them for the \bidiag and \rbidiag algorithms.

\subsection{Tiled algorithms}
\label{sec.tiled}

Tiled algorithms are expressed in terms of tile operations rather than
elementary operations.  Each tile is of size $n_b \times n_b$, where $n_b$ is a
parameter tuned to squeeze the most out of arithmetic units and memory
hierarchy. Typically, $n_b$ ranges from $80$ to $200$ on state-of-the-art
machines~\cite{sc09-agullo}. 

Consider a rectangular tiled matrix $A$ of size $p \times q$.
The actual size of $A$ is thus $(p n_{b}) \times (q n_{b})$, where $n_{b}$ is the tile size.
We number rows from $1$ to $p$, columns from $1$ to $q$, and factorization
steps from $k=1$ to $k=q$. Algorithm~\ref{alg.QR} outlines a tiled QR
algorithm, where loop indices represent tiles:

\begin{algorithm}[htbp]
  \DontPrintSemicolon
  \For{$\textnormal{k} = 1$ to $\min(p,q)$}{
     Step $k$, denoted as $QR(k)$:\\
     \For{$\textnormal{i} = k+1$ to $p$}{
     $\elim(i, piv(i,k), k)$
    }
  }
\caption{$QR(p,q)$ algorithm for a tiled matrix of size $(p,q)$.}
\label{alg.QR}
\end{algorithm}

In Algorithm~\ref{alg.QR}, $k$ is the step, and also the panel index, and $\elim(i, piv(i,k), k)$
is an orthogonal transformation that combines rows $i$ and $piv(i,k)$ to zero
out the tile in position $(i,k)$. We explain below how such orthogonal transformations
can be implemented.

\subsection{Kernels}

To implement a given orthogonal transformation $\elim(i, piv(i,k),
k)$, one can use six different kernels, whose costs are given in
Table~\ref{tab.kernels}.  In this table, the unit of time is the time to
perform $\frac{n_b^3}{3}$ floating-point operations.

\begin{table*}
\centering
\begin{tabular}{|c||c|c||c|c|}
  \hline
  Operation & \multicolumn{2}{|c||}{Panel} & \multicolumn{2}{c|}{Update} \\ \hline
  & Name & Cost & Name & Cost \\ \hline
  Factor square into triangle        & \GEQRT & 4 & \UNMQR & 6  \\ \hline
  Zero square with triangle on top   & \TSQRT & 6 & \TSMQR & 12 \\ \hline
  Zero triangle with triangle on top & \TTQRT & 2 & \TTMQR & 6  \\ \hline
\end{tabular}
\caption{Kernels for tiled QR. The unit of time is $\frac{n_b^3}{3}$, where $n_b$ is the blocksize.}
\label{tab.kernels}
\end{table*}

There are two main possibilities to implement an orthogonal transformation $\elim(i,
piv(i,k), k)$: The first version eliminates tile $(i,k)$ with the \emph{TS (Triangle
on top of square)} kernels, as shown in Algorithm~\ref{alg.elimSQ}:

\begin{algorithm}[htbp]
  \DontPrintSemicolon
  $\GEQRT(piv(i,k), k)$\\
  $\TSQRT(i,piv(i,k), k)$\\
  \For{$\textnormal{j} = k+1$ to $q$}{
     $\UNMQR(piv(i,k), k, j)$\\
     $\TSMQR(i, piv(i,k), k, j)$
  }
\caption{Elimination $\elim(i, piv(i,k), k)$ via \emph{TS (Triangle on top of square)} kernels.}
\label{alg.elimSQ}
\end{algorithm}

Here the tile panel $(piv(i,k), k)$ is factored into a triangle (with \GEQRT).
The transformation is applied to subsequent tiles $(piv(i,k),j)$, $j>k$, in row
$piv(i,k)$ (with \UNMQR). Tile $(i,k)$ is zeroed out (with \TSQRT), and
subsequent tiles $(i,j)$, $j>k$, in row $i$ are updated (with \TSMQR). The flop
count is $4+6+(6+12)(q-k)=10+18(q-k)$ (expressed in same time unit as in
Table~\ref{tab.kernels}).  Dependencies are the following:
\[
\begin{array}{ll}%
\GEQRT(piv(i,k), k) \prec \TSQRT(i, piv(i,k), k)\\%
\GEQRT(piv(i,k), k) \prec \UNMQR(piv(i,k), k, j) & \text{ for } j>k\\%
\UNMQR(piv(i,k) ,k, j) \prec \TSMQR(i, piv(i,k), k, j) & \text{ for } j>k\\%
\TSQRT(i,piv(i,k),k) \prec \TSMQR(i, piv(i,k), k, j) & \text{ for } j>k%
\end{array}%
\]
$\TSQRT(i, piv(i,k), k)$ and $\UNMQR(piv(i,k), k, j)$ can be executed in
parallel, as well as \UNMQR operations on different columns $j, j' >k$. With an
unbounded number of processors, the parallel time is thus $4+6+12=22$
time-units.

\begin{algorithm}[htbp]
  \DontPrintSemicolon
  $\GEQRT(piv(i,k), k)$\\
  $\GEQRT(i, k)$\\
  \For{$\textnormal{j} = k+1$ to $q$}{
     $\UNMQR(piv(i,k), k, j)$\\
     $\UNMQR(i, k, j)$
  }
  $\TTQRT(i, piv(i,k), k)$\\
  \For{$\textnormal{j} = k+1$ to $q$}{
     $\TTMQR(i, piv(i,k), k, j)$\\
  }
\caption{Elimination $\elim(i, piv(i,k), k)$ via \emph{TT (Triangle on top of triangle)} kernels.}
\label{alg.elimTR}
\end{algorithm}

The second approach to implement the orthogonal transformation $\elim(i,
piv(i,k), k)$ is with the \emph{TT (Triangle on top of triangle)} kernels, as shown in
Algorithm~\ref{alg.elimTR}. Here both tiles $(piv(i,k),k)$ and $(i,k)$ are factored into a triangle (with
\GEQRT). The corresponding transformations are applied to subsequent tiles
$(piv(i,k),j)$ and $(i,j)$, $j>k$, in both rows $piv(i,k)$ and $i$ (with \UNMQR).
Tile $(i,k)$ is zeroed out (with \TTQRT), and subsequent tiles $(i,j)$, $j>k$,
in row $i$ are updated (with \TTMQR).  The flop count is
$2(4+6(q-k))+2+6(q-k)=10+18(q-k)$, just as before.  Dependencies are the
following:
\[
\begin{array}{ll}%
\GEQRT(piv(i,k), k) \prec \UNMQR(piv(i,k), k, j) & \text{ for } j>k\\%
\GEQRT(i, k) \prec \UNMQR(i, k, j) & \text{ for } j>k\\%
\GEQRT(piv(i,k), k) \prec \TTQRT(i, piv(i,k), k)\\%
\GEQRT(i, k) \prec \TTQRT(i, piv(i,k), k)\\%
\TTQRT(i, piv(i,k), k) \prec \TTMQR(i, piv(i,k), k, j) & \text{ for } j>k\\%
\UNMQR(piv(i,k), k, j) \prec \TTMQR(i, piv(i,k), k, j) & \text{ for } j>k\\%
\UNMQR(i, k, j) \prec \TTMQR(i, piv(i,k), k, j) & \text{ for } j>k%
\end{array}%
\]%
Now the factor operations in row $piv(i,k)$ and $i$ can be executed in parallel.
Moreover, the \UNMQR updates can be run in parallel with the \TTQRT
factorization.  Thus, with an unbounded number of processors, the parallel time
is $4+6+6=16$ time-units.

In Algorithms~\ref{alg.elimSQ} and~\ref{alg.elimTR}, it is understood that if a
tile is already in triangle form, then the associated \GEQRT and update kernels
do not need to be applied.

\subsection{QR factorization}

Consider a rectangular tiled matrix $A$ of size $p \times q$, with $p \geq q$.
There are many algorithms to compute the QR factorization of $A$, and we refer
to~\cite{c177} for a survey. We use the three following variants:
\begin{itemize}
\item[\textbf{\FlatTS}]  This simple algorithm is the reference algorithm used 
in~\cite{Buttari2008,tileplasma}.
At step $k$, the pivot row is always row $k$, and we perform the eliminations 
$\elim(i, k, k)$ in sequence, for $i=k+1$, $i=k+2$ down to $i=p$. In this algorithm,
\emph{TS} kernels are used.
\item[\textbf{\FlatTT}]  This algorithm is the counterpart of the \FlatTS algorithm with \emph{TT} kernels.
It uses exactly the same elimination operations, but with different kernels.
\item[\textbf{Greedy}] This algorithm is asymptotically optimal, and turns out to be the most efficient 
on a variety of platforms~\cite{henc,j125}. An optimal algorithm is \Grasap~\cite{henc,j125}. The difference
between \Grasap and \Greedy is a small constant term. We work with \Greedy in this paper. 
The main idea is to eliminate many tiles in parallel at
each step, using a reduction tree (see~\cite{c177} for a detailed description).
\end{itemize}

\begin{figure*}[htbp]
  \begin{center}
    \includegraphics[width=.7\textwidth]{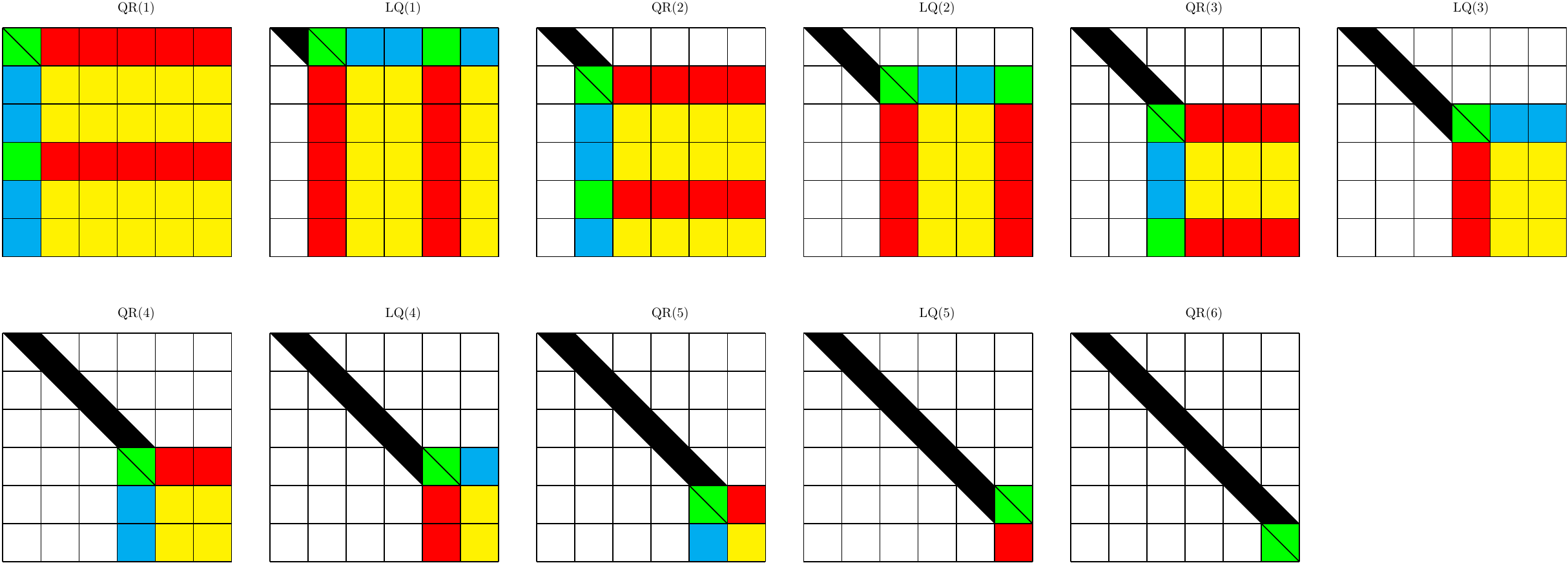}
  \end{center}
  \caption{Snapshots of the bidiagonalization algorithm \bidiag.}
  \label{fig.dessin}
\end{figure*}

\subsection{Bidiagonalization}

Consider a rectangular tiled matrix $A$ of size $p \times q$, with $p \geq q$.
The bidiagonalization algorithm \bidiag proceeds as the QR factorization, but
interleaves one step of LQ factorization between two steps of QR factorization
(see Figure~\ref{fig.dessin}). More precisely,  \bidiag executes the sequence
$$QR(1); LQ(1); QR(2); \dots; QR(q-1); LQ(q-1); QR(q)$$
where $QR(k)$ is the step $k$ of the $QR$ algorithm (see Algorithm~\ref{alg.QR}), 
and $LQ(k)$ is the step $k$ of the $LQ$ algorithm. The latter is a right factorization step
that executes the column-oriented eliminations shown in Algorithm~\ref{alg.LQk}.

\begin{algorithm}[htbp]
  \DontPrintSemicolon
     Step $k$, denoted as $LQ(k)$:\\
     \For{$\textnormal{j} = k+1$ to $q$}{
     $\colelim(j, piv(j,k), k)$
    }
\caption{Step $LQ(k)$ for a tiled matrix of size $p \times q$.}
\label{alg.LQk}
\end{algorithm}

In Algorithm~\ref{alg.LQk}, $\colelim(j, piv(k,j), k)$
is an orthogonal transformation that combines columns $j$ and $piv(k,j)$ to zero
out the tile in position $(k,j)$. It is the exact counterpart to the row-oriented 
eliminations  $\elim(i, piv(i,k), k)$ and be implemented with the very same kernels,
either TS or TT.

\subsection{R-Bidiagonalization}

When $p$ is much larger than $q$, R-bidiagonalization should be preferred,
if minimizing the operation count is the objective. This \rbidiag algorithm
does a QR factorization of $A$, followed by a bidiagonalization of the upper square $q \times q$ matrix.
In other words, given a rectangular tiled matrix $A$ of size $p \times q$, with $p \geq q$,
\rbidiag executes the sequence
$$QR(p,q); LQ(1); QR(2); \dots; QR(q-1); LQ(q-1); QR(q)$$

Let $m=p n_{b}$ and $n = q n_{b}$ be the actual size of $A$
(element wise). The number of arithmetic operations is
$4n^{2}(m-\frac{n}{3})$ for \bidiag and $2n^{2}(m+n)$ for
\rbidiag~\cite[p.284]{Golub}. These numbers show that \rbidiag is less
costly than \bidiag  whenever $m \geq \frac{5n}{3}$, or equivalently, 
whenever $p \geq \frac{5q}{3}$. One major contribution of this paper is to provide a comparison
of \bidiag and \rbidiag in terms of parallel execution time, instead of operation count.

\subsection{Comments}

\paragraph{Case $m \leq n$.} The paper focuses on the $m \geq n$ case but everything holds in the $m \leq n$ case.
One simply can transpose the initial matrix and change switch LQ steps for QR steps and vice versa.

\paragraph{Real/Complex arithmetic.} There is no restriction of our work on whether
the matrices are real or complex. Actually, our codes have been generated to allow for
both real and complex arithmetics, single and double precisions.

\paragraph{Memory usage of our algorithm.} In the standard LAPACK case, GEBRD
takes as input a matrix and overrides the matrix with the Householder
reflectors and the bidiagonal matrix. The algorithm is therefore down in place
and there is no extra storage needed. In the case \rbidiag, it is important to
note that as in the case of the LHC
algorithm~\cite{Chan:1982:IAC:355984.355990}, an extra storage of size $n^2/2$
is needed if one wants to recompute the singular vectors. (If one wants the
singular value, all that matters is the bidiagonal form and any Householder
vectors generated can safely be discarded.)

\paragraph{Any tree is possible.} We note that whether we choose to perform
\bidiag or \rbidiag, any tree is possible at each step of the algorithm. For
example a restriction of our study thereafter is to fix the trees in the QR step and the \bidiag step of \rbidiag 
to be the same. Clearly there is no such need, and one can consider for \rbidiag (for example) a combination like
(\QRTS+\bidiagGreedy). We do not study such combinations. This also simplifies notations. Since 
we consider the same tree for the QR step and the \bidiag step of \rbidiag,
when we write \rbidiagTT (for example), we mean (\QRTT+\bidiagTT).

\paragraph{\bidiagGreedy and \bidiagBinomial are the same.} 
Another name for \bidiagGreedy would be \bidiagBinomial. Since in the \bidiag algorithm, we repeat the 
same tree (in the \Greedy case, a \binomial tree) over and over again. So \bidiagGreedy and \bidiagBinomial are the exact same algorithm.
(Whereas \QRGreedy is not the same as \QRBinomial.)

\section{Critical paths}
\label{sec.critical}
  
In this section, we compute exact or estimated values of the critical paths of the \bidiag and \rbidiag
algorithms with the \FlatTS, \FlatTT, and \Greedy trees. 

\subsection{Bidiagonalization}
\label{subsec.critical.bidiag}

Given a rectangular tiled matrix $A$ of size $p \times q$, with $p \geq q$, 
the bidiagonalization algorithm \bidiag executes the sequence
$$QR(1); LQ(1); QR(2); \dots; QR(q-1); LQ(q-1); QR(q)$$
To compute the critical path, we first observe that there is no overlap
between two consecutive steps $QR(k)$ and $LQ(k)$. To see why, 
consider w.l.o.g. the first two steps $QR(1)$ and $LQ(1)$ on
Figure~\ref{fig.dessin}. Tile $(1,2)$ is used at the end of the
$QR(1)$ step to update the last row of the trailing matrix (whichever it is). In passing, 
note that all columns in this last row are updated in parallel, because we assume unlimited resources when computing critical paths. But tile $(1,2)$
it is the first tile modified by the $LQ(1)$ step, hence there is no possible overlap.
Similarly, there is no overlap between two consecutive steps $LQ(k)$ and $QR(k+1)$.
Consider steps $LQ(1)$ and $QR(2)$ on
Figure~\ref{fig.dessin}.  Tile $(2,2)$ is used at the end of the
$LQ(1)$ step to update the last column of the trailing matrix (whichever it is), and it
is the first tile modified by the $QR(1)$ step.

As a consequence, the critical path length of \bidiag is the sum of the critical path lengths of each QR and LQ step.
And so an optimal \bidiag algorithm is an algorithm which is optimal at each QR and LQ step independently.
We know that an optimal QR step is obtained by a binomial tree.
We know that an optimal LQ step is obtained by a binomial tree.
Therefore the scheme alternating binomial QR and LQ trees, that is \bidiagGreedy, is optimal. This is repeated in the following theorem.

\begin{theorem}
\label{theo:1}
 \bidiagGreedy is optimal, in the sense that it has the shortest critical path length, over all \bidiag algorithms, which are tiled algorithms that alternate a QR step and an
LQ step to obtain a band bidiagonal matrix with a sequence of any QR trees and any
LQ trees.
\end{theorem}

We now compute critical path lengths.
From~\cite{henc,c177,j125} we have the following values for the critical path of one QR step applied to
a tiled matrix of size $(u,v)$:
\begin{itemize}
\item[\textbf{\FlatTS}]~\\
\scalebox{0.9}{%
$
QR-FTS_{1step}(u, v) = \left\{
\begin{array}{rl}
4 + 6(u - 1) & \text{if } v = 1,\\
4 + 6 + 12(u - 1) & \text{otherwise}.
\end{array} \right.
$}
\item[\textbf{\FlatTT}]~\\
\scalebox{0.9}{%
$
QR-FTT_{1step}(u, v) = \left\{
\begin{array}{rl}
4 + 2(u - 1) & \text{if } v = 1,\\
4 + 6 + 6(u - 1) & \text{otherwise}.
\end{array} \right.
$}
\item[\textbf{\Greedy}]~\\
\scalebox{0.9}{%
$
QR-GRE_{1step}(u, v) = \left\{
\begin{array}{rl}
4 + 2 \lceil \log_{2}(u) \rceil & \text{if } v = 1,\\
4 + 6 + 6 \lceil \log_{2}(u) \rceil & \text{otherwise}.
\end{array} \right.
$}
\end{itemize}

\begin{figure*}[htbp]
  \begin{center}
  \begin{subfigure}{0.3\textwidth}
    \includegraphics[width=\textwidth]{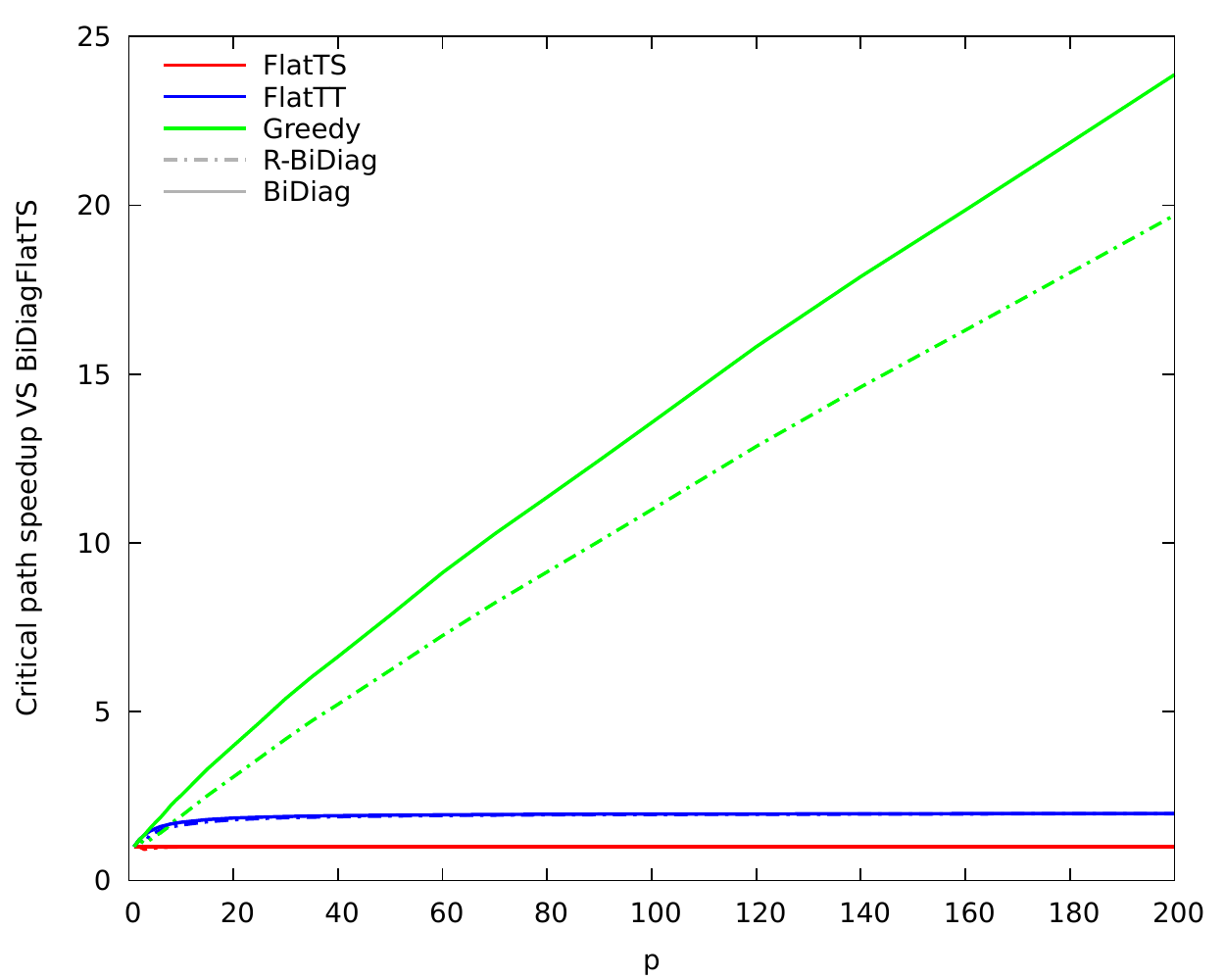}
    \caption{Square ($p=q$)}
    \label{fig:cp-square}
  \end{subfigure}
  \begin{subfigure}{0.3\textwidth}
    \includegraphics[width=\textwidth]{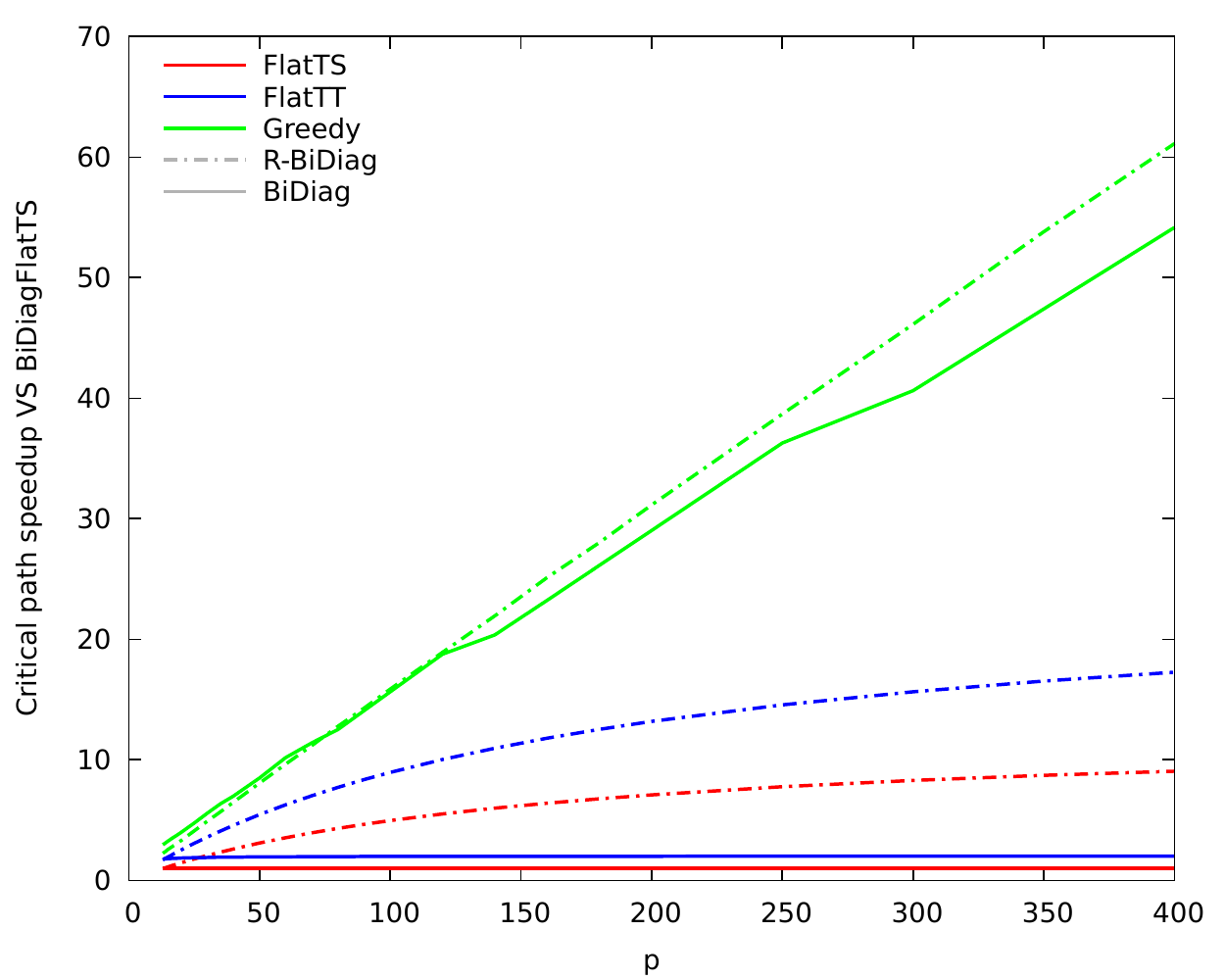}
    \caption{Tall and skinny ($q=13$)}
    \label{fig:cp-q10}
  \end{subfigure}
  \begin{subfigure}{0.3\textwidth}
    \includegraphics[width=\textwidth]{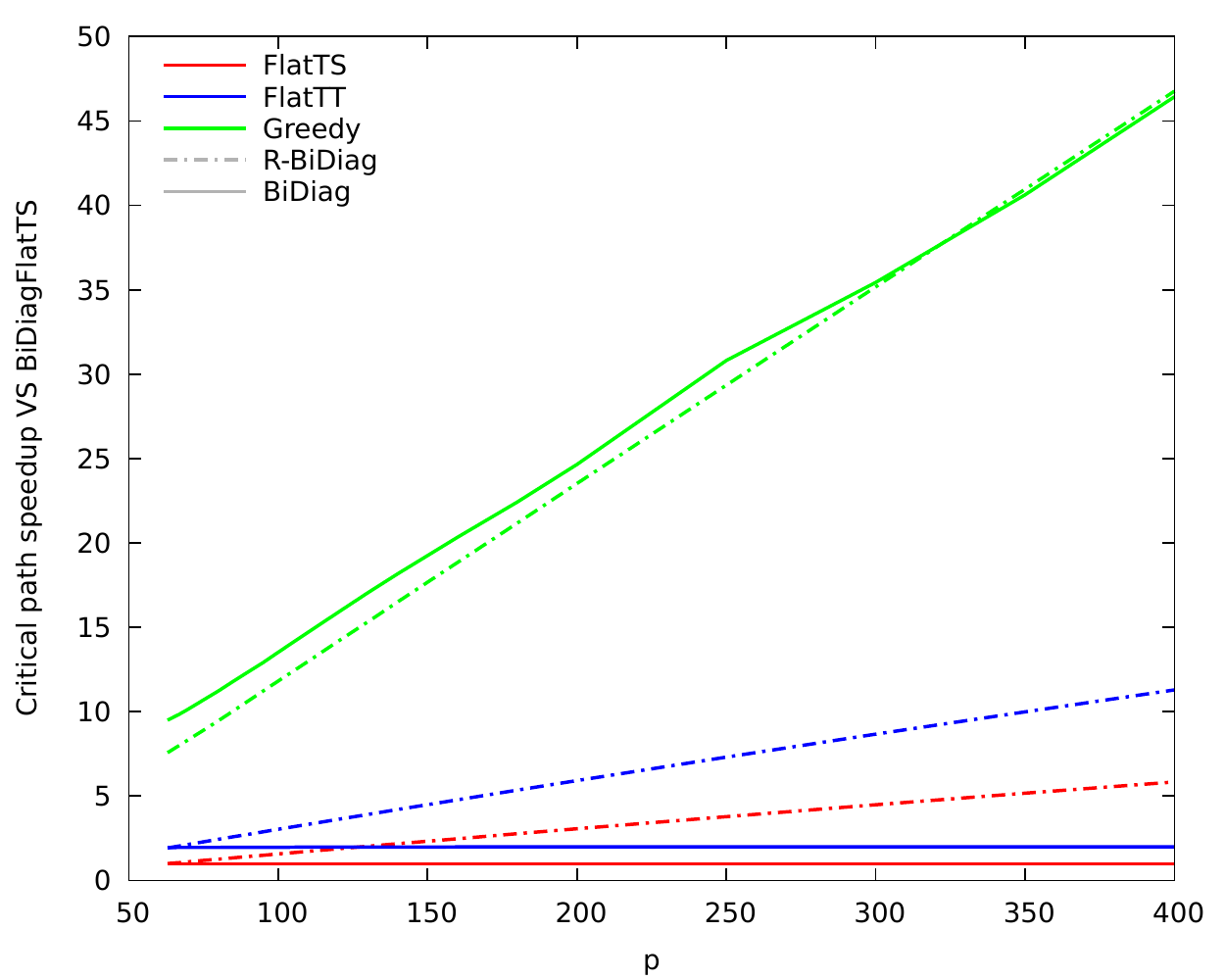}
    \caption{Tall and skinny ($q=63$)}
    \label{fig:cp-q40}
  \end{subfigure}
  \end{center}
  \caption{Critical paths: ratio of the different methods over \bidiag with \FlatTS,
  on three configurations. Solid lines are for \bidiag and dashed lines for \rbidiag.}
  \label{fig:cp}
\end{figure*}

The critical path of one LQ step applied to
a tiled matrix of size $(u,v)$ is $LQ_{1step}(u,v) = QR_{1step}(v,u)$.
Finally, in the \bidiag algorithm, the size of the matrix for step $QR(k)$ is $(p-k+1,q-k+1)$
and the size of the matrix for step $LQ(k)$ is $(p-k+1,q-k)$. Altogether,
we derive the following values:
\begin{itemize}
\item[\textbf{\FlatTS}]~\\
\begin{eqnarray}
\nonumber\bidiagTS(p, q) &=& \sum_{k=1}^{q-1} (10+12(p-k)) + \sum_{k=1}^{q-1} (10+12(q-k-1)) + (4+6(p-q))\\
\nonumber                &=& 12 pq - 6p + 2q - 4
\end{eqnarray}
\item[\textbf{\FlatTT}]~\\
\begin{eqnarray}
\nonumber
\bidiagTT(p, q) & = & \sum_{k=1}^{q-1} (10+6(p-k)) + \sum_{k=1}^{q-1} (10+6(q-k-1)) + (4+2(p-q))\\
\nonumber       & = & 6 p q - 4 p + 12 q - 10
\end{eqnarray}
\end{itemize}

We devote the whole next subsection to the study of the critical path of $\bidiagGreedy(p, q)$,
because this is more complicated.

\subsection{Study of the critical path of $\bidiagGreedy(p, q)$}
\label{subsec.critical.bidiag.path}

\subsubsection{A formula with a summation}

By following the reasoning of the previous subsection, we get
\begin{itemize}
\item[\textbf{\Greedy}]~\\
\begin{eqnarray}
\nonumber \bidiagGreedy(p, q) & = &  \sum_{k=1}^{q-1} (10+6 \lceil \log_{2}(p + 1 - k) \rceil)
            + \sum_{k=1}^{q-1}  (10+6 \lceil \log_{2}(q-k) \rceil) + (4 + 2 \lceil \log_{2}(p+1-q) \rceil\\
\nonumber                     & = &  6 \sum_{k=p-q+2}^{p} \lceil \log_{2}(k) \rceil
            + 6 \sum_{k=1}^{q-1}   \lceil \log_{2}(k) \rceil
+ 20 q + 2 \lceil \log_{2}(p+1-q) \rceil -16\\
\label{eq:1}                  & = &  
              6 \sum_{k=1}^{p} \lceil \log_{2}(k) \rceil
            - 6 \sum_{k=1}^{p-q+1} \lceil \log_{2}(k) \rceil
            + 6 \sum_{k=1}^{q-1}   \lceil \log_{2}(k) \rceil
+ 20 q + 2 \lceil \log_{2}(p+1-q) \rceil -16
\end{eqnarray}
\end{itemize}

The latter formula is exact and is easy enough to compute with a small computer code. However
it does not provide us with much insight.
To get a better grasp on $\bidiagGreedy(p, q)$, we develop some more Equation~\eqref{eq:1}.

\subsubsection{An exact formula when $p$ and $q$ are powers of two}

First we note that:

\begin{eqnarray}
\nonumber \textmd{Let }r\textmd{ be a power of 2},\quad \sum_{k=1}^{r} \lceil \log_2(k)   \rceil
& = & 1 + 2 +2 + 3 +3+3+3+4+4+4+4+4+4+4+4+5\ldots\\
\label{eq:2} & = & r \log_2(r)-r + 1
\end{eqnarray}
One way to understand the previous relation is to remember that ``the'' antiderative 
of  $\log(x)$ is $x\log(x) - x$. The previous relation reminds us of this result.
The general case ($r$ is not a power of 2) of Equation~(\ref{eq:2}) is
\begin{equation}
\label{eq:3}
\textmd{Let }r\textmd{ be an integer},\quad
\sum_{k=1}^{r} \lceil \log_2(k)   \rceil 
= 
\left(\lfloor \log_2(r) \rfloor-1\right)2^{ \lfloor \log_2(r) \rfloor } +1  
+ \left( r- 2^{ \lfloor \log_2(r) \rfloor } \right)\lceil \log_2(r) \rceil
\end{equation}
From Equations~(\ref{eq:1}) and~(\ref{eq:2}), we derive that if $q$ is a power of two:
$$
\textmd{Let }q\textmd{ be a power of 2},\quad 
\bidiagGreedy(q, q) = 12 q \log_2(q) + 8 q - 6 \log_2(q) - 4$$
From Equations~(\ref{eq:1}),~(\ref{eq:2}) and~(\ref{eq:3}), we derive that,
if both $p$ and $q$ are powers of two, with $p> q$:
$$\textmd{Let }p\textmd{ and }q\textmd{ be powers of 2},\quad 
\textmd{with }p > q,\quad 
\bidiagGreedy(p, q) = 6 q \log_2(p) + 6 q \log_2(q)
+  14 q - 4 \log_2(p)  - 6 \log_2(q) - 10.$$

\begin{theorem}
\textmd{Let }$p$\textmd{ and }$q$\textmd{ be powers of 2},
\begin{eqnarray}
\nonumber
\textmd{If }p= q,\quad &&
\bidiagGreedy(q, q) = 12 q \log_2(q) + 8 q - 6 \log_2(q) - 4,\\
\nonumber
\textmd{Else if }p > q,\quad &&
\bidiagGreedy(p, q) = 6 q \log_2(p) + 6 q \log_2(q)
+  14 q - 4 \log_2(p)  - 6 \log_2(q) - 10.
\end{eqnarray}
\end{theorem}

\subsubsection{Using Stirling's formula}
\label{subsub:assymptot1}

We can also consider Equation~(\ref{eq:1}) again and obtain
simpler bounds by rounding down and up the ceiling function in the logarithms
and then using asymptotic analysis.
We consider the bounds:
$$
\textmd{Let }k\textmd{ be an integer},\quad 
 \log_{2}(k) \leq \lceil \log_{2}(k) \rceil \leq \log_{2}(k) + 1.$$
Applied to Equation~(\ref{eq:1}),
this
leads to
\begin{eqnarray}
\nonumber 
&   6\log_2( p! ) - 6\log_2( (p-q+1)! ) +6\log_2( (q-1)! ) + 20q + 2  \log_2(p+1-q) -16\\
\nonumber 
&   \leq \bidiagGreedy(p, q) \leq \\
\nonumber 
&   6\log_2( p! ) - 6\log_2( (p-q+1)! ) +6\log_2( (q-1)! ) + 32q + 2  \log_2(p+1-q) -26.
\end{eqnarray}

We note that the difference between the lower bound (left side) and the
upper bound (right side) is $12q - 10$. 

We now consider ``large'' $p$ and $q$, and we use Stirling's formula as follows:
$$ \log(p!) = p \log(p) - p +\mathcal{O}\left(\log p\right).$$
And with one more term we get:
$$ \log(p!) = p \log(p) - p + \frac{1}{2}\log(2\pi p) + \mathcal{O}\left(\frac{1}{p}\right).$$

In base 2, 
$$ \log_2(p!) = p \log_2(p) - \log_2(e) p + \frac{1}{2} \log_2(p) + \log_2( \sqrt{2\pi} ) +\frac{\log_2(e)}{12p}+\mathcal{O}\left(\frac{1}{p^3}\right).$$

We obtain that
\begin{eqnarray}
&
\nonumber 
\begin{array}{lll}
    6 p \log_2(p)
    - 6 (p-q)\log_2(p-q+1)
    + 6 q\log_2(q-1) & ~~~~~~~~&   [ x.\log_2x\textmd{ terms} ] \\
    ~~~~+ ( 20 - 12 \log_2(e) ) (q) && [ \textmd{linear terms} ] \\
    ~~~~+ 3\log_2(p) 
    - 3\log_2(q-1)
    - 7 \log_2(p-q+1) && [ \log_2 \textmd{linear terms} ] \\
    ~~~~+  6 \log_2( \sqrt{2\pi} ) 
    + 12 \log_2(e) 
    - 16 && [ \textmd{constant terms} ] \\
    ~~~~+ \frac{1}{2}\log_2(e)\frac{1}{p}
    + \frac{1}{2}\log_2(e)\frac{1}{q-1}
    - \frac{1}{2}\log_2(e)\frac{1}{p-q+1}\\
    ~~~~+\mathcal{O}\left(\max(\frac{1}{(p-q)^3}, \frac{1}{q^3})\right)\\
\end{array}\\
\nonumber 
&   \leq \bidiagGreedy(p, q) \leq \\
\nonumber 
&
\begin{array}{lll}
    6 p \log_2(p)
    - 6 (p-q)\log_2(p-q+1)
    + 6 q\log_2(q-1) & ~~~~~~~~&   [ x.\log_2x\textmd{ terms} ] \\
    ~~~~+ ( 32 - 12 \log_2(e) ) (q) && [ \textmd{linear terms} ] \\
    ~~~~+ 3\log_2(p) 
    - 3\log_2(q-1)
    - 7 \log_2(p-q+1) && [ \log_2 \textmd{linear terms} ] \\
    ~~~~+  6 \log_2( \sqrt{2\pi} ) 
    + 12 \log_2(e) 
    - 26 && [ \textmd{constant terms} ] \\
    ~~~~+ \frac{1}{2}\log_2(e)\frac{1}{p}
    + \frac{1}{2}\log_2(e)\frac{1}{q-1}
    - \frac{1}{2}\log_2(e)\frac{1}{p-q+1}\\
    ~~~~+\mathcal{O}\left(\max(\frac{1}{(p-q)^3}, \frac{1}{q^3})\right).\\
\end{array}
\end{eqnarray}

We stop the expansion a little earlier.
\begin{eqnarray}
&
\nonumber 
\begin{array}{lll}
    6 p \log_2(p)
    - 6 (p-q)\log_2(p-q+1)
    + 6 q\log_2(q-1) & ~~~~~~~~&   [ x.\log_2x\textmd{ terms} ] \\
    ~~~~+ ( 20 - 12 \log_2(e) ) (q) && [ \textmd{linear terms} ] \\
    ~~~~+ 3\log_2(p) 
    - 3\log_2(q-1)
    - 7 \log_2(p-q+1) && [ \log_2 \textmd{linear terms} ] \\
    ~~~~+\mathcal{O}\left(1\right)\\
\end{array}\\
\nonumber 
&   \leq \bidiagGreedy(p, q) \leq \\
\nonumber 
&
\begin{array}{lll}
    6 p \log_2(p)
    - 6 (p-q)\log_2(p-q+1)
    + 6 q\log_2(q-1) & ~~~~~~~~&   [ x.\log_2x\textmd{ terms} ] \\
    ~~~~+ ( 32 - 12 \log_2(e) ) (q) && [ \textmd{linear terms} ] \\
    ~~~~+ 3\log_2(p) 
    - 3\log_2(q-1)
    - 7 \log_2(p-q+1) && [ \log_2 \textmd{linear terms} ] \\
    ~~~~+\mathcal{O}\left(1\right).\\
\end{array}
\end{eqnarray}

We now study the term: $ p \log_2(p) - (p-q)\log_2(p-q+1)$, this gives:
\begin{eqnarray}
\nonumber p\log_2( p ) - (p-q)\log_2( p-q+1 )
\nonumber &=& q\log_2( p ) - (p-q)\log_2( 1-\frac{q-1}{p} ).
\end{eqnarray}

We now study the term: $ (p-q)\log_2( 1-\frac{q-1}{p} ) $. 
We can set $p=q+\alpha q$ with $\alpha \geq 0$. We get
$$ (p-q)\log_2( 1-\frac{q-1}{p} )= q \left( \alpha\log_2( \frac{ \alpha }{1+\alpha} +\frac{1}{p} ) \right).$$
We have that
$$
\textmd{for }\alpha \geq 0,\quad
 \alpha \log_2( \frac{ \alpha }{1+\alpha} ) \leq \log_2( \frac{ \alpha }{1+\alpha} +\frac{1}{p} ) \leq 0 .$$
We are therefore drawn to study $\textmd{for }\alpha \geq 0, ~~ \alpha \log_2( \frac{ \alpha }{1+\alpha} ) $.
We see that this is a decreasing function of $\alpha$. 
We are interested in the limit value when $\alpha$ goes to $\infty$. For large  values of $\alpha$, we have
 %\textmd{( we use l'H{\^o}pital's rule here ) }
\begin{eqnarray}
\nonumber 
\alpha \log_2( \frac{ \alpha }{1+\alpha} )
&=&
\alpha \log_2(e) \log\left( 1- \frac{ 1 }{1+\alpha} \right),\\
\nonumber 
&=&
\alpha \log_2(e) \log \left( - \frac{ 1 }{1+\alpha} +\mathcal{O}( \frac{ 1 }{\alpha^2} ) \right),\\
\nonumber 
&=&
 \log_2(e) \log \left( - \frac{ \alpha }{1+\alpha} +\mathcal{O}( \frac{ 1 }{\alpha} ) \right).
\end{eqnarray}
So that 
$$
\lim_{\alpha\to +\infty}
\left( \alpha \log_2( \frac{ \alpha }{1+\alpha} ) \right)
= - \log_2(e).
$$

So all in all, we have that
$$
\textmd{for }\alpha \geq 0,\quad
 - \log_2(e) \leq \alpha \log_2( \frac{ \alpha }{1+\alpha} ) .
$$

And so 
$$ - \log_2(e) q \leq (p-q)\log_2( 1-\frac{q-1}{p} ) \leq 0 .$$

Thus
$$ q\log_2( p ) \leq p\log_2( p ) - (p-q)\log_2( p-q+1 ) \leq q\log_2( p ) + \log_2(e) q .$$

So we get
\begin{eqnarray}
&
\nonumber 
\begin{array}{lll}
    6 q\log_2( p )
    + 6 q\log_2(q-1) & ~~~~~~~~&   [ x.\log_2x\textmd{ terms} ] \\
    ~~~~+ ( 20 - 12 \log_2(e) ) (q) && [ \textmd{linear terms} ] \\
    ~~~~+ 3\log_2(p) 
    - 3\log_2(q-1)
    - 7 \log_2(p-q+1) && [ \log_2 \textmd{linear terms} ] \\
    ~~~~+\mathcal{O}\left(1\right)\\
\end{array}\\
\nonumber 
&   \leq \bidiagGreedy(p, q) \leq \\
\label{eq:top}
&
\begin{array}{lll}
    6 q\log_2( p )
    + 6 q\log_2(q-1) & ~~~~~~~~&   [ x.\log_2x\textmd{ terms} ] \\
    ~~~~+ ( 32 - 6 \log_2(e) ) (q) && [ \textmd{linear terms} ] \\
    ~~~~+ 3\log_2(p) 
    - 3\log_2(q-1)
    - 7 \log_2(p-q+1) && [ \log_2 \textmd{linear terms} ] \\
    ~~~~+\mathcal{O}\left(1\right).\\
\end{array}
\end{eqnarray}

We note that the difference between the lower bound (left side) and the
upper bound (right side) is $( 12 + 6 \log_2(e) )q - 10$. 

We now get the following theorem
\begin{theorem}
$$
\bidiagGreedy(p, q)  = 6 q\log_2( p ) + 6 q\log_2(q) + \mathcal{O}\left(\max(\log_2(p),q)\right).
$$
\end{theorem}
This theorem matches the power of 2 case. 
Also we understand that we perform $q$ QR steps that each are done 
with a binomial tree of length $\log_2( p )$. This explains the $ q\log_2( p )$ term;
we perform $q$ LQ steps that each are done 
with a binomial tree of length $\log_2( q )$. This explains the $ q\log_2( q )$ term.
The coefficient 6 in front is due to the weights use for the kernels.
This theorem holds for any $p$ and $q$.

In the $p=q$ case, we get that 
$$
\bidiagGreedy(q, q)  = 12 q\log_2( q ) + \mathcal{O}\left(q\right).
$$

In the $p=\beta q^{1+\alpha}$ case, we get that 
$$
\bidiagGreedy(\beta q^{1+\alpha}, q)  = (12 + 6 \alpha) q\log_2( q ) + \mathcal{O}\left(q\right).
$$

We use Equation~(\ref{eq:top}) and get
\begin{eqnarray}
&&\nonumber 
\textmd{ for }q
\textmd{ constant, and }p
\textmd{ going to infinity, }\\
&&\nonumber 
   ~~~~~~~~~~~
\bidiagGreedy(p, q) = ( 6 q - 4 )\log_2(p) +\mathcal{O}\left(1\right)
\end{eqnarray}
We note that this formula agrees with the case when $p$ and $q$ are powers of 2.
Also note that
the $\mathcal{O}\left(1\right)$ includes 
$q\log_2(q)$ terms, $q$ terms, etc since $q$ is a constant.

\subsubsection{An exact formula for all $p$ and $q$}

We conclude the subsection on the 
study of $\bidiagGreedy(p, q)$ by giving an exact formula. 
The first-order term in the exact formula is the same as in Subsubsection~\ref{subsub:assymptot1}
with asymptotic analysis. However one lower-order term is harder to analyze.
So we prefer to present this formula last. This is nevertheless an exact
formula for our problem so it has some merit in itself.

For the general case, (either $p$ or $q$ is not a power of 2,) using Equations~(\ref{eq:1}),~(\ref{eq:2}) and~(\ref{eq:3}),
the expression gets more complicated: letting
\begin{eqnarray*}
\nonumber \alpha_p &=& \lceil \log_2(p) \rceil - \log_2(p) \\
\nonumber \alpha_k &=& \lceil \log_2(p-q+1) \rceil - \log_2(p-q+1) \\
\nonumber \alpha_q &=& \lceil \log_2(q-1) \rceil - \log_2(q-1); \\
\nonumber \beta_p &=& \alpha_p-2^{\alpha_p}-\alpha_k+2^{\alpha_k};\\
\nonumber \beta_q &=& \alpha_q-2^{\alpha_q}+\alpha_k-2^{\alpha_k};
\end{eqnarray*}
we derive that
$$
\bidiagGreedy(p,q) = 6 p \log_2(p)
- 6 (p-q+1)\log_2(p-q+1)
+ 6 (q-1)\log_2(q-1)  
+ 20 q + 6 \beta_p p  + 6 \beta_q (q-1)  + 2 \lceil \log_2(p+1-q)  \rceil - 10 .
$$

We see that we have a term in $p$. Namely $6 \beta_p p$. This is
counter-intuitive since we expect an expansion with the term $ p \log_2(p) -
(p-q+1)\log_2(p-q+1) + (q-1)\log_2(q-1)  $ as we already have seen in
Subsubsection~\ref{subsub:assymptot1}, and then, for the next terms after, we
expect a term in $q$ and no term in $p$. The next term in $p$ should at most be
a $\log_2(p)$.  The reason is that the term $6 \beta_p p$ behaves has a linear
function of $q$.  (Although this is not trivial to understand.)

If 
further analysis is sought from this formula,
it is important to note that $\beta_p$ and $\beta_q$, while being complicated functions of $p$ and $q$, are bounded. 
Indeed, since $\alpha_p$, $\alpha_k$, and $\alpha_q$ are between $0$ and $1$, we have
that $\alpha-2^{\alpha}$ is between
 $-(1+\log\log 2))/\log2$ and -1, 
(that is between: $-0.9139$ and $-1$,) and so
\begin{eqnarray}
\nonumber - \left( 1-(1+\log(\log(2)))/\log(2) \right) & \leq & \beta_p \leq + \left( 1-(1+\log(\log(2)))/\log(2) \right)  \\
\nonumber - 0.0861 & \leq & \beta_p \leq + 0.0861
\end{eqnarray}
and 
\begin{eqnarray}
\nonumber - 2 & \leq & \beta_q \leq  -2\left( (1+\log(\log(2)))/\log(2) \right)\\
\nonumber - 2 & \leq & \beta_q \leq - 1.8278
\end{eqnarray}

\subsection{R-Bidiagonalization}
\label{subsec.critical.rbidiag}

\subsubsection{Formula for tiled QR factorization using various trees}

For a tiled matrix of size $(p,q)$, \rbidiag executes the sequence
$$QR(p,q); LQ(1); QR(2); \dots; QR(q-1); LQ(q-1); QR(q)$$
From~\cite{c177,henc,j125}, we have the following critical paths for the QR factorization of a tiled $(p,q)$ matrix, which can be computed using the three variants described in Section~\ref{sec.algos}:
\begin{itemize}
\item[\textbf{\FlatTS}]~\\
$QR-FTS(p, q) = \left\{
\begin{array}{rl}
6p - 2 & \text{if } q = 1,\\
30q - 34 & \text{if } q = p,\\
12p+18q -32 & \text{otherwise}.
\end{array} \right.$
\item[\textbf{\FlatTT}]~\\
$QR-FTT(p, q) = \left\{
\begin{array}{rl}
2p + 2 & \text{if } q = 1,\\
22q - 24 & \text{if } q = p,\\
6p + 16q - 22 & \text{otherwise}.
\end{array} \right.$
\end{itemize}

For $QR-GRE$ we have no closed-form formula. By combining~\cite[Theorem 3.5]{henc}
with~\cite[Theorem 3]{j12} we derive that
$$QR-GRE(p, q) =22q + o(q)$$
whenever $p =o(q^{2})$, which includes the case where $p=\ratio q^{1+\alpha}$, with $0 \leq \alpha < 1$.

\subsubsection{Critical path length of \rbidiag with various trees}

Computing the critical path of  \rbidiag  is more difficult than for \bidiag, because kernels partly overlap.
For example, there is no need to wait for the end of the (left) $QR$ factorization to start the first (right) factorization step $LQ(1)$. In fact, this step can start as soon as the first step $QR(1)$ is over  because the
first row of the matrix is no longer used throughout the whole $QR$ factorization at this point. 
However, the
interleaving of the following kernels gets quite intricate. 
For example we note that $\rbidiagGreedy(65,5)=262$ while $\rbidiagGreedy(66,5)=260$.
So we factorize a larger matrix in a shorter critical path.
The reason is that, even though the critical path of $QR-GRE$(66,5) is longer than 
the critical path of $QR-GRE$(65,5), 
the DAG of $QR-GRE$(66,5) interleaves with the DAG of \bidiag(5,5)
more than the  DAG of $QR-GRE$(65,5) interleaves with the DAG of \bidiag(5,5).
There are many such examples. For example 
$\rbidiagGreedy(134,10)=680$ while $\rbidiagGreedy(133,10)=682$; 
$\rbidiagGreedy(535,50)=4946$ while $\rbidiagGreedy(534,50)=4948$; etc.

Taking into account the interleaving, or not, 
does not change the higher-order terms, in the following we simply sum up
the values obtained without overlap, adding the cost of the QR factorization of size $(p,q)$
to that of the bidiagonalization of the top square $(q,q)$ matrix, and subtracting step $QR(1)$ as discussed above.
In other words
$$ \rbidiag(p,q) \leq QR(p,q) + \bidiag(q,q) - QR_{1step}(q).$$

This leads us to the following upper bounds:

\begin{itemize}
\item[\textbf{\FlatTS}]~\\
$\rbidiagTS(p, q) \leq (12p+18q-32) + (12q^{2}-4q-4) - (12q-14)=\\
12q^{2}+12p+2q-22
$
\item[\textbf{\FlatTT}]~\\
$\rbidiagTT(p, q) \leq (6p+16q-22) + (6q^{2}+8q-10) - (6q-2)=\\
6q^{2}+6p+ 18q-30
$
\item[\textbf{\Greedy}]~\\
$\rbidiagGreedy(p, q) \leq (22q + o(q)) + (12 q \log_2(q) +(20-12\log_{2}(e))q + o(q)) - o(q)=
12q \log_2(q) + (42-12\log_{2}(e))q +o(q)
$
whenever $p =o(q^{2})$.
\end{itemize}

For the sake of completeness, here are the exact values (obtained using the \parsec programming tool,
see Section~\ref{sec.expes}) for the critical paths of \rbidiag with the \FlatTS and \FlatTT variants:

\begin{center}
\scalebox{0.8}{%
$\rbidiagTS(p, q) = \left\{
\begin{array}{ll}
6 p - 2                  & \text{if } q = 1,\\
36                       & \text{if } q = 2 \text{ and } q = p ,\\
12 p + 4                 & \text{if } q = 2 \text{ and } q \neq p,\\
12 q^2 - 16 q + 12 p + 4 & \text{if } q = 3 \text{ and } q = p,\\
12 q^2 - 16 q + 12 p + 6 & \text{otherwise}.
\end{array} \right.
$}

\scalebox{0.8}{%
$\rbidiagTT(p, q) = \left\{
\begin{array}{ll}
2 p + 2                  & \text{if } q = 1,\\
30                       & \text{if } q = 2 \text{ and } q = p ,\\
6 p + 20                 & \text{if } q = 2 \text{ and } q \neq p,\\
6 q^2 + 18 q - 36        & \text{if } q = 3 \text{ and } q = p ,\\
6 q^2 + 12 q + 6 p - 34  & \text{if } q = 3 \text{ and } q \neq p,\\
6 q^2 +  6 q + 6 p - 16  & \text{otherwise}.
\end{array} \right.
$}
\end{center}
In both cases, we check that the exact and approximate values differ by a factor $O(q)$
and have same higher-order term $O(q^{2})$.

\subsubsection{Comparison between \bidiagGreedy and \rbidiagGreedy}
\label{subsub:switch}

Again, we are interested in the asymptotic analysis of \rbidiagGreedy, and in the comparison with \bidiagGreedy.
In fact, when $p =o(q^{2})$, say $p=\ratio q^{1+\alpha}$, with $0 \leq \alpha < 1$, the 
cost of the QR factorization $QR(p,q)$  is negligible in front of the cost of the
bidiagonalization $\bidiagGreedy(q,q)$, so that $\rbidiagGreedy(p,q)$ is asymptotically equivalent
to $\bidiagGreedy(q,q)$, and we derive that:
\begin{equation}
\label{eq.rbidiagasympt}
\textmd{ for }0 \leq \alpha < 1,\quad
\lim_{q \rightarrow \infty} \frac{\bidiagGreedy(\beta q^{1+\alpha},q)}{\rbidiagGreedy(\beta q^{1+\alpha},q)} = 1+\frac{\alpha}{2}
\end{equation}
Asymptotically, \bidiagGreedy is at least as costly (with equality is $p$ and $q$ are proportional) and at most $1.5$ times as costly as \rbidiagGreedy (the maximum ratio being reached when $\alpha= 1 -\varepsilon$ for small values of $\varepsilon$.

Just as before, \rbidiagGreedy  is asymptotically optimal among all possible reduction trees,
and we have proven the following result, where for notation convenience
we let $\bidiag(p,q)$ and $\rbidiag(p,q)$ denote the optimal critical path lengths of the algorithms:

\begin{theorem}
For $p=\ratio q^{1+\alpha}$, with $0 \leq \alpha < 1$:\\
$$\lim_{q \rightarrow \infty} \frac{\bidiag(p,q)}{(12 + 6 \alpha) q \log_2(q)} = 1$$
$$\lim_{q \rightarrow \infty} \frac{\bidiag(p,q)}{\rbidiag(p,q)} = 1+\frac{\alpha}{2}$$
\end{theorem}
When $p$ and $q$ are proportional ($\alpha = 0$, $\beta \geq 1$), both algorithms have same asymptotic cost $12 q log_2(q)$. On the contrary, for very elongated matrices with fixed $q \geq 2$,  the ratio of the critical path lengths of \bidiag and \rbidiag gets high asymptotically: the cost of the QR factorization is
equivalent to $6 \log_{2}(p)$ and that of $\bidiag(p,q)$ to $6q \log_{2}(p)$. Since the cost of
$\bidiag(q,q)$ is a constant for fixed $q$, we get a ratio of $q$.
Finally, to give a more practical insight,
Figure~\ref{fig:cp} reports the ratio of the critical paths of all schemes over that of \bidiag
with \FlatTS. We observe the superiority of \rbidiag for tall and skinny matrices.

\subsubsection{Switching from $\bidiag$ and $\rbidiag$}

\begin{figure*}[hbtp]
  \begin{center}
  \begin{subfigure}{0.49\textwidth}
    \includegraphics[width=\textwidth]{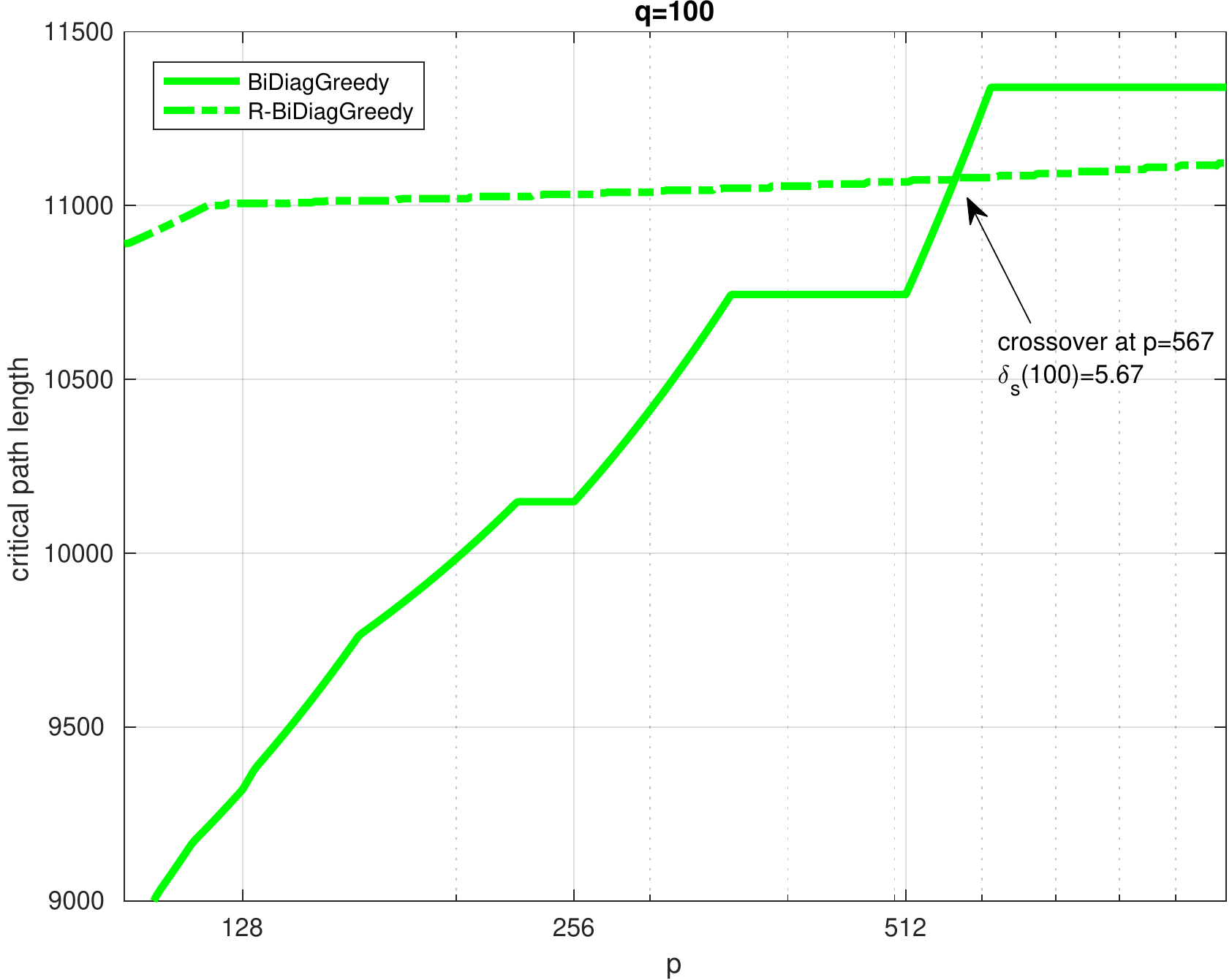}
    \caption{Critical path lengths for $\bidiagGreedy$ and $\rbidiagGreedy$ for $q=100$ as a function of $p$. We see that $\delta_s(100)=5.67$.}
    \label{fig:cpjulien:1}
  \end{subfigure}
  \begin{subfigure}{0.49\textwidth}
    \includegraphics[width=\textwidth]{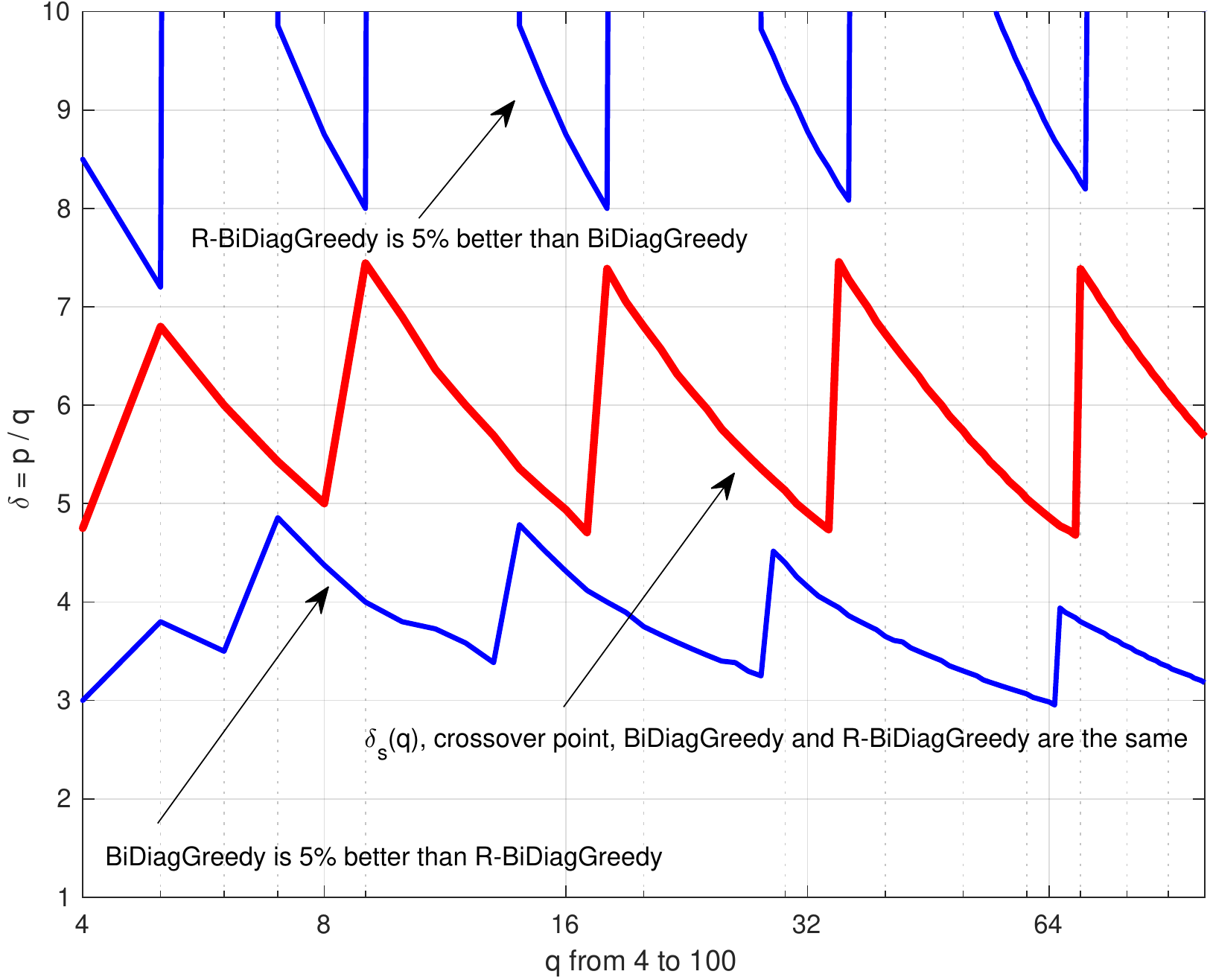}
    \caption{Values of $\delta_{s}$ where to switch as a function of $q$.}
    \label{fig:cpjulien:2}
  \end{subfigure}
  \end{center}
  \caption{Switching between \bidiag and \rbidiag.}
  \label{fig:cpjulien}
\end{figure*}

For square matrices, $\bidiag$ is better than $\rbidiag$.  For tall and skinny
matrices, this is the opposite. For a given $q$, what is the ratio
$\delta=p/q$ for which we should switch between $\bidiag$ and $\rbidiag$?
Let  $\delta_s$ denote this crossover ratio.
The question was answered by Chan~\cite{Chan:1982:IAC:355984.355990} when 
considering the operation count. The question has multiple
facets. The optimal switching point is not the same whether one wants singular
values only, right singular vectors, left, or both (left and right).
Chan~\cite{Chan:1982:IAC:355984.355990} answers all these facets. For example,
Chan~\cite{Chan:1982:IAC:355984.355990} shows that the optimal switching point between $\bidiag$ and $\rbidiag$ when
singular values only are sought is $\delta = \frac{5}{3}$.
Here, we consider a similar question but when critical path length (instead of
number of flops) is the objective function.
Since $\bidiagGreedy$ is optimal and $\rbidiagGreedy$ is asymptotically optimal, 
we only focus on these two algorithms. 
It turns out that $\delta_s$ is a complicated function of $q$. The function
of $\delta_s(q)$ oscillates between $5$ and $8$ (see Figure~\ref{fig:cpjulien:2})
To find $\delta_s$, we coded snippets that explicitly compute the critical
path lengths for given $p$ and $q$, and find the intersection for a given $q$. See Figure~\ref{fig:cpjulien} for an
example with $q=100$. We see that, in this case, $\delta_s(100)=5.67$.

\subsection{Critical path length study}

In Figures~\ref{fig:cp-square},~\ref{fig:cp-q10} and~\ref{fig:cp-q40}, we present the critical path length speedup taking
for base \bidiagTS.  The higher the better.  We see that the \Greedy base \bidiag
and \rbidiag algorithms (green curves) are much better any other trees as
expected by the theory. 
In the square case, we do not present \rbidiag (since there is no interest in \rbidiag  in the square case).

We set the dimension of the problem $p$ and $q$ so as to reproduce the 
dimension in our experimental section.
In the experimental section, we will use a tile size
of $160$. 
Figure~\ref{fig:cp-square} studies the square case, 
it is the same condition as Figure~\ref{fig:gebrd-square}.
Figure~\ref{fig:cp-q40} studies the rectangular case when $n=2,000$, so when $q=13$,
it is the same condition as Figure~\ref{fig:gebrd-n2k}.
Figure~\ref{fig:cp-q40} studies the rectangular case when $n=10,000$, so when $q=63$,
it is the same condition as Figure~\ref{fig:gebrd-n10k}.

From Figure~\ref{fig:cp-square}, 
we see that the critical path of \bidiagGreedy for a square 40x40 tile matrix is more than 6 times shorter than for \bidiagTS.

From Figure~\ref{fig:cp-q10}, 
we see that the critical path of \bidiagGreedy and \rbidiagGreedy for a square 400x13 tile matrix is more than 60 times shorter than for \bidiagTS.
We also the crossover point between \bidiagGreedy and \rbidiagGreedy.
For ``very tall and skinny'' matrix ($p\gg q$), \rbidiagGreedy is better than \bidiagGreedy.
For ``not-too tall and skinny'' matrix ($p\approx q$), \bidiagGreedy is better than \rbidiagGreedy.

\subsection{Comments}

\paragraph{When singular vectors are requested.} It is important to understand
that the optimality in Theorem~\ref{theo:1} assumes that we do not request
singular vectors. The whole analysis is significantly changed if one requests
singular vectors.  In this paper, we explain that, since there is no overlap
between QR and LQ steps, then the shortest trees at each step makes up an
optimal algorithm. When we apply the trees in reverse to obtain the singular
vectors, then we apply all the LQ steps in sequence (without the QR steps) to
obtain $W$, and we apply all the QR steps in sequence (without the LQ steps) to
obtain $U$. In this case, we can clearly pipeline the trees as in a regular QR
factorization.  It is therefore interesting to use a ``full'' greedy tree or
even a standard \FlatTT tree where a pipeline happens. The \bidiagGreedy algorithm
becomes less interesting.  This paper focuses only on the bidiagonal phase.
This means in some sense that we are interested in the singular value problem
where we only request the singular values, we do not request the singular
vectors. Another study should be done when the singular vectors are requested.
However, we believe that, despite the absence of pipeline in the construction of the singular vectors phase, \bidiagGreedy 
is still an algorithm of choice since the overhead in the construction is of order $\log_2p$
while the overhead of the other algorithms in the bidiagonalization is of order $p$.

\begin{figure*}[hbtp]
\begin{center}
    \includegraphics[width=.5\textwidth]{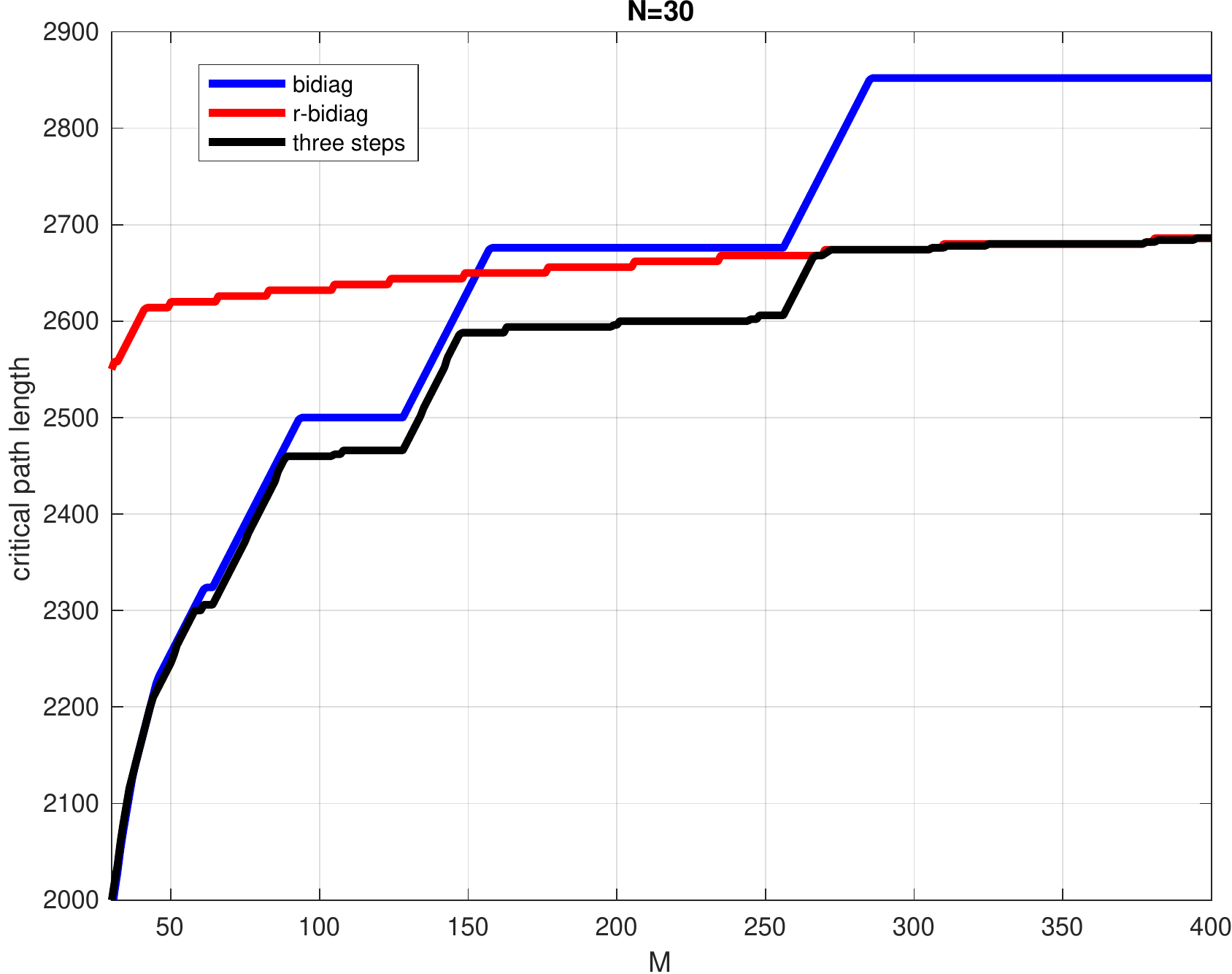}
    \caption{\bidiag, \rbidiag and three-step algorithm for $q=30$ and $p$ varying from 30 to 400.}
    \label{fig:threesteps}
\end{center}
\end{figure*}

\paragraph{Three-step algorithm: partialGE2BD+preQR+GE2BD.} In 1997, Trefethen and
Bau~\cite{Trefethen:1997:NLA} present a ``three-step'' approach.
The idea is that having a cut-off when to switch from \bidiag to \rbidiag algorithm seems 
unnatural and leads to a continuous but not smooth function. Smooth is better and seems more natural.
So instead of doing either \rbidiag or \bidiag, we consider the following algorithm.
We start with \bidiag until the matrix is tall and skinny enough. 
(We note that \bidiag makes the matrix more and more tall and skinny.)
Once the matrix is tall and skinny enough, we switch to \rbidiag.
Trefethen and
Bau~\cite{Trefethen:1997:NLA} studied the  ``three-step'' approach where they want to minimize the 
number of flops. 
Here we experimentally study their ``three-step'' approach  when we want to minimize the 
critical path length.
We do not have an exact formula to know when to stop the first bidiagonalization. 
Therefore, for a first study, we consider the brute force approach and consider all possible stops for the first 
bidiagonalization, and then we take the one stop that gives the shortest critical path length. 
In fine, we obtain for example a figure like Figure~\ref{fig:threesteps}.

We fix $q=30$ and consider $p$ varying from $30$ to $400$.
We see that three-step algorithm (black curve) is always the best.
(This is expected because we consider all possibilities including \bidiag and \rbidiag, and take the minimum over all possibilities.
So clearly, three-step will always be better than \bidiag and \rbidiag.)
Initially  three-step is the same as \bidiag. Then, in the middle region $p=60$ to $260$, 
we see that three-step is better than both \bidiag and \rbidiag. 
This is when indeed we do not perform \bidiag, we do not perform \rbidiag, we perform a ``true'' three-step 
algorithm where we start a bidiagonalization algorithm, then stop the bidiagonalization, switch to a QR 
factorization of the remaining submatrix, and finish with a bidiagonalization. Finally after $p=260$, 
three-step is same as \rbidiag.

Once more,  the idea behind three-step is that: (1) \rbidiag is better for tall-and-skinny matrices;
(2) the \bidiag process makes the trailing submatrix more and more tall-and-skinny. 
So the idea is to start \bidiag until the matrix is tall-and-skinny enough and switch to \rbidiag.
If we consider the minimum of \bidiag (blue curve) and \rbidiag (red curve) then we have a non-smooth function.
This is not natural. When we consider the three-step algorithm (black curve), we have a nice smooth function.
This is more natural.

We see that three-step is better than \bidiag and \rbidiag. However when we look at the gain (y-axis), 
we do not have staggering gain. So, while three-step is an interesting and elegant idea, we did not push its study further.

\section{Implementation}
\label{sec.dague}

To evaluate experimentally the impact of the different reduction trees on the
performance of the GE2BND and GE2VAL algorithms, we have implemented both
the \bidiag and \rbidiag algorithms in the \dplasma library~\cite{dplasma-6008998}, 
which runs on top of
the \parsec runtime system~\cite{dague-engine-Bosilca201237}.
\parsec is a high-performance fully-distributed scheduling environment
for generic data-flow algorithms. It takes as input a
problem-size-independent, symbolic representation of a Direct Acyclic
Graph in which each node represents a tasks, and each edge a
dependency, or data movement, from one task to another.
\parsec schedules those tasks on distributed
parallel machine of multi-cores, potentially heterogeneous, while
complying with the dependencies expressed by the programmer.
At runtime, tasks executions trigger data movements, and create new
ready tasks, following the dependencies defined by the DAG representation.
The runtime engine is responsible for actually moving the data from
one machine (node) to another, if necessary, using an underlying
communication mechanism, like MPI. 
Tasks that are ready to compute are scheduled
according to a data-reuse heuristic: each core will try to execute close
successors of the last task it ran, under the assumption that these
tasks require data that was just touched by the terminated one. This
policy is tuned by the user through a priority function: among the
tasks of a given core, the choice is done following this function. To
balance load between the cores, tasks of a same cluster in the
algorithm (reside on a same shared memory machine) are shared between
the computing cores, and a NUMA-aware job stealing policy is
implemented.
The user is then responsible only to provide the algorithm, the
initial data distribution, and potentially the task distribution. The
last one is usually correlated to the data distribution when the
(default) owner-compute rule is applied. In our case, we use a $2D$
block-cyclic data distribution as used in the \scalapack library, and
we map the computation together with the data.
A full description of \parsec can be found in~\cite{dague-engine-Bosilca201237}.

The implementation of the \bidiag and \rbidiag algorithms have then
been designed as an extension of our previous work on HQR
factorization~\cite{j125} within the \dplasma library.
The HQR algorithm proposes to perform the tiled QR
factorization of a $(p \times q)$-tile matrix, with $p \geq q$, by using a variety of 
 trees that are optimized for both the target
architecture and the matrix size. It relies on multi-level reduction
trees. The highest level is a tree of size $R$, where $R$ is the number of rows
in the $R \times C$ two-dimensional grid distribution of the matrix, and it is
configured by default to be a flat tree if $p \geq 2q$, and a Fibonacci
tree otherwise. The second level, the domino level, is an optional
intermediate level that enhances the pipeline of the lowest levels
when they are connected together by the highest distributed tree. It
is by default disabled when $p \geq 2q$, and enabled otherwise.
Finally,  the last two levels of trees are use to create
parallelism within a node and work only on local tiles. They correspond to
a composition of one or multiple \FlatTS trees that are connected
together with an arbitrary tree of TT kernels. The bottom \FlatTS tree
enables highly efficient kernels while the TT tree on top of it
generates more parallelism to feed all the computing resources from
the architecture. The default is to have \FlatTS trees of size $4$
that are connected by a \Greedy tree in all cases.
This design is for QR trees, a similar design exists for LQ trees.
Using these building blocks,
we have crafted an implementation of  \bidiag and \rbidiag
within the abridged representation
used by \parsec to represent algorithms. This implementation is independent of 
the type of trees selected for the computation, thereby allowing the user to test a large
spectrum of configuration without the harassment of rewriting all the
algorithm variants.

\begin{figure*}[!htbp]
  \begin{center}
  \begin{subfigure}{0.3\textwidth}
    \includegraphics[width=\textwidth]{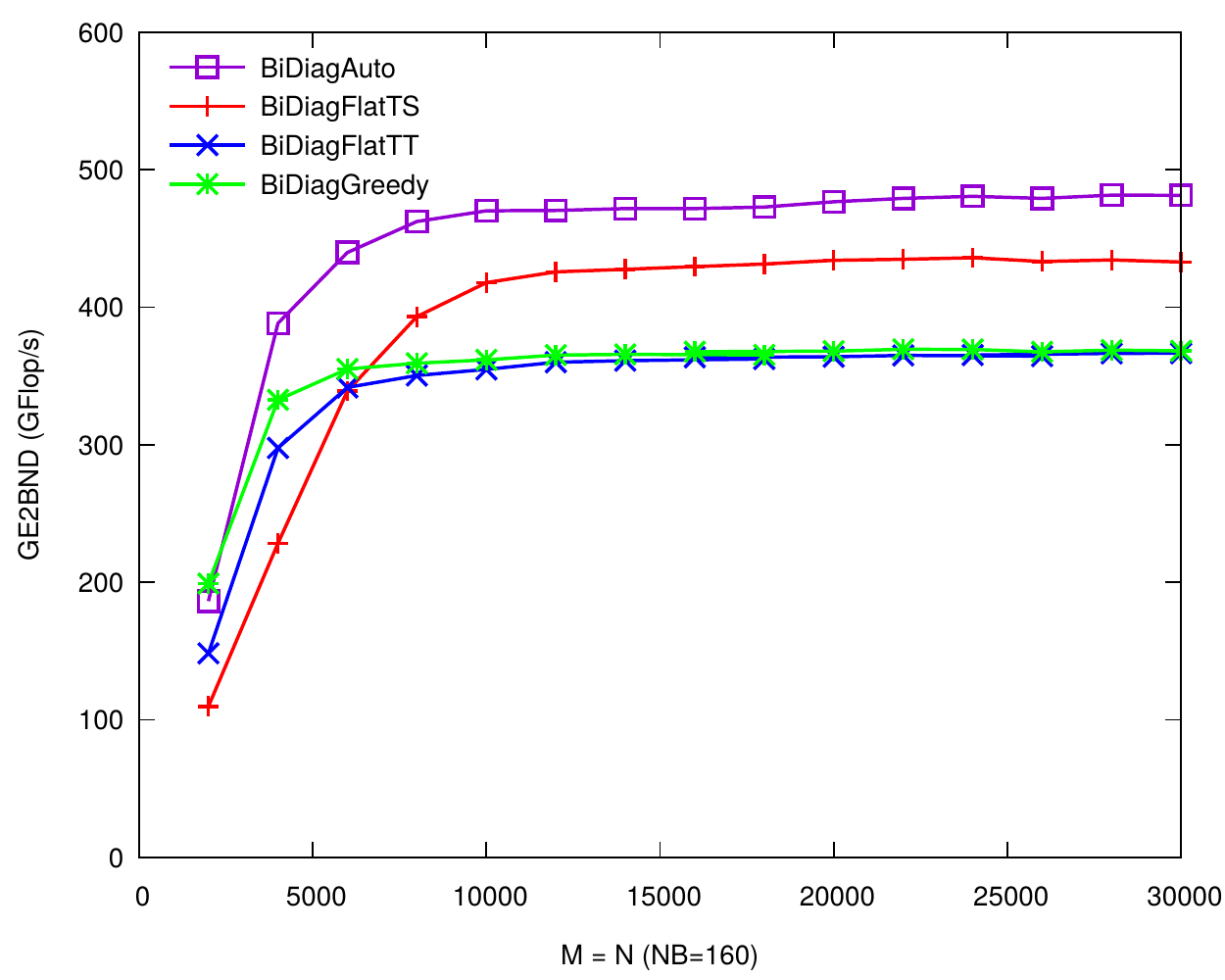}
    \caption{Square ($M=N$)}
    \label{fig:gebrd-square}
  \end{subfigure}
  \begin{subfigure}{0.3\textwidth}
    \includegraphics[width=\textwidth]{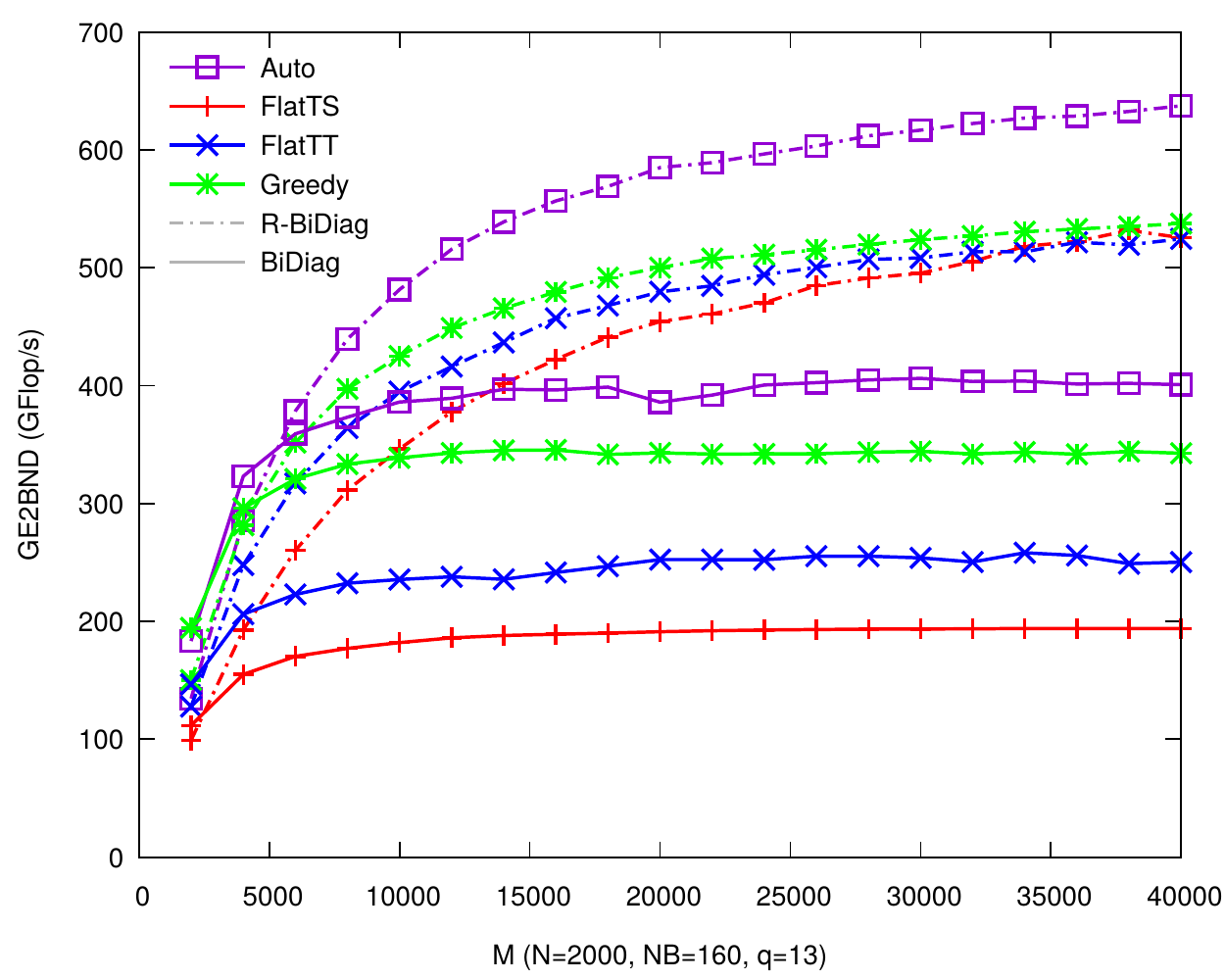}
    \caption{Tall and skinny ($N=2000$)}
    \label{fig:gebrd-n2k}
  \end{subfigure}
  \begin{subfigure}{0.3\textwidth}
    \includegraphics[width=\textwidth]{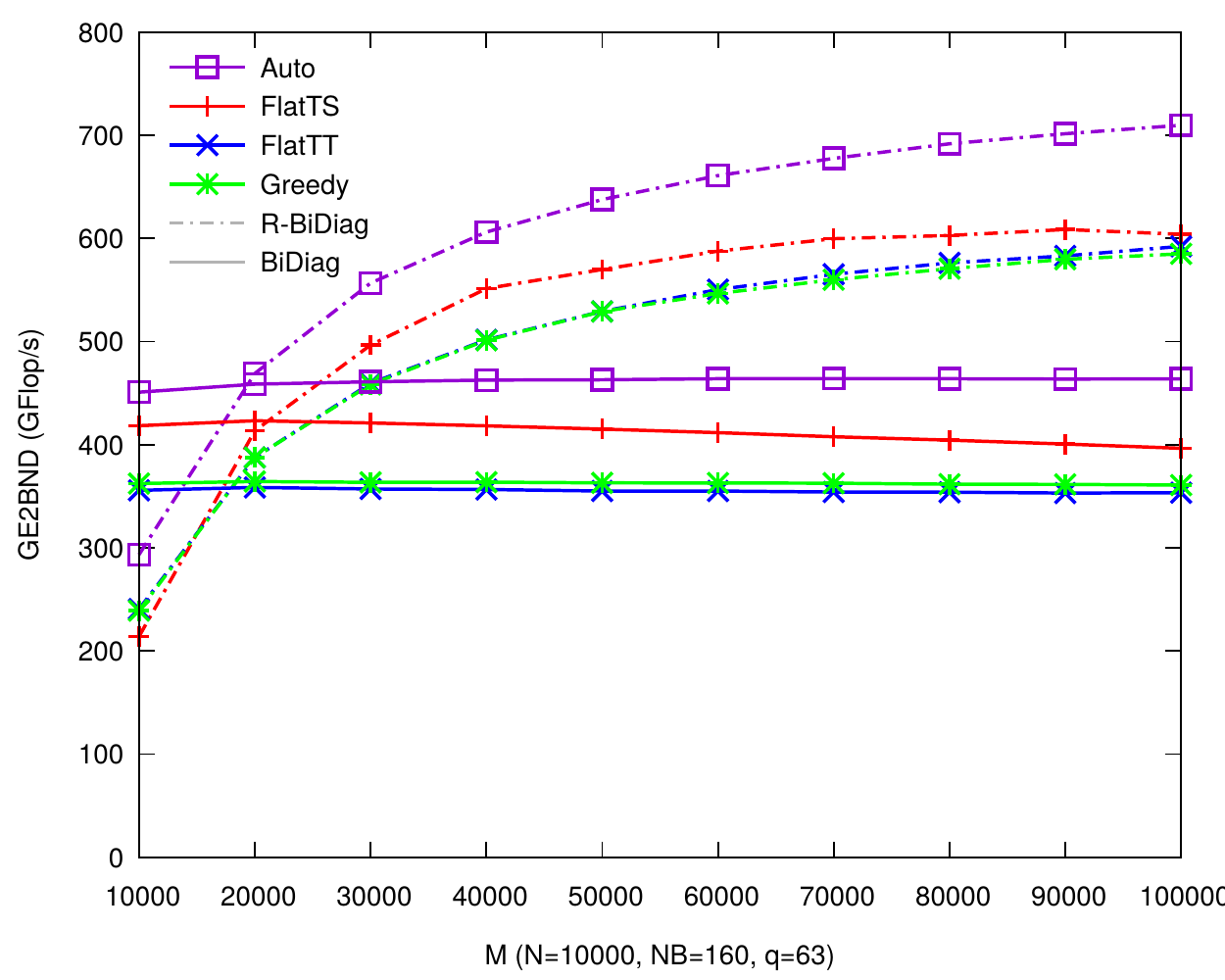}
    \caption{Tall and skinny ($N=10000$)}
    \label{fig:gebrd-n10k}
  \end{subfigure}

  \begin{subfigure}{0.3\textwidth}
    \includegraphics[width=\textwidth]{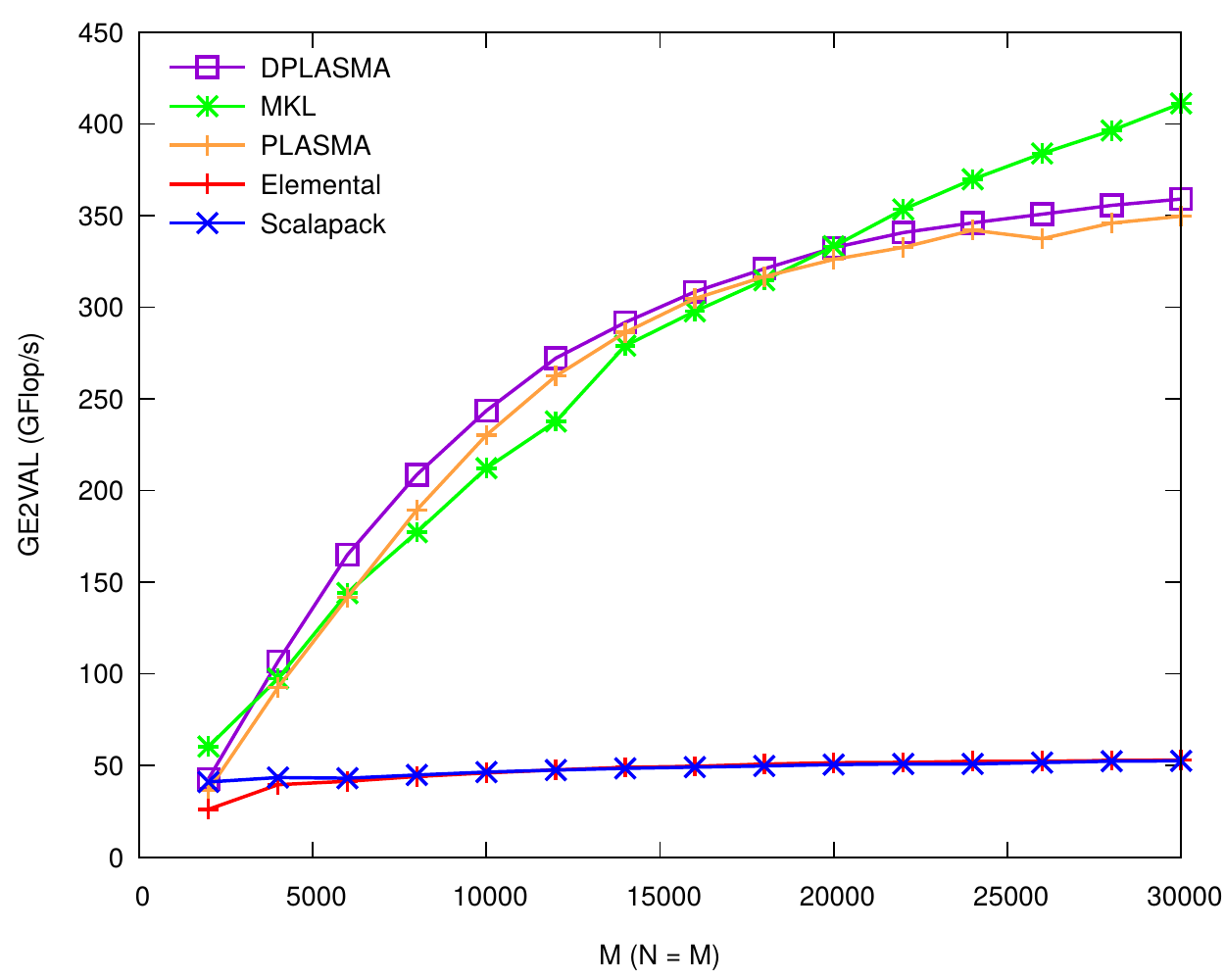}
    \caption{Square ($M=N$)}
    \label{fig:gesvd-square}
  \end{subfigure}
  \begin{subfigure}{0.3\textwidth}
    \includegraphics[width=\textwidth]{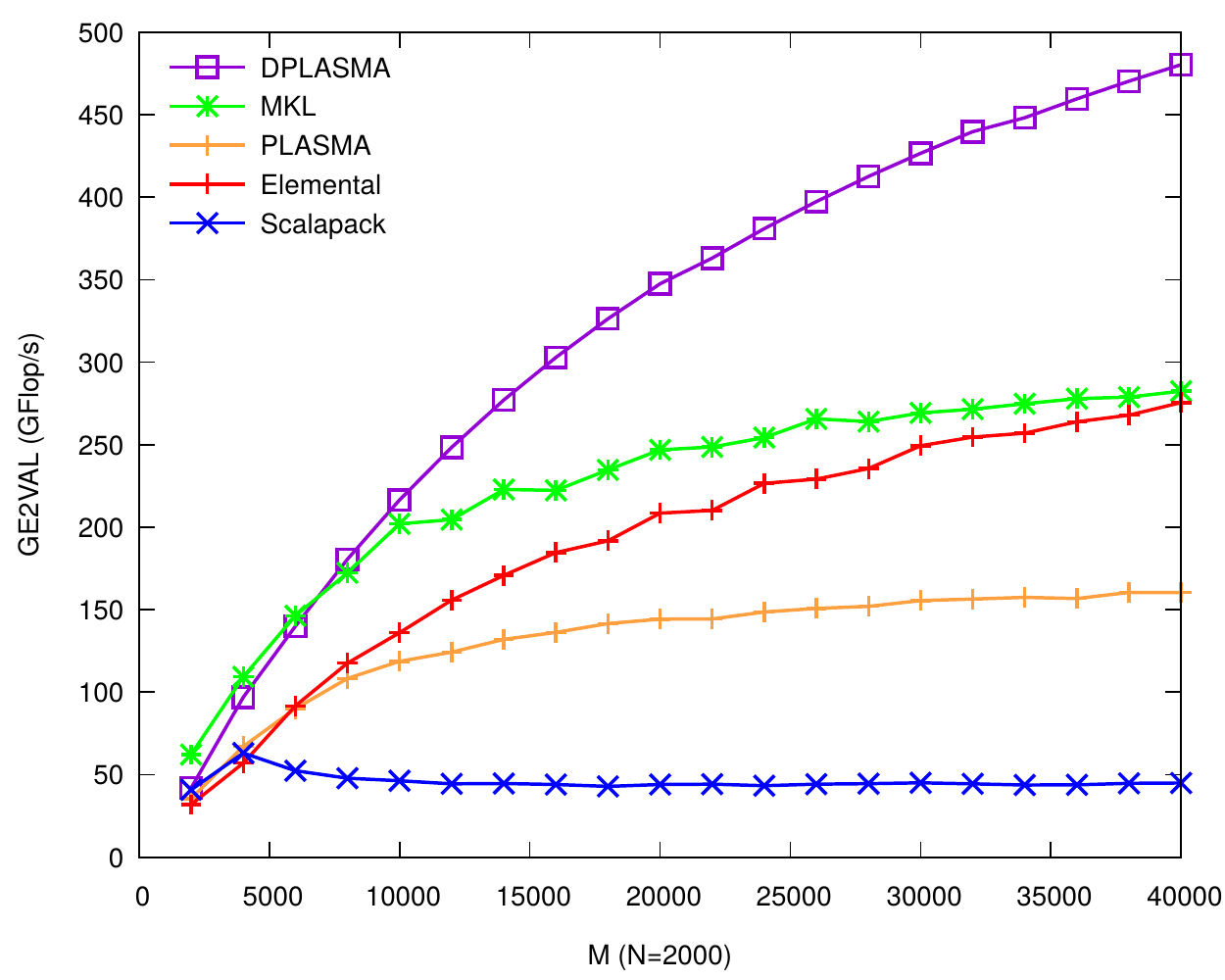}
    \caption{Tall and skinny ($N=2000$)}
    \label{fig:gesvd-n2k}
  \end{subfigure}
  \begin{subfigure}{0.3\textwidth}
    \includegraphics[width=\textwidth]{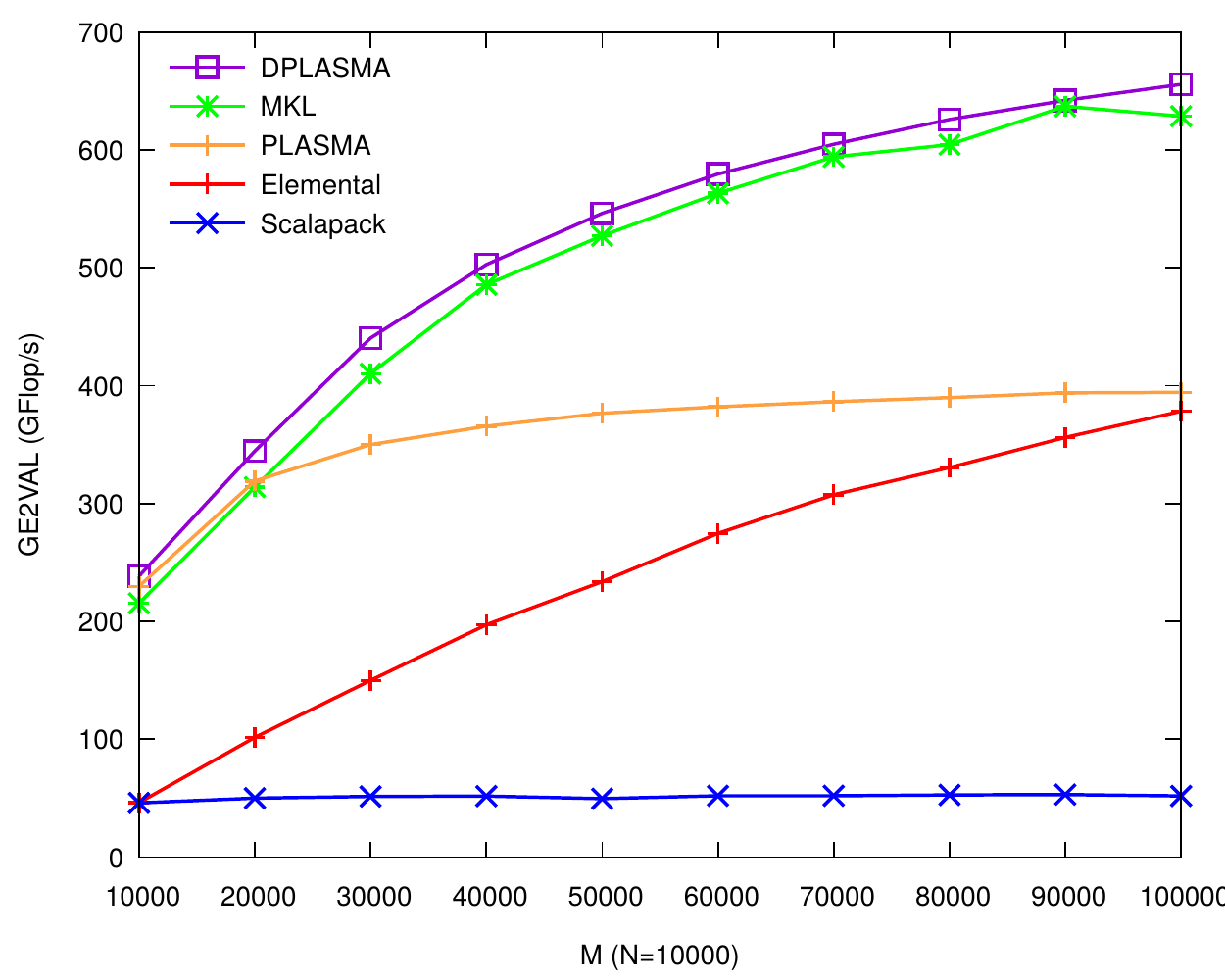}
    \caption{Tall and skinny ($N=10000$)}
    \label{fig:gesvd-n10k}
  \end{subfigure}

  \end{center}
  \caption{Shared memory performance of the multiple variants for the
    GE2BND algorithm on the first row, and for the GE2VAL algorithm
    on the second row, using a single  $24$ core node of the \texttt{miriel} cluster.}
  \label{fig:shm}
\end{figure*}

One important contribution is the introduction of two new tree structures
dedicated to the \bidiag algorithm.
The first tree, \Greedy, is a binomial tree which reduces a panel in the minimum amount of steps. 
The second tree, \Auto,
is an adaptive tree which automatically adapts to the size of the
local panel and number of computing resources.
We developed the auto-adaptive tree to take advantage of (i) the
higher efficiency of the TS kernels with respect to the TT kernels, (ii)
the highest degree of parallelism of the \Greedy tree with respect to
any other tree, and (iii) the complete independence of each step of the
\bidiag algorithm, which precludes any possibility of pipelining.
Thus, we propose to combine in this configuration a set of \FlatTS
trees connected by a \Greedy tree, and to automatically adapt the
number of \FlatTS trees, and by construction their sizes, $a$, to provide
enough parallelism to the available computing resources.
Given a matrix of size $p \times q$, at each step $k$, we need to
apply a QR factorization on a matrix of size $(p-k-1) \times (q-k-1)$, the
number of parallel tasks available at the step beginning of the step
is given by $\lceil (p-k-1) / a \rceil * (q-k-1)$. Note that we
consider the panel as being computed in parallel of the update, which
is the case when $a$ is greater than $1$, with an offset of one time
unit. Based on this formula, we compute $a$ at each step of the
factorization such that the degree of parallelism is greater than
a quantity $\rratio \times \nbcores$, where \rratio is a parameter and \nbcores is the number of cores. For the experiments, we set $\rratio=2$.
Finally, we point out that \Auto is defined for a resourced-limited platform, hence computing its critical path would have no meaning, which explains a posteriori that
it was not studied in Section~\ref{sec.critical}.

\section{Experiments}
\label{sec.expes}

In this section, we evaluate the performance of the proposed algorithms
for the GE2BND kernel against existing competitors.

\subsection{Architecture}

Experiments are carried out using the PLAFRIM
experimental testbed\footnote{Inria PlaFRIM
development action with support from Bordeaux INP, LABRI and IMB and
other entities: Conseil R\'egional d'Aquitaine, Universit\'e de Bordeaux
and CNRS, see \url{https://www.plafrim.fr/}.}.
We used up to $25$ nodes of the \texttt{miriel}
cluster, each equipped with $2$ Dodeca-core Haswell Intel Xeon
E5-2680 v3 and 128GB of memory. The nodes are interconnected with an
Infiniband QDR TrueScale network with provides a bandwidth of 40Gb/s.
All the software are compiled with gcc 4.9.2, and linked against the
sequential BLAS implementation of the Intel MKL 11.2 library. For the
distributed runs, the MPI library used is OpenMPI 2.0.0.
The practical GEMM performance is of $37$ GFlop/s on one core, and
$642$ GFlop/s when the 24 cores are used.
For each experiment, we generated a matrix with prescribed singular values
using \lapack \texttt{LATMS} matrix generator and checked that the computed 
singular values were satisfactory up to machine precision.

\subsection{Competitors}

This paper presents new parallel distributed algorithms and implementations for
GE2BND using DPLASMA.  To compare against competitors on GE2VAL, we follow up
our DPLASMA GE2BND implementation with the PLASMA multi-threaded BND2BD algorithm, and then use
the Intel MKL multi-threaded BD2VAL implementation.  We thus obtain GEVAL by
doing GE2BND+BND2BD+BD2VAL.

It is important to note that we do not use parallel distributed implementations
neither for BND2BD nor for BD2VAL.  We only use shared memory implementations for these two
last steps.  Thus, for our distributed memory runs, after the GE2BND step in
parallel distributed using DPLASMA, the band is gathered on a single node, and
BND2BD+BD2VAL is performed by this node while all all other nodes are left idle. 
We will show that, despite this current limitation for parallel distributed, 
our implementation outperforms its competitors.

On the square test cases, only $23$ cores of a 24-core node were used for
computation, and the $24^{th}$ core was left free to handle MPI communications progress.

The implementation of the algorithm is available in a public fork of the
DPLASMA library at \url{https://bitbucket.org/mfaverge/parsec}.

{\bf PLASMA} is the closest alternative to our proposed solution but 
it is only using \FlatTS as its reduction tree, and
is limited to single-node platform, and
is supported by a different runtime.
For both our
code, and PLASMA, the tile size parameter is critical to get good
performance: a large tile size will get an higher kernel efficiency
and a faster computation of the band, but it will increase the number
of flops of the BND2BD step which is heavily memory bound. On the
contrary, a small tile size will speed up the BND2BD step by fitting
the band into cache memory, but decreases the efficiency of the kernels
used in the GE2BND step. We tuned the $n_{b}$ (tile size) and $i_{b}$ (internal 
blocking in \emph{TS} and \emph{TT} kernels) parameters to get
the better performance on the square case $m=n=20000$, and
$m=n=30000$ on the PLASMA code. The selected values are $n_{b}=160$, and $i_{b}=32$.
We used the same parameters in the
DPLASMA implementation for both the shared memory runs ant the
distributed ones. The PLASMA 2.8.0 library was used.

{\bf Intel MKL} proposes an multi-threaded implementation of the GE2VAL
algorithm which gained an important speedup while switching from
version 11.1 to 11.2~\cite{intel}.
While it is unclear which algorithm is used beneath, the speedup
reflects the move to a multi-stage algorithm. 
{\bf Intel MKL} is limited to single-node platforms.

{\bf \scalapack} implements the parallel distributed version of the \lapack
GEBRD algorithm which interleaves phases of memory bound BLAS2 calls
with computational bound BLAS3 calls. It can be used either with one
process per core and a sequential BLAS implementation, or with a
process per node and a multi-threaded BLAS implementation. The latter
being less efficient, we used the former for the experiments. The
blocking size $n_{b}$  is critical to get performances since it impacts
the phase interleaving. We tuned the $n_{b}$ parameter to get the better
performance on a single node with the same test cases as for PLASMA,
and $n_{b}=48$ was selected.

{\bf Elemental} implements an algorithm similar to \scalapack, but it
automatically switches to Chan's algorithm~\cite{Chan:1982:IAC:355984.355990} 
when $m \geq 1.2 n$. As for
\scalapack, it is possible to use it as a single MPI implementation,
or an hybrid MPI-thread implementation. The first one being
recommended, we used this solution. Tuning of the $n_{b}$ parameter
similarly to previous libraries gave us the value $nb=96$. A better 
algorithm developed on top of the LibFLAME~\cite{flame-Gunnels:2001:FFL:504210.504213} is provided by Elemental, but this one is used only when singular vectors are
sought.

In the following, we compare all these implementation on the
\texttt{miriel} cluster with $3$ main configurations: (i) square
matrices; (ii) tall and skinny matrices with $n=2,000$; this choice restricts the
level of parallelism induced by the number of panels to half the
cores; and (iii) tall and skinny matrices with $n=10,000$: this choice enables for more parallelism.
For all performance comparisons, we use the same operation count
as in~\cite[p. 123]{lawn41} for the GE2BND and GE2VAL algorithms.
The BD2VAL step
has a  negligible cost $O(n^2)$. For \rbidiag, we use the same number of flops as for
\bidiag; in other words,  we do not  assess the
absolute performance of \rbidiag,  instead we provide a direct comparison with \bidiag.

\subsection{Shared Memory}
\begin{figure*}[!htbp]
  \begin{center}
  \begin{subfigure}{0.32\textwidth}
    \includegraphics[width=\textwidth]{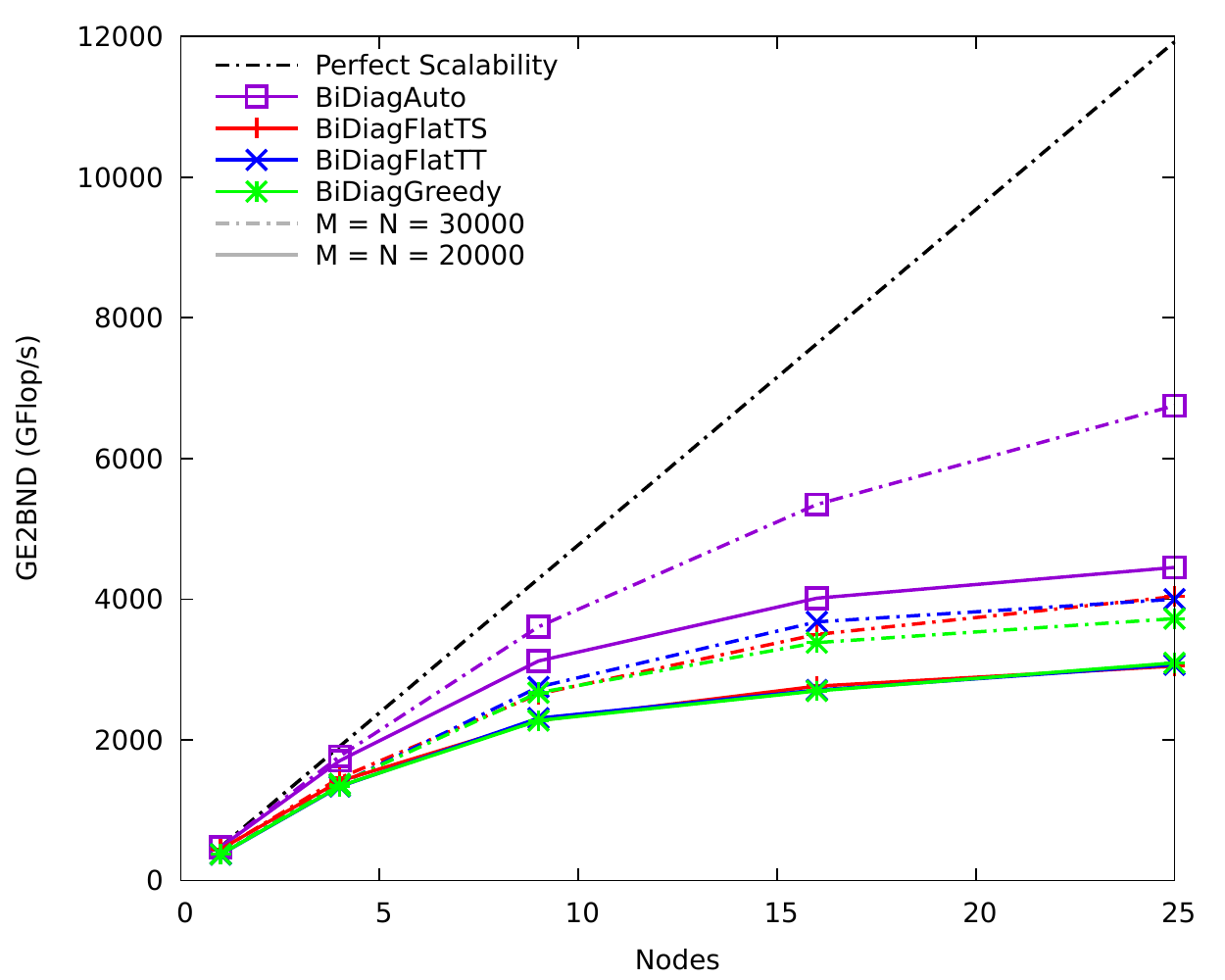}
    %% \caption{Square ($M=N$)}
    %% \label{fig:dist-gebrd-square}
  \end{subfigure}
  \begin{subfigure}{0.32\textwidth}
    \includegraphics[width=\textwidth]{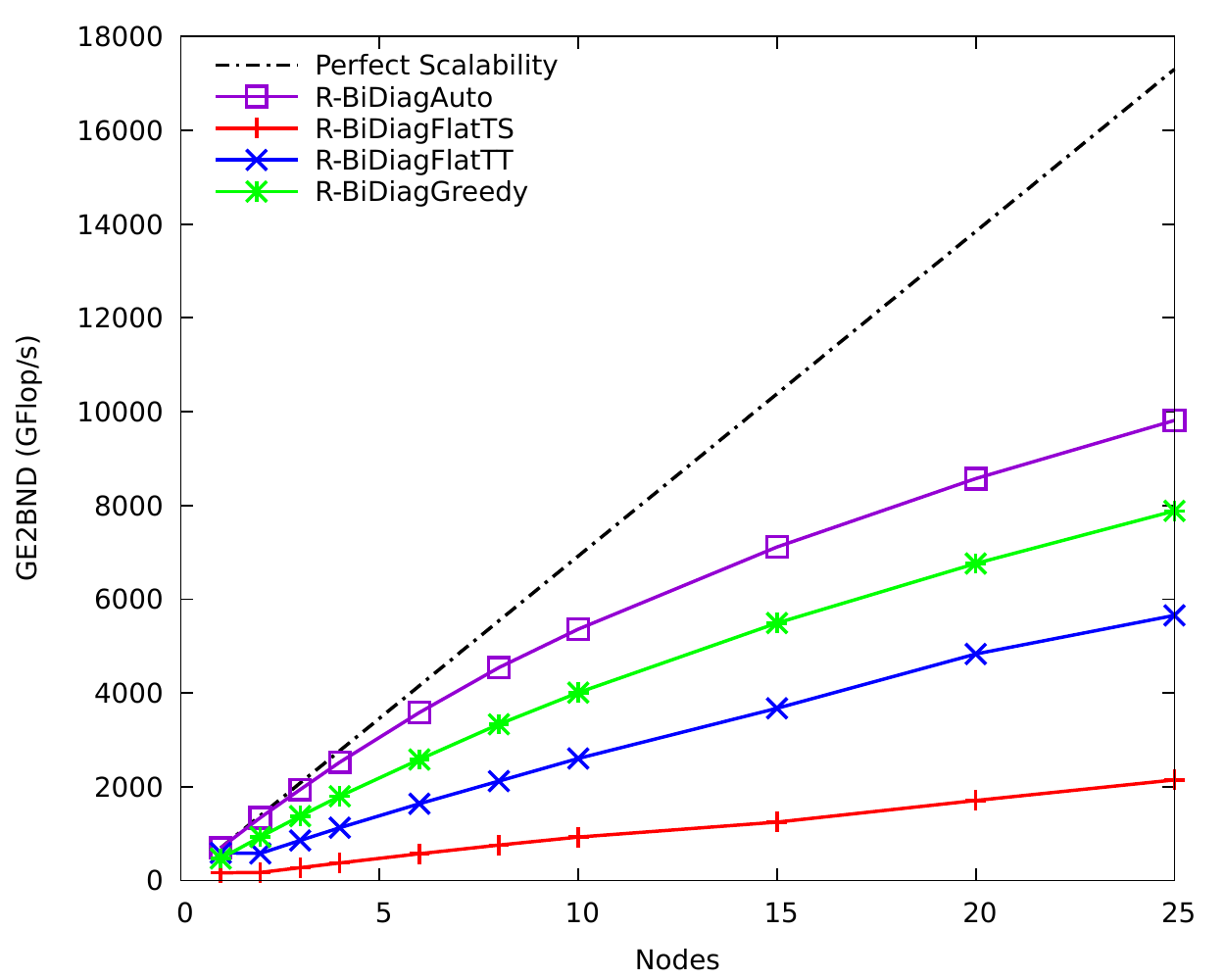}
    %% \caption{Tall and skinny ($N=2000$)}
    %% \label{fig:dist-gebrd-n2k}
  \end{subfigure}
  \begin{subfigure}{0.32\textwidth}
    \includegraphics[width=\textwidth]{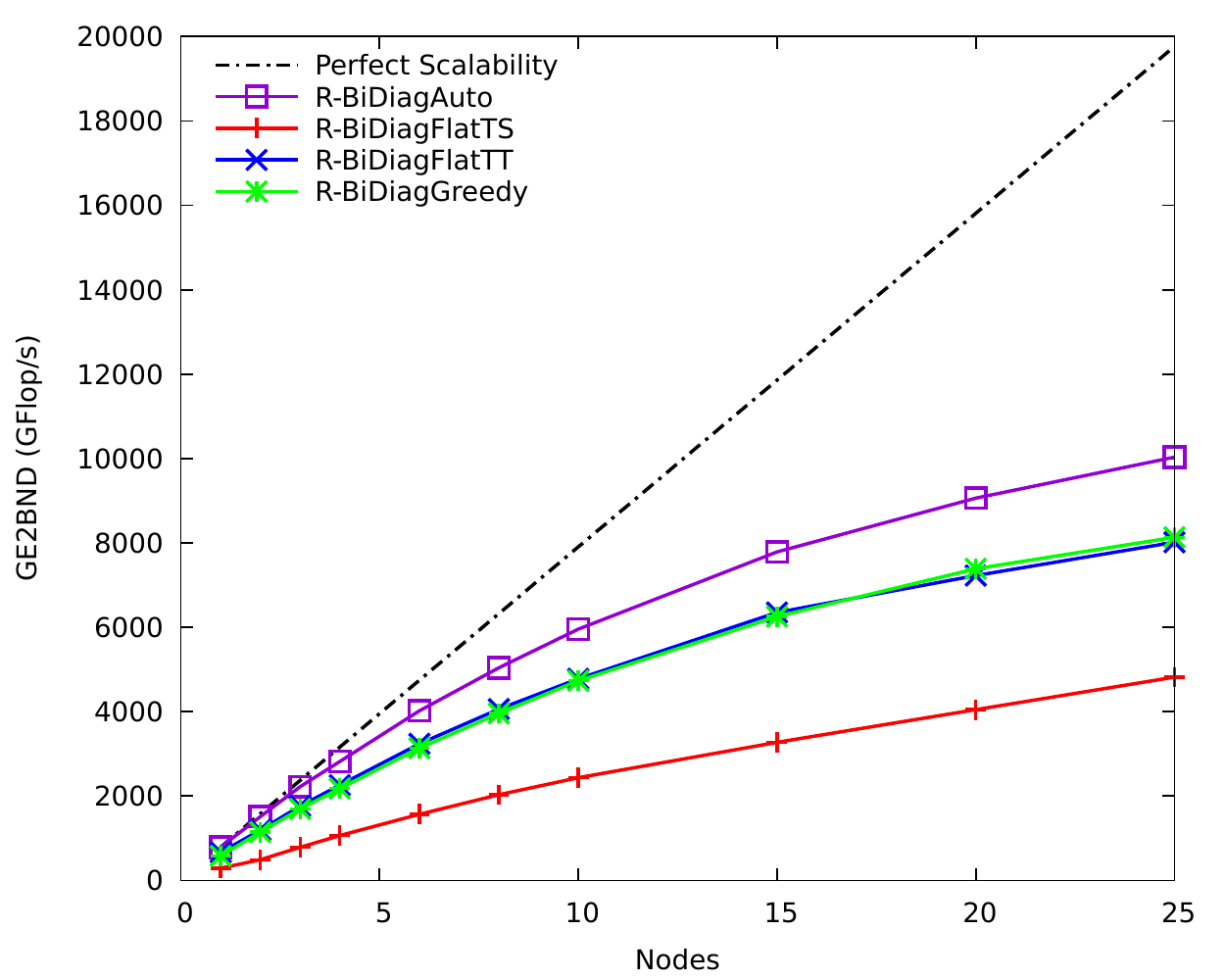}
    %% \caption{Tall and skinny ($N=10000$)}
    %% \label{fig:dist-gebrd-n10k}
  \end{subfigure}

  \begin{subfigure}{0.32\textwidth}
    \includegraphics[width=\textwidth]{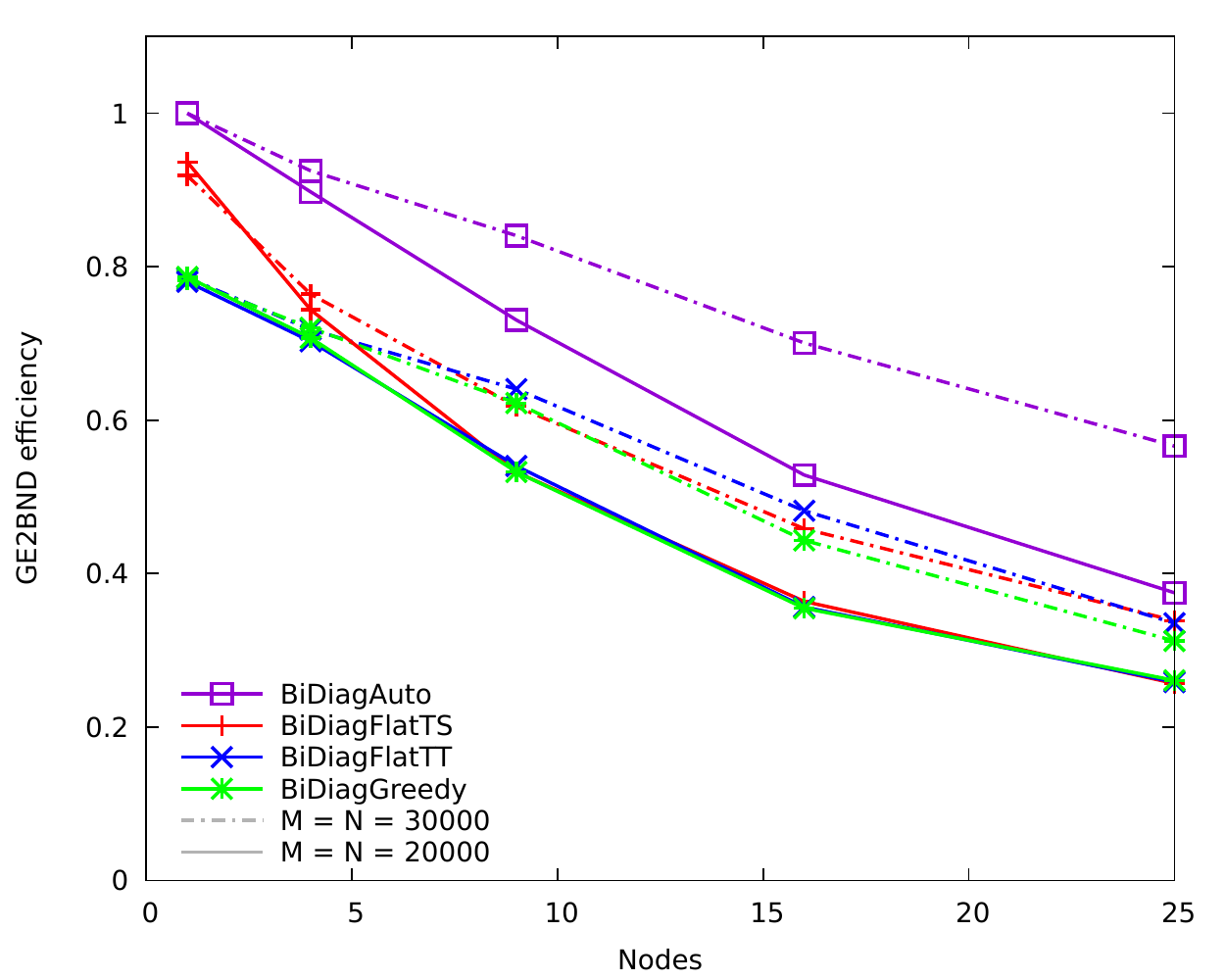}
    \caption*{Square ($m=n$)}
    %% \label{fig:dist-gebrd-square}
  \end{subfigure}
  \begin{subfigure}{0.32\textwidth}
    \includegraphics[width=\textwidth]{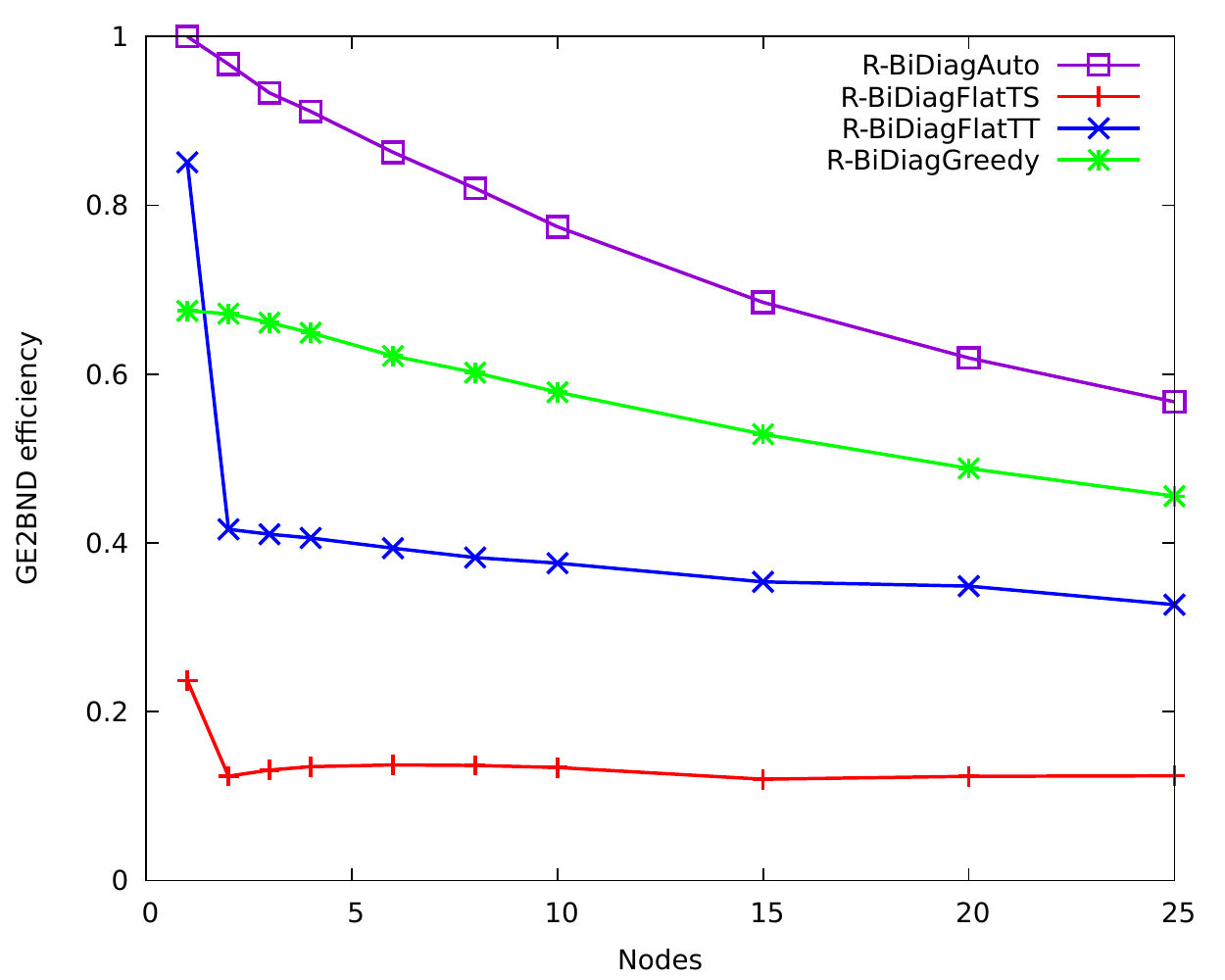}
    \caption*{Tall \& skinny \footnotesize{($m\!=\!2,000,000, n\!=\!2,000$)}}
    %% \label{fig:dist-gebrd-n2k}
  \end{subfigure}
  \begin{subfigure}{0.32\textwidth}
    \includegraphics[width=\textwidth]{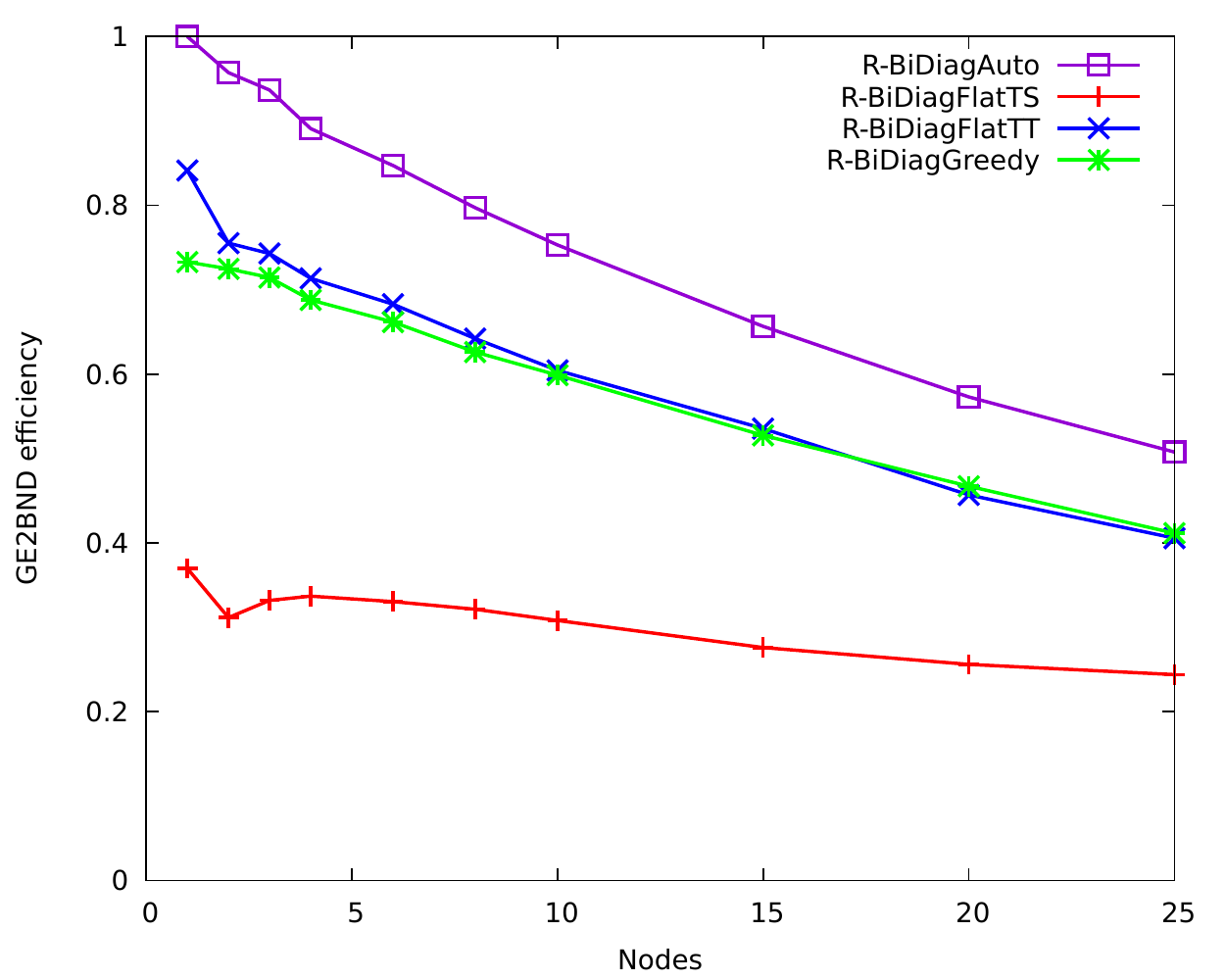}
    \caption*{Tall and skinny \footnotesize{($m\!=\!1,000,000, n\!=\!10,000$)}}
    %% \label{fig:dist-gebrd-n10k}
  \end{subfigure}

  \end{center}
  \caption{Distributed memory performance (first row) and efficiency (second row) of the multiple
    variants for the GE2BND algorithm on the \texttt{miriel}
    cluster. Grid data distributions are
    $\sqrt{\nbnodes} \times \sqrt{\nbnodes}$ for square matrices, and $\nbnodes \times 1$
    for tall and skinny matrices. For the square case, solid lines are for $m=n=20,000$ and dashed lines
    for $m=n=30,000$.}
  \label{fig:dist-strong-gebrd}
\end{figure*}

The top row of Figure~\ref{fig:shm} presents the performance of
the three configurations selected for our study of GE2BND. On
the top left, the square case perfectly illustrates the strengths and
weaknesses of each configuration. On small matrices, \FlatTT in
blue and \Greedy in green illustrate the importance of
creating algorithmically more parallelism to feed all 
resources. However, on large size problems, the performance is limited
by the lower efficiency of the TT kernels. The \FlatTS tree behaves at
the opposite: it provides better asymptotic
performance thanks to the TS kernels, but lacks parallelism when
the problem is too small to feed all cores. \Auto is able to benefit from the advantages of both
\Greedy and \FlatTS trees to provide a significant improvement on small
matrices, and a $10\%$ speedup on the larger matrices.

For the tall and skinny matrices, we observe that the \rbidiag
algorithm (dashed lines) quickly outperforms the \bidiag algorithm,
and is up to $1.8$ faster. On the small case ($n=2,000$), the crossover
point is immediate, and both  \FlatTT and \Greedy, exposing more
parallelism,  are able to get better performances than  \FlatTS. On
the larger case ($n=10,000$), the parallelism from the larger matrix size
allows \FlatTS to perform better, and to postpone the crossover
point due to the ratio in the number of flops. In both cases, \Auto 
provides the better performance with an extra
$100$ GFlop/s.

On the bottom row of Figure~\ref{fig:shm}, we compare our best
solutions, namely \Auto
tree with \bidiag for square cases and with \rbidiag on tall and
skinny cases, to the competitors on the GE2VAL
algorithm. The difference between our solution and PLASMA, which is
using the \FlatTS tree, is not as impressive due to the additional
BND2BD and BD2VAL steps which have limited parallel efficiency. Furthermore, in our
implementation, due to the change of runtime, we cannot pipeline the GE2BND and
BND2BD steps to partially overlap the second step. However these two
solutions still provide a good improvement over MKL
which is slower on the small cases but overtakes at larger sizes.
For such sizes, Elemental and \scalapack are note able to scale and reach
up a maximum of $50$ Gflop/s due to their highly memory bound algorithm.

On the tall and skinny cases, differences are more emphasized. 
We see the limitation of using only the \bidiag algorithm on MKL,
PLASMA and \scalapack, while our solution and elemental keep scaling up
with matrix size. We also observe that MKL behaves correctly on
the second test case, while it quickly saturates in the first one where
the parallelism is less important. In that case, we are able to reach twice the
MKL performance.

\subsection{Distributed Memory}
\paragraph{Strong Scaling} Figure~\ref{fig:dist-strong-gebrd} presents a
scalability study of the three variants on $4$ cases: two square
matrices with \bidiag, and two tall and skinny matrices
with \rbidiag. For all of them, we couple high-level
distributed trees, and low-level shared memory trees. \FlatTS and
\FlatTT configuration are coupled with a high level flat tree, while
\Greedy and \Auto are coupled with a high
level \Greedy tree. The configuration of the preQR step is setup
similarly, except for \Auto which is using the
automatic configuration described previously.

On all cases, performances are as expected. \FlatTS, 
which is able to provide higher efficient kernels, hardly  behaves better on
the large square case; \Greedy, which provides better
parallelism,  is the best solution out of the three on the first tall and
skinny case. We also observe the impact of the high level tree: 
\Greedy doubles the number of
communications on square cases~\cite{j125}, which impacts its performance and
gives an advantage to the flat tree which performs half the
communication volume.  Overall, \Auto keeps taking benefit
from its flexibility, and scales well despite the fact that
local matrices are less than $38 \times 38$ tiles, so less than $2$
columns per core.

\begin{figure*}[t]
  \begin{center}
  \begin{subfigure}{0.32\textwidth}
    \includegraphics[width=\textwidth]{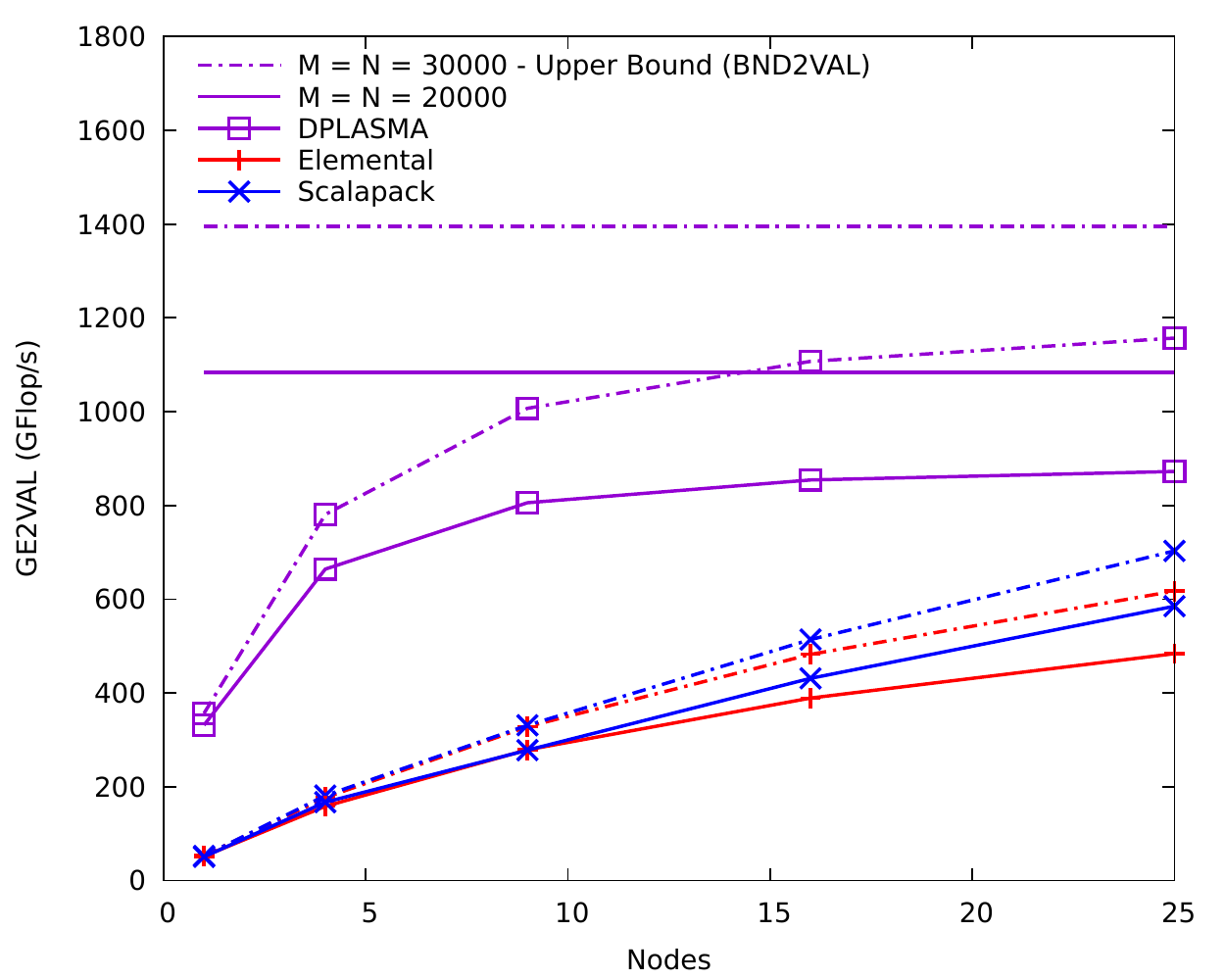}
    %% \caption{Square ($M=N$)}
    %% \label{fig:dist-gesvd-square}
  \end{subfigure}
  \begin{subfigure}{0.32\textwidth}
    \includegraphics[width=\textwidth]{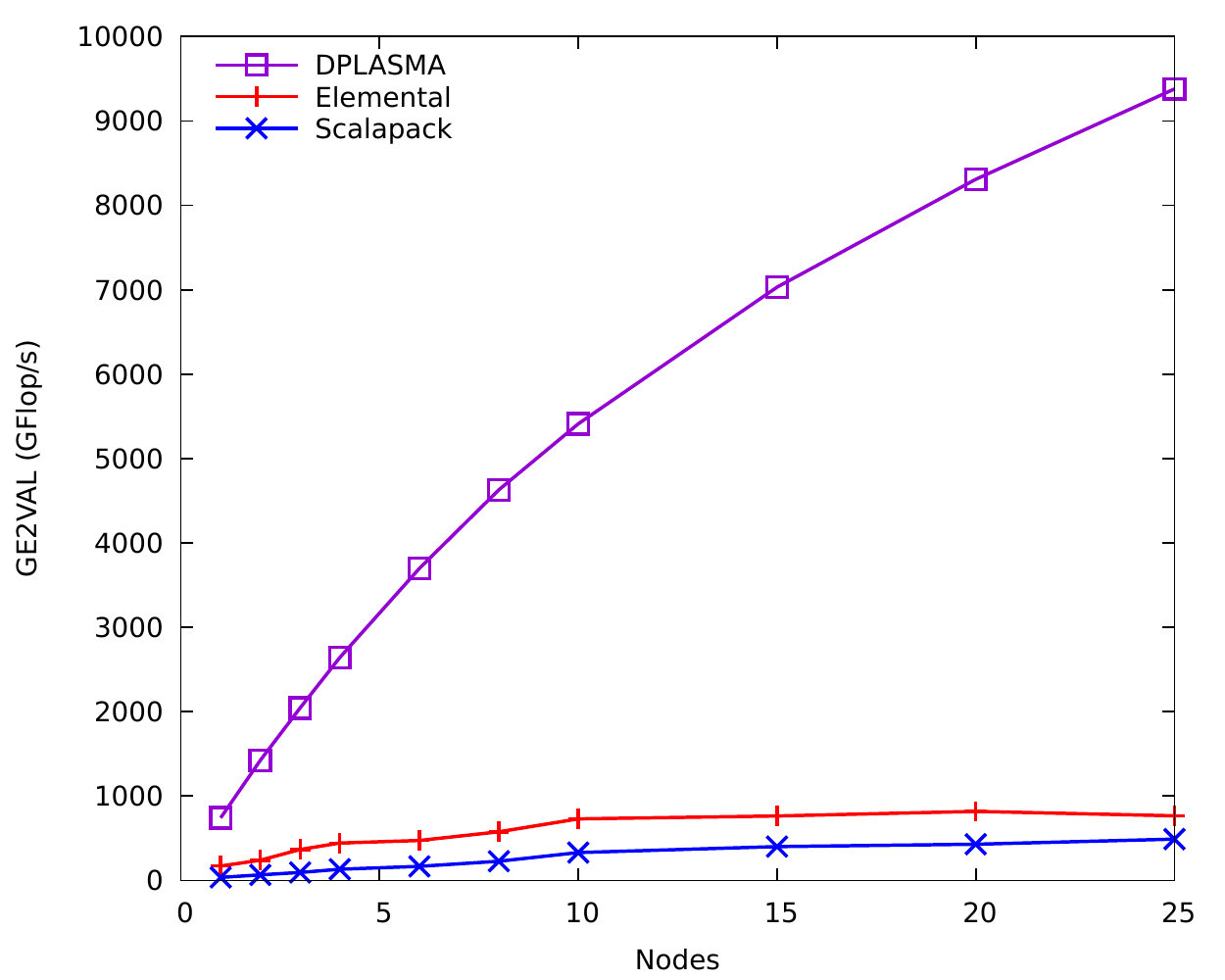}
    %% \caption{Tall and skinny ($N=2000$)}
    %% \label{fig:dist-gesvd-n2k}
  \end{subfigure}
  \begin{subfigure}{0.32\textwidth}
    \includegraphics[width=\textwidth]{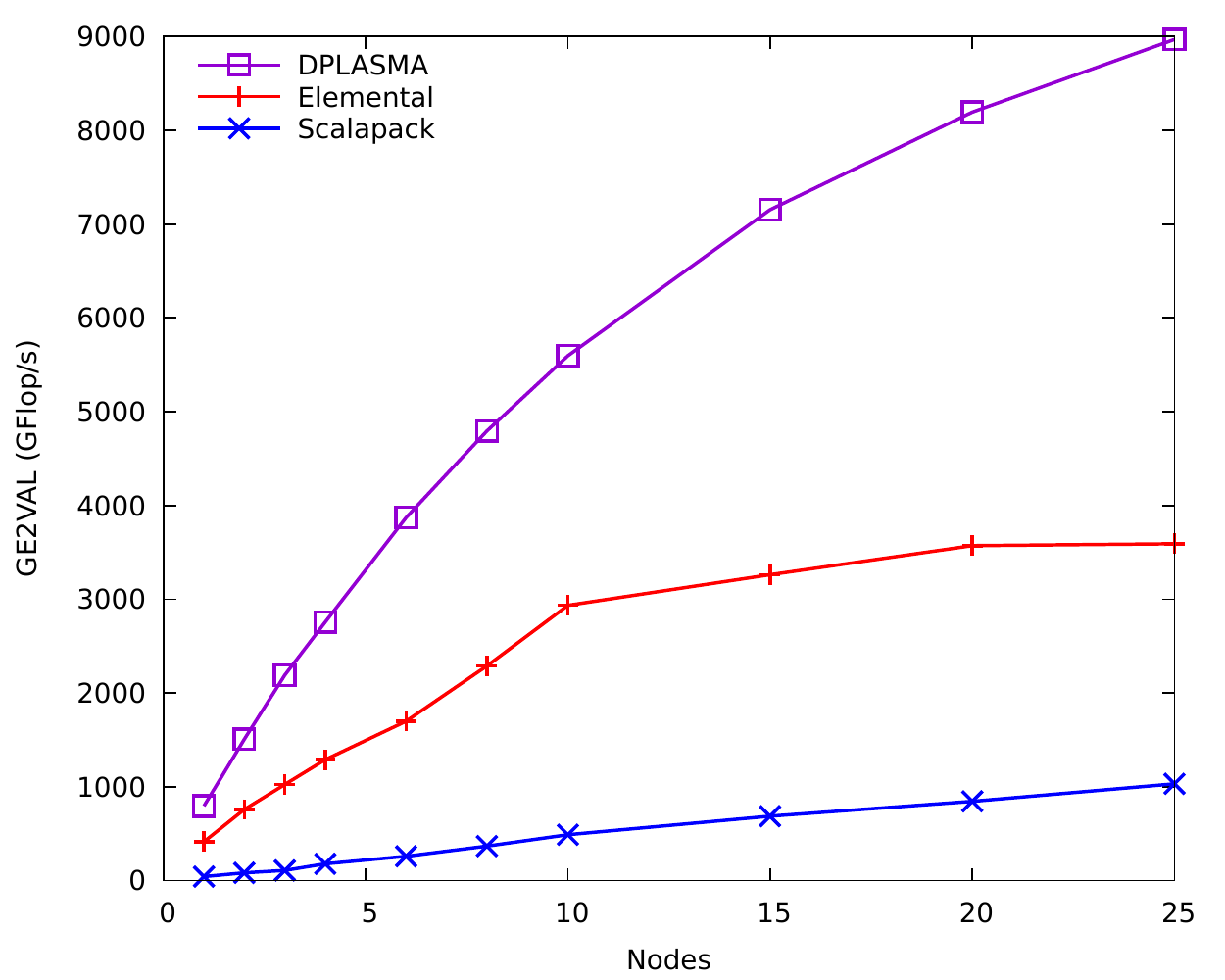}
    %% \caption{Tall and skinny ($N=10000$)}
    %% \label{fig:dist-gesvd-n10k}
  \end{subfigure}

  \begin{subfigure}{0.32\textwidth}
    \includegraphics[width=\textwidth]{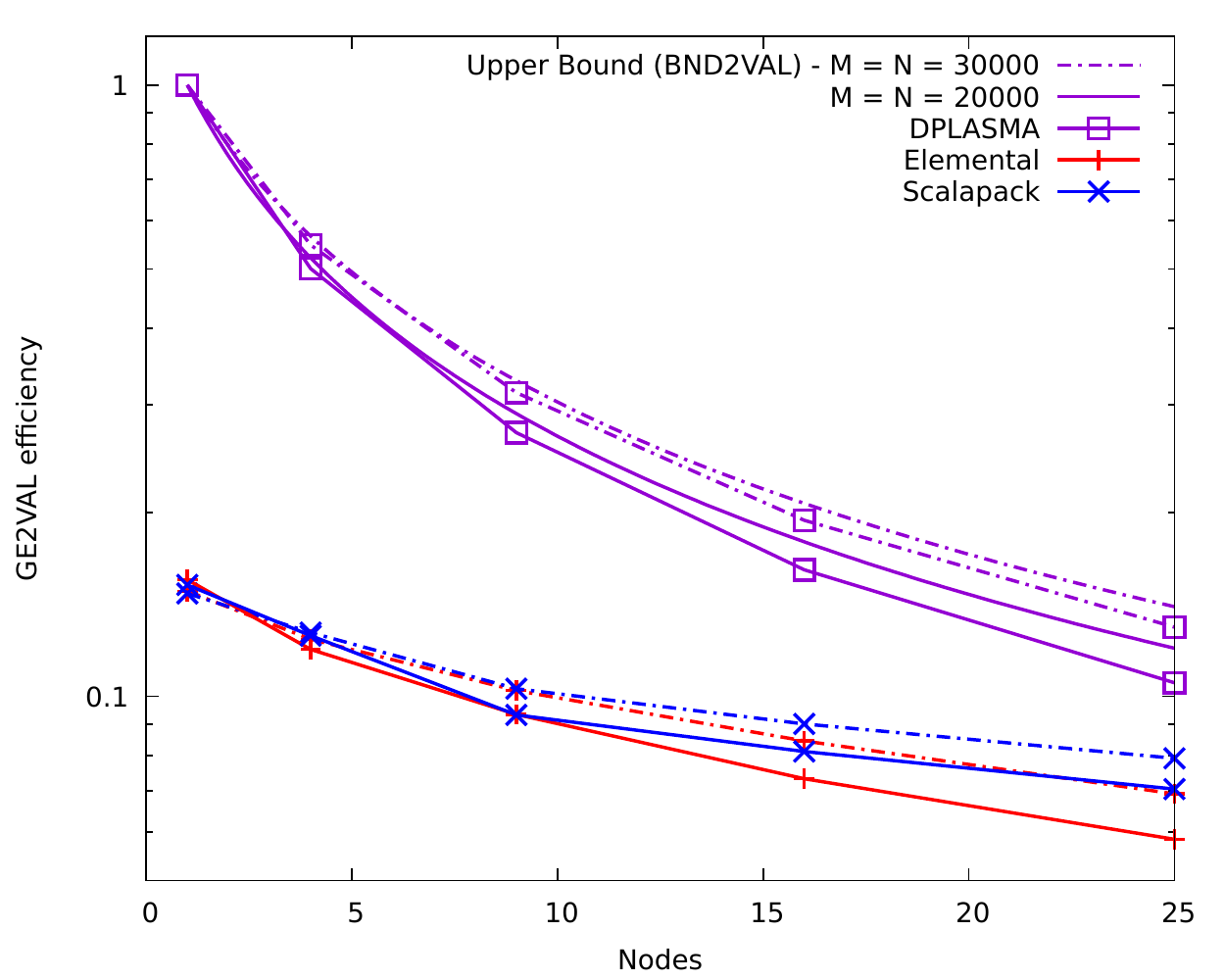}
    \caption*{Square ($m=n$)}
    %% \label{fig:dist-gesvd-square}
  \end{subfigure}
  \begin{subfigure}{0.32\textwidth}
    \includegraphics[width=\textwidth]{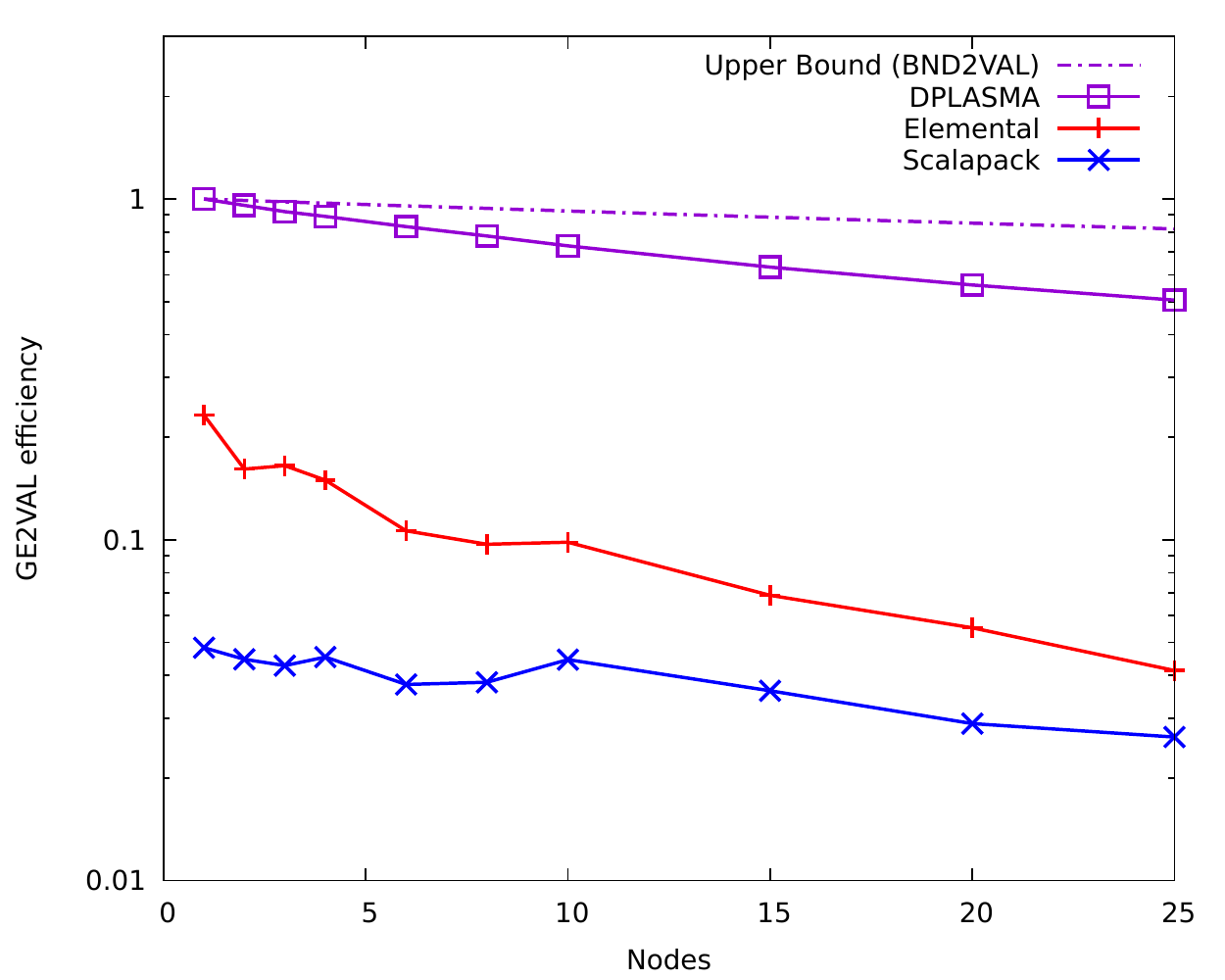}
    \caption*{Tall and skinny \footnotesize{($m\!=\!2,000,000, n\!=\!2,000$)}}
    %% \label{fig:dist-gesvd-n2k}
  \end{subfigure}
  \begin{subfigure}{0.32\textwidth}
    \includegraphics[width=\textwidth]{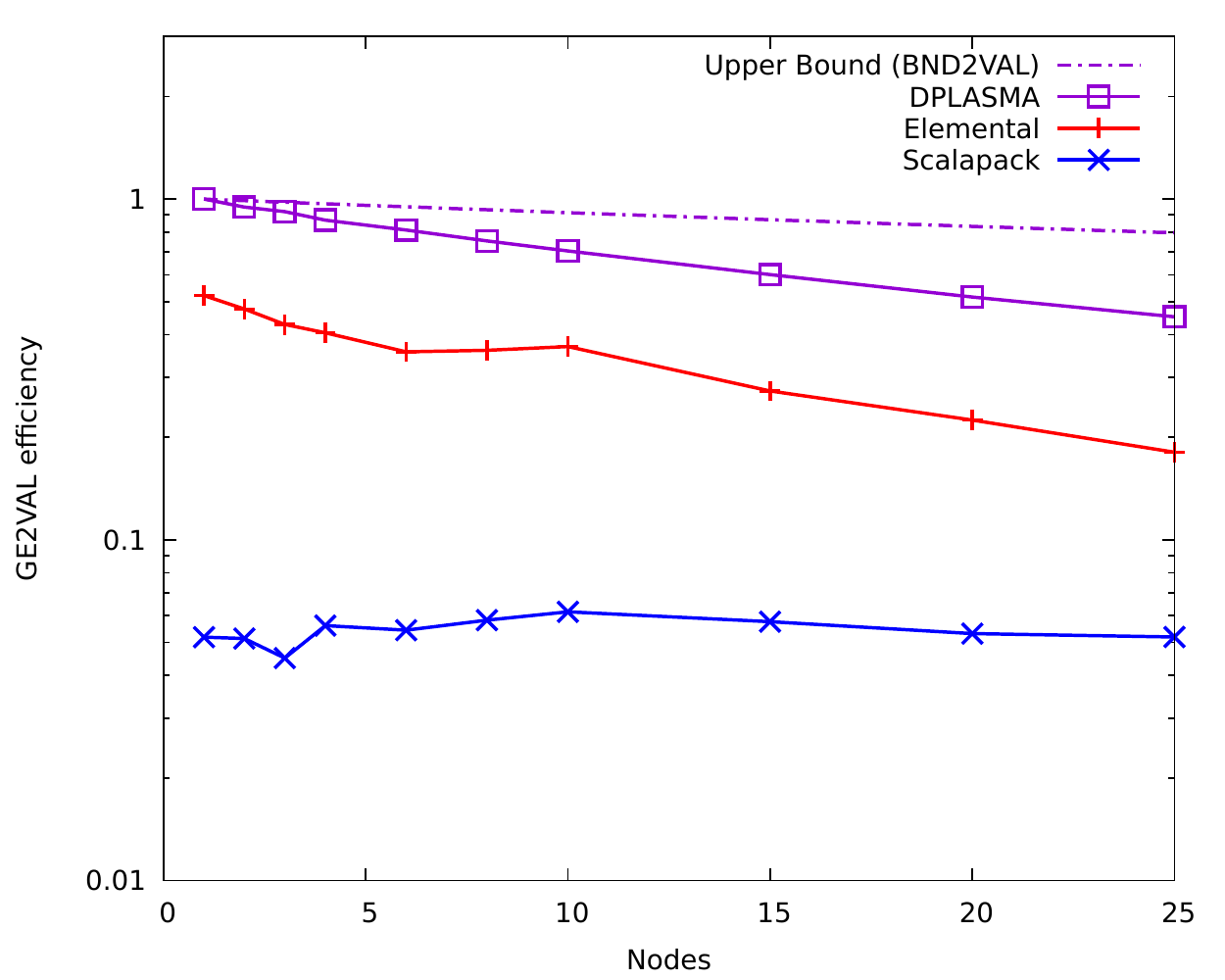}
    \caption*{Tall and skinny \footnotesize{($m\!=\!1,000,000, n\!=\!10,000$)}}   %% \label{fig:dist-gesvd-n10k}
  \end{subfigure}

  \end{center}
  \caption{Distributed memory performance (first row) and efficiency (second row) of the GE2VAL
    algorithm on the \texttt{miriel} cluster. Grid data distributions are
    $\sqrt{\nbnodes} \times \sqrt{\nbnodes}$ for square matrices, and $\nbnodes \times 1$
    for tall and skinny matrices.}
  \label{fig:dist-strong-gesvd}
\end{figure*}

When considering the full GE2VAL algorithm on
Figure~\ref{fig:dist-strong-gesvd}, we observe a huge drop in the overall
performance. This is due to the integration of the shared memory
BND2BD and BD2VAL steps which do not scale when adding more nodes. For the
the square case, we added the upper bound that we cannot beat due to
those two steps. However, despite this limitation, our solution
brings an important speedup to algorithms looking for the
singular values, with respect to the competitors presented here. 
Elemental again benefits from the automatic switch to the
\rbidiag algorithm, which allows a better scaling on tall and
skinny matrices. However, it surprisingly reaches a plateau after $10$
nodes where the performance stops increasing significantly. Our
solution automatically adapts to create more or fewer parallelism, and
reduces the amount of communications, which allows it to sustain a good
speedup up to $25$ nodes ($600$ cores).

\paragraph{Weak Scaling}

\begin{figure*}[t]
  \begin{center}
  \begin{subfigure}{0.3\textwidth}
    \includegraphics[width=\textwidth]{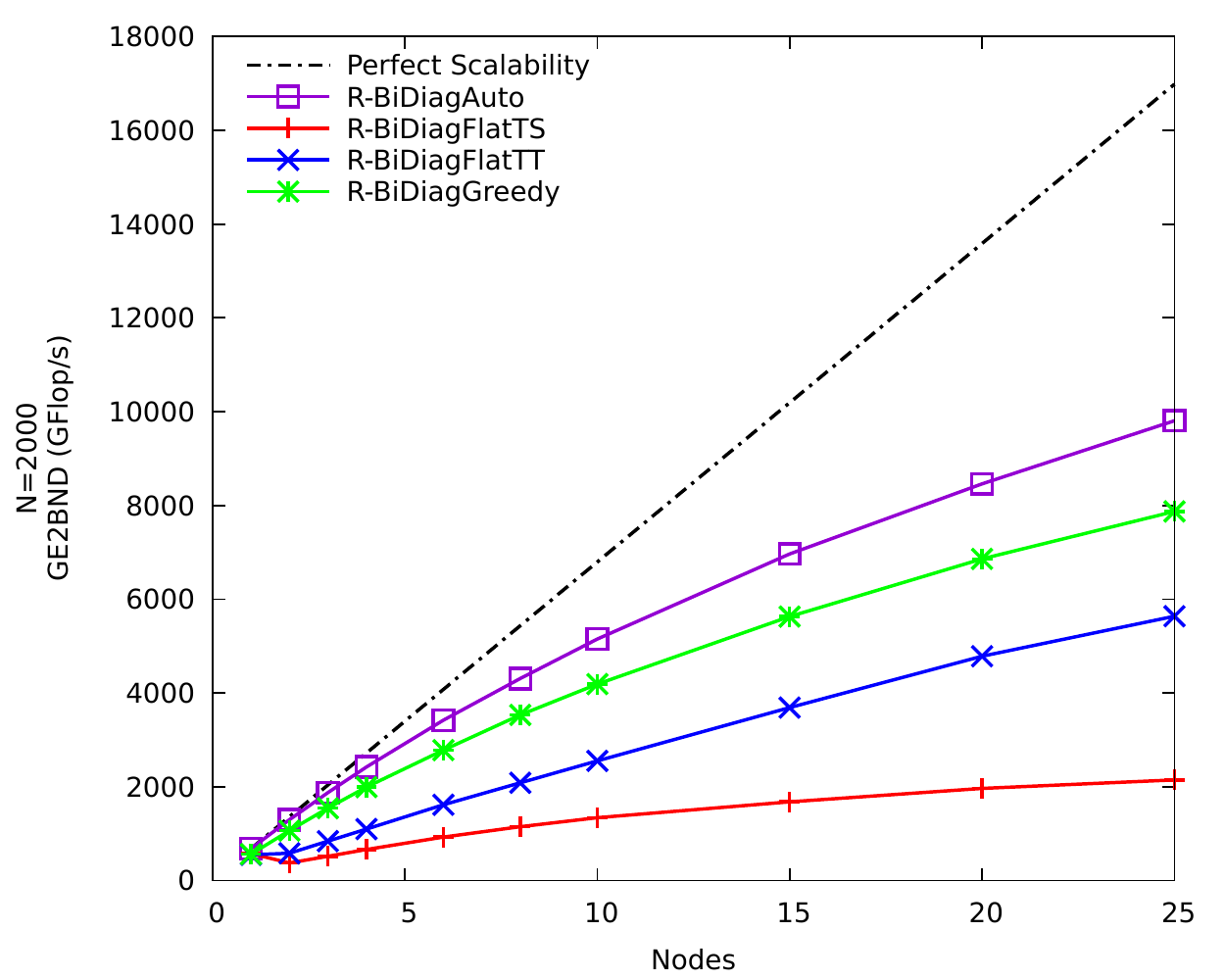}
    %% \caption{Tall and skinny ($N=2000$)}
    %% \label{fig:dist-gebrd-n2k}
  \end{subfigure}
  \begin{subfigure}{0.3\textwidth}
    \includegraphics[width=\textwidth]{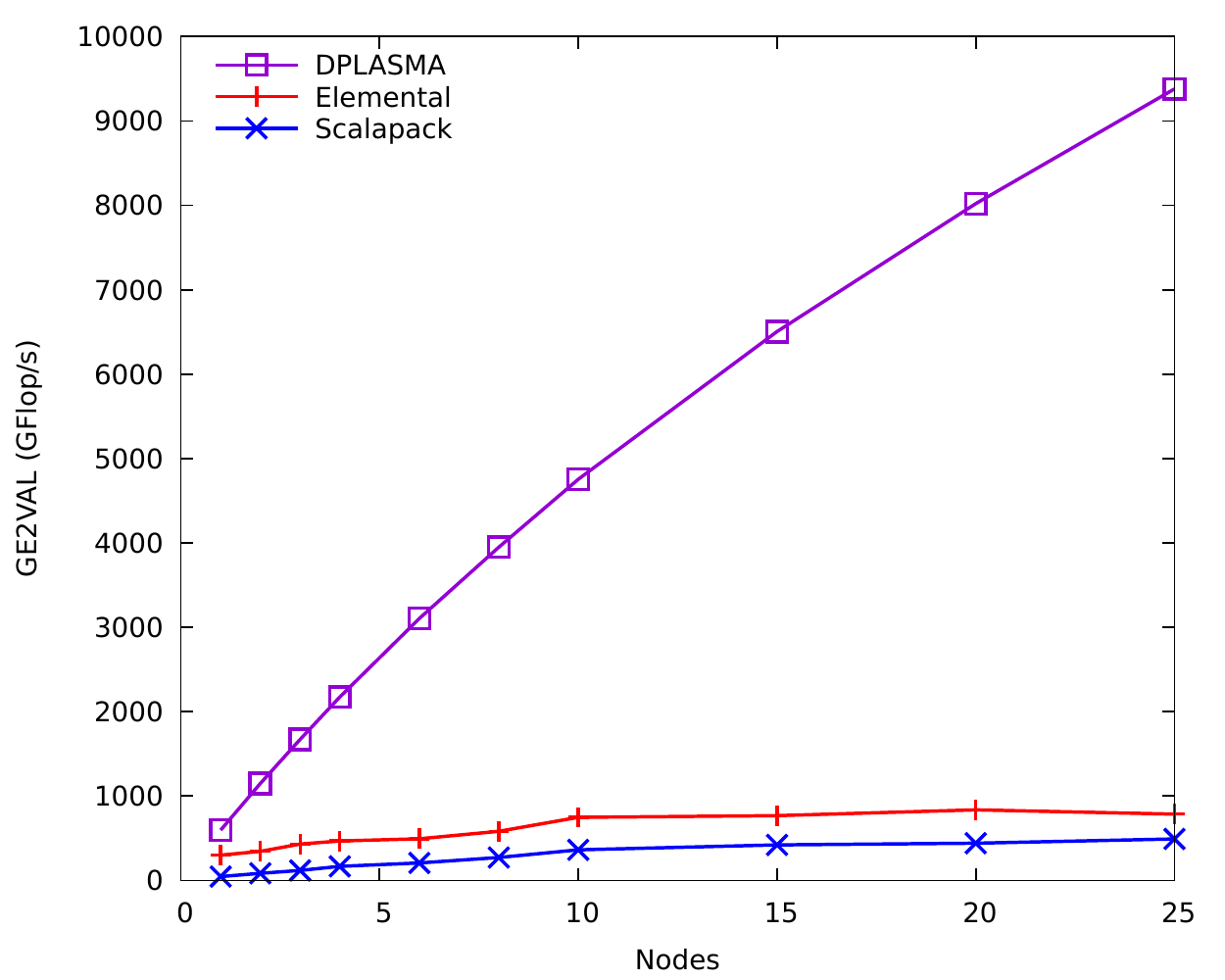}
    %% \caption{Tall and skinny ($N=2000$)}
    %% \label{fig:dist-gesvd-n2k}
  \end{subfigure}
  \begin{subfigure}{0.3\textwidth}
    \includegraphics[width=\textwidth]{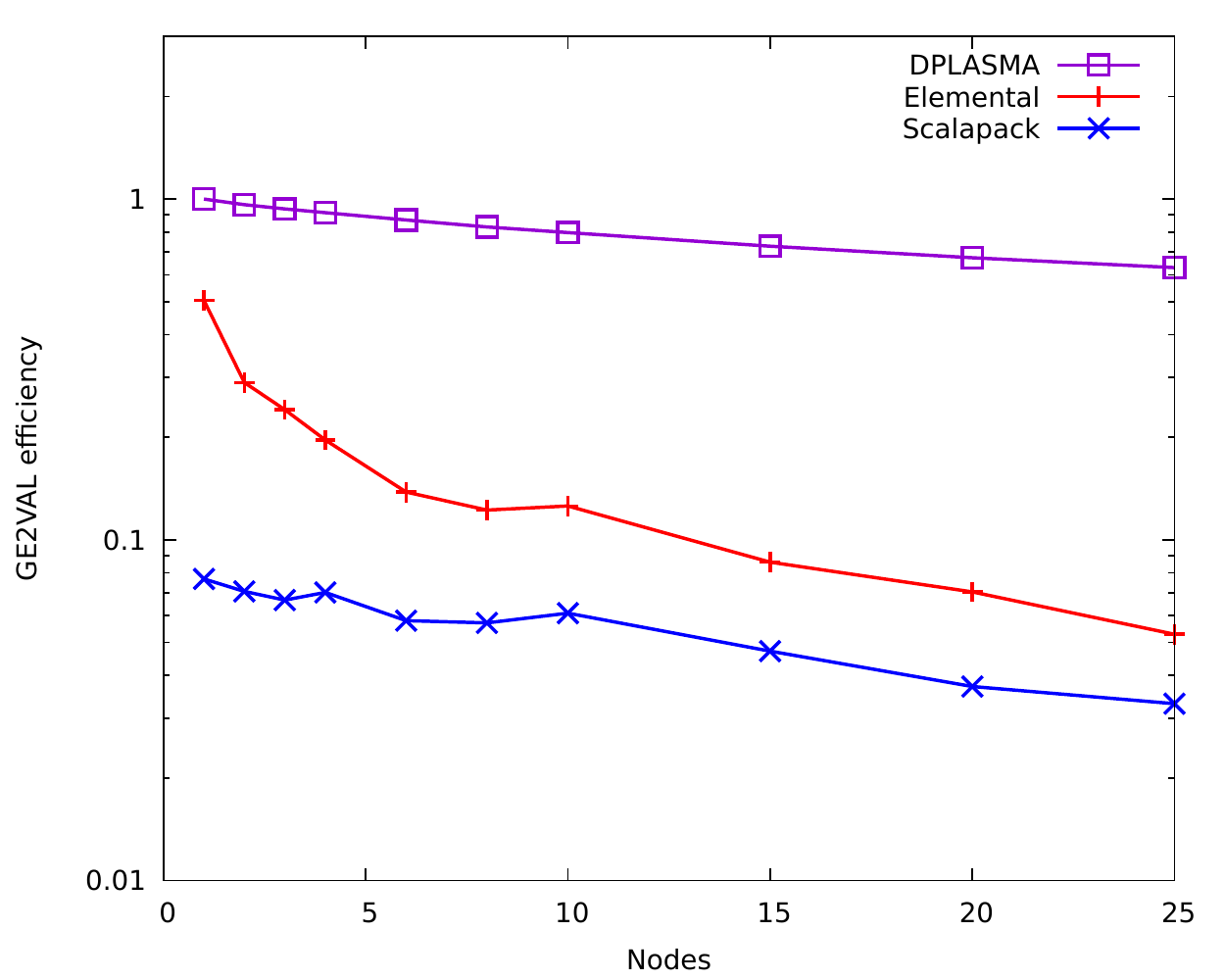}
    %% \caption{Tall and skinny ($N=2000$)}
    %% \label{fig:dist-gesvd-n2k}
  \end{subfigure}

  \begin{subfigure}{0.3\textwidth}
    \includegraphics[width=\textwidth]{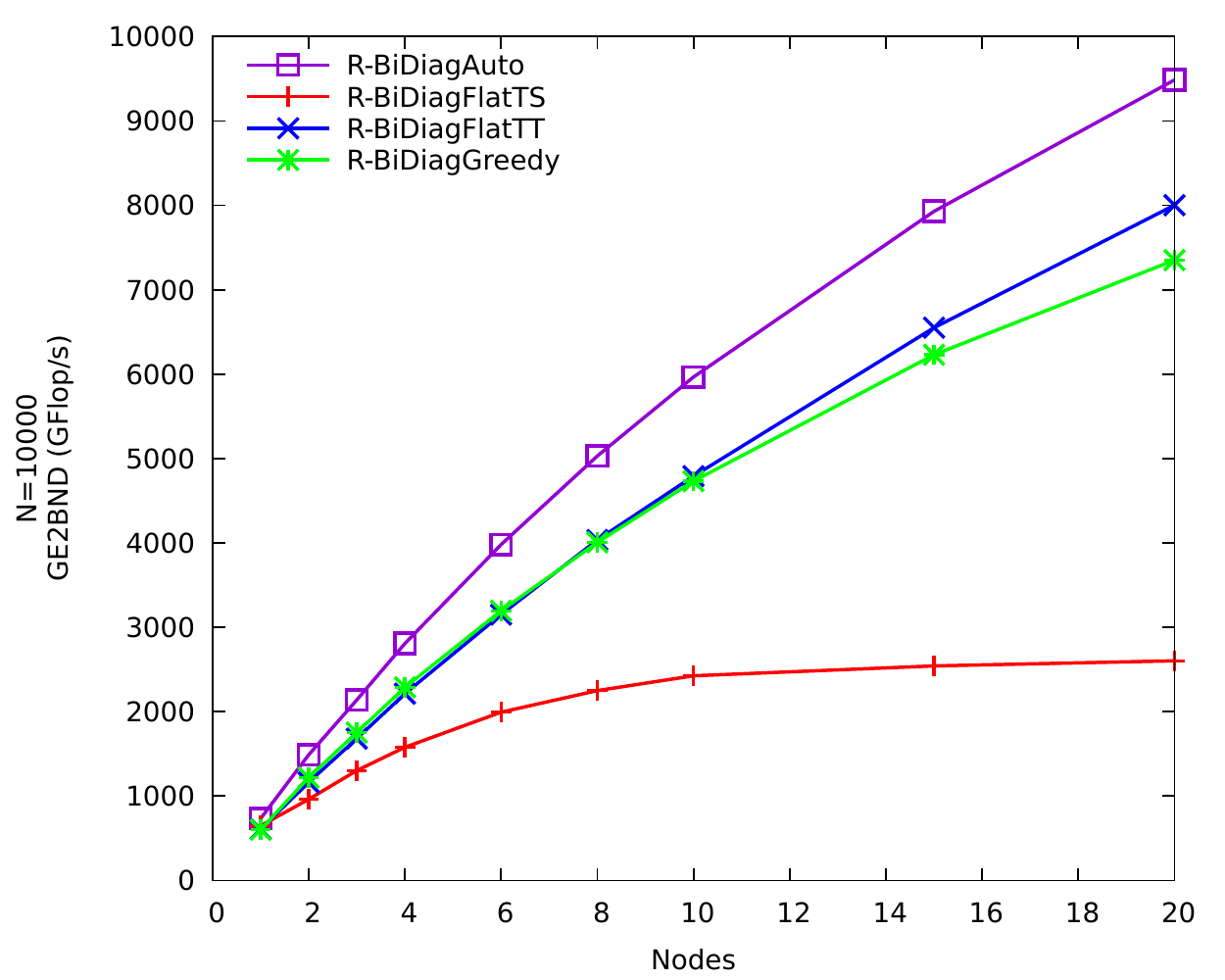}
    %% \caption{Tall and skinny ($N=10000$)}
    %% \label{fig:dist-gebrd-n10k}
  \end{subfigure}
  \begin{subfigure}{0.3\textwidth}
    \includegraphics[width=\textwidth]{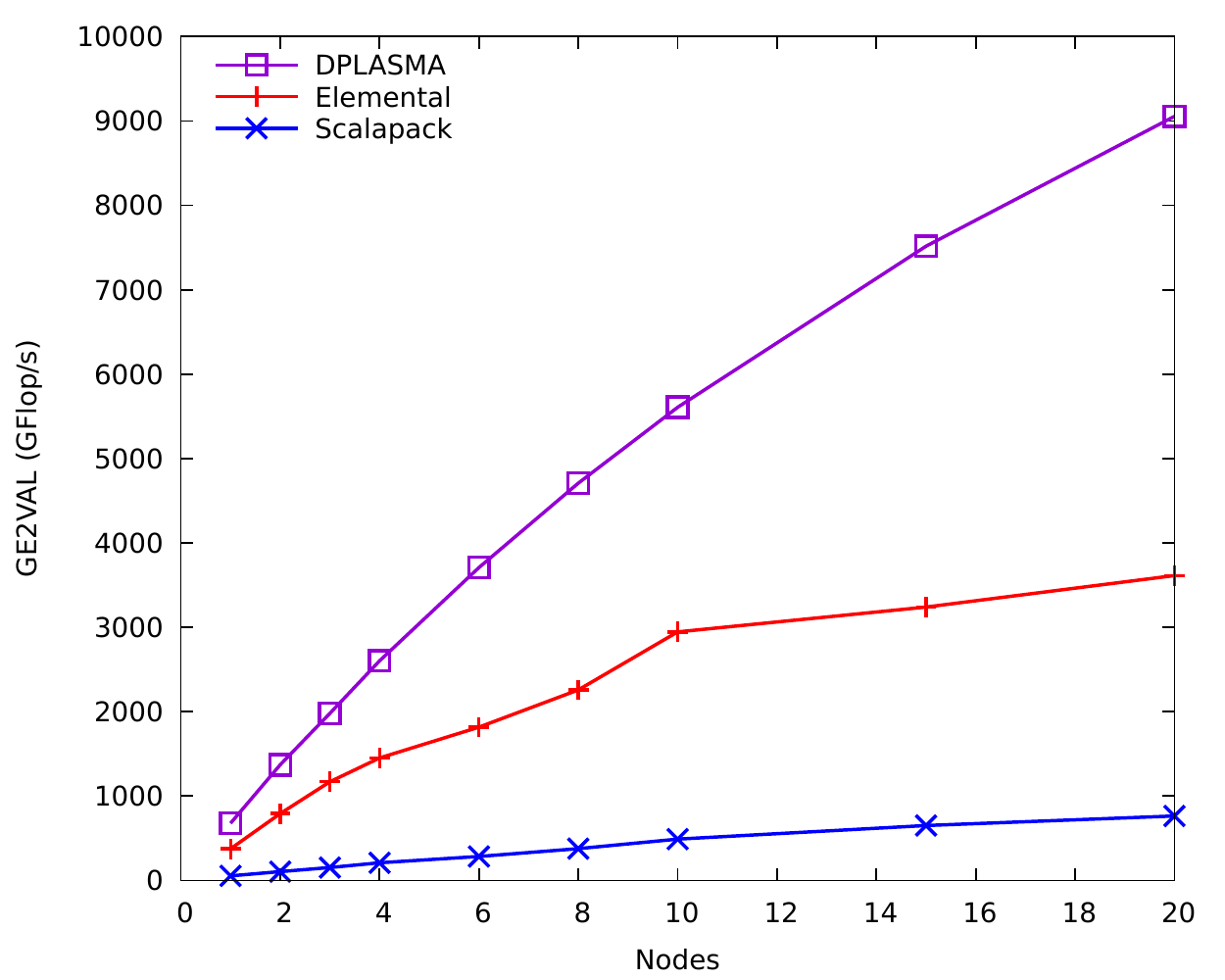}
    %% \caption{Tall and skinny ($N=10000$)}
    %% \label{fig:dist-gesvd-n10k}
  \end{subfigure}
  \begin{subfigure}{0.3\textwidth}
    \includegraphics[width=\textwidth]{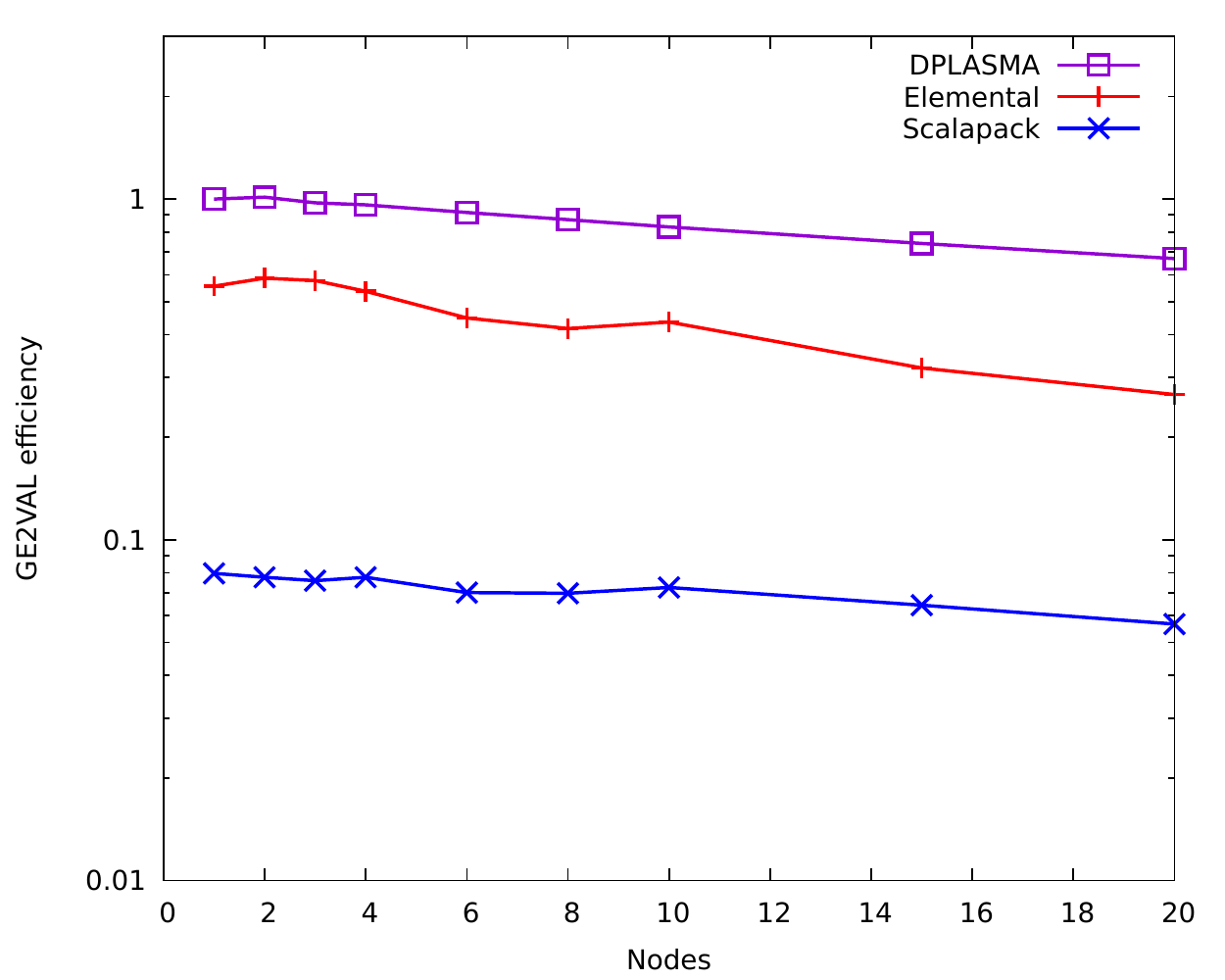}
    %% \caption{Tall and skinny ($N=10000$)}
    %% \label{fig:dist-gesvd-n10k}
  \end{subfigure}

  \end{center}
  \caption{Study of the distributed weak scalability on tall and
    skinny matrices of size $(80,000 \;\nbnodes) \times 2,000$ on the first
    row, and $(100,000 \; \nbnodes) \times 10,000$ on the second row.
    First column presents the GE2BND performance, second column the
    GE2VAL performance, and third column the GE2VAL scaling efficiency.}
  \label{fig:dist-weak}
\end{figure*}

Figure~\ref{fig:dist-weak} presents a weak scalability study with tall
and skinny matrices of width $n=2,000$ on the first row, and $n=10,000$ on
the second row\footnote{Experiments for the $n=10,000$
  case stop at $20$ nodes due to the 32 bit integer default interface
  for all libraries}. As previously, \FlatTS quickly saturates
due to its lack of parallelism. \FlatTT  is able to
compete with, and even to outperform, \Greedy on the larger case due to
its lower communication volume. \Auto offers a better
scaling and is able to reach $10$ TFlop/s which represents $400$ to
$475$ GFlop/s per node. 
When comparing to Elemental and \scalapack on the GE2VAL algorithm,
the proposed solution offers a much better scalability. Both Elemental
and \scalapack suffer from their memory bound \bidiag algorithm.
With the switch to a \rbidiag algorithm, Elemental is able to provide
better performance than \scalapack, but the lack of scalability of the
Elemental QR factorization compared to the HQR implementation quickly
limits the overall performance of the GE2VAL implementation.

\section{Conclusion}
\label{sec.conclusion}

In this paper, we have presented the use of many reduction trees for tiled
bidiagonalization algorithms.  We proved that, during the bidiagonalization
process, the alternating QR and LQ reduction trees cannot overlap. Therefore,
minimizing the time of each individual tree will minimize the overall time.
Consequently,  if one considers an unbounded number of cores and no
communication, one will want to use a succession of greedy trees.  We show that such an
approach (\bidiagGreedy) is asymptotically much better than previously
presented approach (\FlatTS).  In practice, in order to have an effective
solution, one have to take into account load balancing and communication, hence we propose trees
that adapt to the parallel distributed topology (highest level tree) and enable more sequential but
faster kernels on a node (\Auto).

We have also studied R-bidiagonalization in the context of tiled algorithms. While
R-bidiagonalization is not new, it had never been used in the context of tiled
algorithms. Previous work was comparing bidiagonalization and
R-bidiagonalization in term of flops, while our comparison is conducted in term of
critical path lengths. We show that bidiagonalization has
a shorter critical path than R-bidiagonalization, that this is the opposite for tall and skinny matrices,
and provide
an asymptotic analysis.
Along all this work, we give detailed critical path lengths for many of the
algorithms under study.

Our implementation is the first parallel distributed tiled algorithm
implementation for bidiagonalization.
We show the benefit of our approach (DPLASMA) on a multicore node against
existing software (PLASMA, Intel MKL, Elemental and ScaLAPACK) for various
matrix sizes, for computing the singular values of a matrix. We also  show the
benefit of our approach (DPLASMA) on a few multicore nodes against existing
software (Elemental and ScaLAPACK) for various matrix sizes, for computing the
singular values of a matrix. 

Future work will be devoted to gain access to a large distributed platform with
a high count of multicore nodes, and to assess the efficiency and scalability
of our parallel distributed \bidiag and \rbidiag algorithms. Other research
directions are the following: (i) investigate the trade-off of our approach when singular
vectors are requested; a previous study~\cite{Haidar:2013:IPS:2503210.2503292}
in shared memory was conclusive for \FlatTS and no \rbidiag (square matrices
only); the question is to study the problem on parallel distributed platforms, with or without
\rbidiag, for various shapes of matrices and various trees; and (ii) develop a
scalable parallel distributed BND2BD step; for now, for parallel distributed
experiments on many nodes, we are limited in scalability by the BND2BD step,
since it is performed using the shared memory library PLASMA on a single node.

\section*{Acknowledgments}
Work by J. Langou was partially supported by NSF award 1054864 and NSF award 1645514.
Yves Robert is with Institut Universitaire de France.

\bibliographystyle{abbrv}
\bibliography{biblio}

\clearpage
\end{document}